\documentclass[twocolumn,fleqn]{svjour2}    
\smartqed 
\usepackage{amscd}
\usepackage{amsfonts}
\usepackage{amsmath}
\usepackage{amssymb}
\usepackage{color}
\usepackage{graphicx}
\usepackage{graphics}
\usepackage{hyperref}
\usepackage{wrapfig}
\usepackage[T1]{fontenc}
\usepackage[latin1]{inputenc}
\usepackage{revsymb}
\newcommand{\ar}{\rangle}
\newcommand{\al}{\langle}
\newcommand{\n}{\nonumber}

\newcommand{\id}{\openone}
\begin{document}

\title{Heat transport through a quantum Brownian harmonic chain beyond the weak-coupling regime:
An exact treatment\thanks{The author acknowledges financial support
provided by the U.S. Army Research Office (Grant No.
W911NF-13-1-0323).}
}
\titlerunning{Heat transport through a quantum Brownian harmonic chain}
\author{Ilki Kim}
\institute{I. Kim \at
              Center for Energy Research and Technology, North Carolina A$\&$T State University, Greensboro, NC 27411, USA\\
              Tel.: -1-336-285-4972\\
              Fax: -1-336-256-2240\\
              \email{hannibal.ikim@gmail.com}}
\date{\today}
\maketitle
\begin{abstract}
We rigorously consider a linear chain of quantum harmonic
oscillators, in which the number of the individual oscillators is
given by an {\em arbitrary} number $N$, and each oscillator is
coupled at an {\em arbitrary} strength $\kappa$ to its nearest
neighbors (``intra-coupling''), as well as the two end oscillators
of the chain are coupled at an {\em arbitrary} strength $c_{\nu}$ to
two separate baths at {\em arbitrarily} different temperatures,
respectively. We derive an exact closed expression for the
steady-state heat current flowing from a hot bath through the chain
to a cold bath, in the Drude-Ullersma damping model going beyond the
Markovian damping. This allows us to explore the behavior of heat
current relative to the intra-coupling strength as a control
parameter, especially in pursuit of the heat power amplification.
Then it turns out that in the weak-coupling regime ($\kappa, c_{\nu}
\ll 1$), the heat current is small, as expected, and almost
independent of chain length $N$, hence violating Fourier's law of
heat conduction; this is consistent to the earlier results obtained
within the rotating wave approximation for the intra-coupling as
well as in the Born-Markov approximation for the chain-bath
coupling. Beyond the weak-coupling regime, on the other hand, we
typically observe that with increase of the intra-coupling strength
the heat current is gradually amplified, and reaches its maximum
value at some specific coupling strength $\kappa_{\mbox{\tiny R}}$
``resonant'' to a given chain-bath coupling strength. Also, the
behavior of heat current versus chain length appears typically in
such a way that the magnitude of current reaches its maximum with $N
= 1$ and then gradually decreases with increase of the chain length,
being in fact almost $N$-independent in the range of $N$ large
enough. As a result, Fourier's law proves violated also in this
regime.
\end{abstract}

\section{Introduction}\label{sec:introduction}
%
The study of heat transport through small-scale quantum systems has
recently attracted considerable interest due to an increasing demand
for an understanding of the fundamental limit and efficiency of
energy harvesting from a thermal machine at the quantum level
\cite{MAH04}. One of the fundamental physical quantities considered
in this subject is the heat current in the (non-equilibrium) steady
state flowing from a hot bath through the quantum object of interest
to a cold bath.

The steady-state heat flux has been believed, for a long while, to
obey Fourier's law of heat conduction stating that the heat flux is
proportional to the gradient of temperature along its path,
explicitly expressed as ${\mathcal J} = -\kappa_{{\mbox{\tiny
F}}}\,\nabla T$ \cite{GRO84}; here the proportional constant
$\kappa_{{\mbox{\tiny F}}}$ denotes the heat conductivity of the
system in consideration, which is, typically for bulk materials,
independent of the system size $N$ and its shape, so giving rise to
${\mathcal J} \propto 1/N$. In their seminal work, however, Rieder
et al. discovered \cite{RIE67} that the steady-state heat flux
through a one-dimensional classical harmonic chain is given by
${\mathcal J} \propto \Delta T$ and so independent of the chain
length (representing a novel form of energy flow), which accordingly
deviates from Fourier's law. Since then, the validity (or not) of
Fourier's law has come under scrutiny in various classical (e.g.,
\cite{RUB71}-\cite{DHA08}) and quantum systems (e.g.,
\cite{BRI13}-\cite{MAN12}). Over the last few decades, in fact, it
has turned out that Fourier's law may be violated in low-dimensional
lattices whereas there is evidence that Fourier's law is still valid
even for some one-dimensional classical and quantum systems.
Therefore it remains an open question to rigorously determine the
system-size dependence of the heat current.

The rigorous analysis of heat transport through a small-scale
quantum object has been carried out more recently \cite{COM13}. One
of the interesting works is, e.g., the study, given in \cite{GAU07},
of steady-state heat current through a disordered harmonic chain
coupled to two baths at different temperatures, which was discussed
mainly numerically, but giving rise to no clear conclusion regarding
the system-size dependence of the heat current. Next, there was
another interesting analytical treatment of this topic by Asadian
{\em et al.}, given in \cite{MAN12,BRI13}, with a concrete
conclusion that the heat current is independent of the system size
accordingly violating Fourier's law of heat conduction. This was
explicitly discussed in a harmonic chain coupled to baths as well as
in a chain of two-level systems coupled to baths. In this treatment,
they applied the Lindblad master equation formalism (within the
Born-Markov approximation) as well as the rotating wave
approximation neglecting all energy non-conserving terms induced by
the intra-coupling and so considering only the hopping of a single
excitation between two nearest neighboring chain elements. As such,
their analysis is restricted to the weak-coupling regime both in the
chain-bath coupling and in the intra-coupling, as is typically the
case for most studies. Accordingly, no sufficiently large energy
flow can be anticipated to obtain. In addition, they demonstrated
interestingly that Fourier's law can be recovered with chain length
$N \to \infty$ by adding, into the original Lindblad master
equation, a superoperator representing the (phonon-induced)
dephasing appearing in the condensed matter system. Needless to say,
however, this additional dephasing Lindbladian cannot be derived
from the original Hamiltonian describing the coupled chain plus
baths under our current consideration.

Therefore we are now demanded to study the steady-state heat flux
beyond the weak-coupling regime, which has so far remained
extensively unexplored. By looking into this problem, it is possible
to examine behaviors of the heat flow relative to the coupling
strengths as control parameters. This examination could stimulate
the possibilities for increasing the efficiency of energy harvesting
by providing the amplified heat flow followed by some additional
novel quantum control of thermodynamic processes. In fact, the
effects of dissipative environments due to the system-bath coupling,
which are normally negligible in macroscopic systems, become
``detrimental'' to low-dimensional quantum objects, and so the
resultant noise is a major challenging factor to the control of,
e.g., NEMS systems, as well-known \cite{ROU05}. Consequently, this
subject is worthwhile to pay attention to, not only from the
viewpoint of challenge in the quantum statistical physics but also
from the viewpoint of quantum engineering.

In the present paper, we consider a linear chain of quantum harmonic
oscillators coupled at an arbitrary strength to two separate baths
at different temperatures (``quantum Brownian harmonic chain''), in
which each individual chain element is intra-coupled at another
arbitrary strength to its nearest neighbors. We intend to provide an
exact closed expression for the steady-state heat current through
the harmonic chain as our central result [cf. Eqs.
(\ref{eq:heat-current0811})-(\ref{eq:heat-current08120})]. The
treatment of this physical quantity with rigor is mathematically
manageable due to the linear structure of our system. We approach
this open problem by applying the quantum Langevin equation
formalism to the Caldeira-Leggett type Hamiltonian. By doing this,
we can go beyond the aforementioned weak-coupling approach. Our
result may be straightforwardly generalized into a model of heat
transport through a three-dimensional harmonic chain beyond the
weak-coupling regime in case that heat diffusion perpendicular to
the direction of the heat flux is neglected.

The general layout of this paper is the following. In Sect.
\ref{sec:bath_correlation_fkt} we briefly review and refine the
general results regarding the quantum Brownian harmonic chain to be
needed for our discussion and derive an exact closed expression of
the bath correlation function. In Sect. \ref{sec:heat-current1} we
rigorously introduce a formal expression of the steady-state heat
current. In Sect. \ref{sec:heat-current3} we apply this formal
expression to the simplest case of chain length $N=1$ and derive an
exact closed expression of the heat current. This result will be
used as a basis for our discussion of the subsequent cases. In Sect.
\ref{sec:heat-current4} we give an exact expression of the
steady-state heat current for $N=2$. In Sect.
\ref{sec:heat-current5} the same analysis will be carried out for an
arbitrary chain length $N \geq 3$, giving rise to our central
result. Finally we give the concluding remarks of this paper in
Sect. \ref{sec:conclusion}.

\section{Exact expression for the bath correlation function}\label{sec:bath_correlation_fkt}
The linear chain of quantum Brownian oscillators under consideration
is described by the model Hamiltonian of the Caldeira-Leggett type
\cite{WEI08}
\begin{equation}\label{eq:total_hamiltonian1}
    \hat{H}\; =\; \hat{H}_s\, +\, \hat{H}_{b_1-sb_1}\, +\, \hat{H}_{b_{{\scriptscriptstyle N}}-sb_{{\scriptscriptstyle N}}}\,,
\end{equation}
where the isolated chain of $N$ coupled linear oscillators, denoted
by ``system'',
\begin{align}\tag{\ref{eq:total_hamiltonian1}a}\label{eq:chain1}
    \hspace*{-0cm}\hat{H}_s\, =\, \sum_{j=1}^N \left(\frac{\hat{P}_j^2}{2 M} +
    \frac{M}{2}\,\Omega_j^2\,\hat{Q}_j^2\right)\, +\,
    \sum_{j=1}^{N-1} \frac{\kappa_j}{2} \left(\hat{Q}_j - \hat{Q}_{j+1}\right)^2\,,
\end{align}
and two surrounding baths coupled to the first and the last
oscillators of the chain are given by
\begin{align}
    \hspace*{-.9cm}\hat{H}_{b_1-sb_1}\; &=\; \sum_{\nu=1}^{N_b}\,\left\{\frac{\hat{p}_{1,\nu}^2}{2 m_{\nu}} +
    \frac{m_{\nu}}{2} \omega_{\nu}^2 \left(\hat{x}_{1,\nu} -
    \frac{c_{\nu}}{m_{\nu}\,\omega_{\nu}^2}\,\hat{Q}_1\right)^2\right\}\tag{\ref{eq:total_hamiltonian1}b}\label{eq:total_hamiltonian2}\\
    \hspace*{-.9cm}\hat{H}_{b_{{\scriptscriptstyle N}}-sb_{{\scriptscriptstyle N}}}\; &=\;
    \sum_{\nu=1}^{N_b}\,\left\{\frac{\hat{p}_{{\scriptscriptstyle N},\nu}^2}{2 m_{\nu}} +
    \frac{m_{\nu}}{2} \omega_{\nu}^2 \left(\hat{x}_{{\scriptscriptstyle N},\nu} -
    \frac{c_{\nu}}{m_{\nu}\,\omega_{\nu}^2}\,\hat{Q}_{{\scriptscriptstyle N}}\right)^2\right\}\,,\tag{\ref{eq:total_hamiltonian1}c}
\end{align}
respectively. Each of the two baths can split into the isolated bath
and the system-bath coupling such as
\begin{align}
    \hat{H}_{b_{\mu}}\; &=\; \sum_{\nu=1}^{N_b}
    \left(\frac{\hat{p}_{\mu,\nu}^2}{2 m_{\nu}} + \frac{m_{\nu}}{2}
    \omega_{\nu}^2\,\hat{x}_{\mu,\nu}^2\right)\n\\
    \hat{H}_{sb_{\mu}}\; &=\;
    -\hat{Q}_{\mu} \sum_{\nu=1}^{N_b} c_{\nu}\,\hat{x}_{\mu,\nu}\, +\, \hat{Q}_{\mu}^2
    \sum_{\nu=1}^{N_b} \frac{c_{\nu}^2}{2 m_{\nu}\,\omega_{\nu}^2}\,,\tag{\ref{eq:total_hamiltonian1}d}\label{eq:chain11}
\end{align}
where the subscript $\mu = 1, N$. Here the constant $\kappa_j$ is
the (positive-valued) intra-coupling strength between two nearest
neighboring oscillators of the chain with $N \geq 2$, and the set of
constants $\{c_{\nu}\}$ denotes the (positive-valued) coupling
strength between chain and each bath. Then the total system denoted
by $\hat{H}$ is assumed initially in the separable state given by
$\hat{\rho}(0) = \hat{\rho}_s(0) \otimes
\hat{\rho}_{{\scriptscriptstyle \beta_1}} \otimes
\hat{\rho}_{{\scriptscriptstyle \beta_{\scriptscriptstyle N}}}$. The
local density matrix $\hat{\rho}_s(0)$ is an (arbitrary) initial
state of the isolated chain $\hat{H}_s$ only, and the density matrix
$\hat{\rho}_{{\scriptscriptstyle \beta_{\mu}}} = \exp(-\beta_{\mu}
\hat{H}_{b_\mu})/Z_{{\scriptscriptstyle \beta_{\mu}}}$ is the
canonical thermal equilibrium state of the isolated bath
$\hat{H}_{b_{\mu}}$, where $\beta_{\mu} = 1/(k_{\mbox{\tiny B}}
T_{\mu})$ and the partition function $Z_{{\scriptscriptstyle
\beta_{\mu}}}$. Without any loss of generality, the two bath
temperatures are assumed to meet $T_1 \geq T_{{\scriptscriptstyle
N}}$.

We apply the Heisenberg equation of motion to the Hamiltonian given
in (\ref{eq:total_hamiltonian1}), which can straightforwardly give
rise to $\hat{P}_j(t) = M \dot{\hat{Q}}_j(t)$ and then\\the quantum
Langevin equation \cite{GAU07}
\begin{eqnarray}\label{eq:langevin-equation1}
    &&M \ddot{\hat{Q}}_j(t)\, +\, M \int_0^t d\tau\,\gamma(t - \tau)\, \Delta_{jk}\,\dot{\hat{Q}}_k(\tau)\, +\n\\
    &&M C_{jk}\,\hat{Q}_k(t)\; =\; \hat{\xi}_j(t)\,,
\end{eqnarray}
where for the sake of simplicity in form, the Einstein convention is
applied in dealing with the subscripts $j,k = 1, 2, \cdots, N$. Here
the damping kernel and the shifted noise operator (representing a
fluctuating force) are explicitly given by
\begin{align}
    \gamma(t)\; =&\; \frac{1}{M} \sum_{\nu=1}^{N_b}
    \frac{c_{\nu}^2}{m_{\nu}\,\omega_{\nu}^2}\,\cos(\omega_{\nu}\,t)\tag{\ref{eq:langevin-equation1}a}\label{eq:damping_kernel1}\\
    \hat{\xi}_j(t)\; =&\;
    \left\{- M \gamma(t)\,\hat{Q}_1(0)\, +\, \hat{\xi}_{b_1}(t)\right\} \delta_{1j}\, +\n\\
    &\; \left\{- M \gamma(t)\,\hat{Q}_{\scriptscriptstyle N}(0)\, +\, \hat{\xi}_{b_{\scriptscriptstyle N}}(t)\right\}
    \delta_{{\scriptscriptstyle N}j}\,,\tag{\ref{eq:langevin-equation1}b}\label{eq:damping_kernel11}
\end{align}
respectively, where the fluctuating force of either isolated bath
$\mu$,
\begin{equation}\tag{\ref{eq:langevin-equation1}c}\label{eq:noise-op}
    \hat{\xi}_{b_{\mu}}(t)\, =\, \sum_{\nu=1}^{N_b}\,c_{\nu} \left\{\hat{x}_{\mu,\nu}(0)\,\cos(\omega_{\nu}\,t)\,
    +\, \frac{\hat{p}_{\mu,\nu}(0)}{m_{\nu}\,\omega_{\nu}}\,\sin(\omega_{\nu}\,t)\right\}\,.
\end{equation}
It is instructive to note that due to its linearity, the equation of
motion (\ref{eq:langevin-equation1}) can also be understood
classically for the corresponding classical quantities, and so the
Ehrenfest theorem straightforwardly follows for the position or
momentum operator; the quantum behaviors of the second moments
involving the position and momentum operators, such as the
steady-state heat current (to be discussed in the following
sections), are ascribed entirely to the quantum nature of the bath
correlation function to be introduced below.

Then it is easy to verify that the average
value\\$\al\hat{\xi}_{b_{\mu}}(t)\ar_{{\scriptscriptstyle
\beta_{\mu}}} =
\mbox{Tr}\{\hat{\xi}_{b_{\mu}}(t)\,\hat{\rho}_{{\scriptscriptstyle
\beta_{\mu}}}\}$ vanishes at any bath temperature $\beta_{\mu}$, as
required. Next, the diagonal matrix $\Delta_{jk} =
\delta_{jk}\,(\delta_{1k} + \delta_{{\scriptscriptstyle N}k})$
connects the damping kernel to the two end oscillators directly
coupled to the two separate baths. And the tridiagonal matrix of the
isolated chain, $C_{jn} = \{\Omega_n^2 + (\kappa_n +
\kappa_{n-1})/M\}\,\delta_{jn} - (\kappa_n\,\delta_{j-1,n} +
\kappa_{n-1}\,\delta_{j+1,n})/M = C_{nj}$, where we let $\kappa_0 =
\kappa_{\scriptscriptstyle N} \equiv 0$. It is also worthwhile to
point out that this symmetric matrix $\hat{C}$ of real numbers is
positive-definite, which can easily be shown by ${\bf
v}^t\,\hat{C}\,{\bf v}
> 0$ for any non-zero vector ${\bf v}$ of real numbers \cite{MAR88}, thus all
its eigenvalues being positive-valued, as physically required in
(\ref{eq:langevin-equation1}).

\begin{figure}[htb]
\centering\hspace*{-.7cm}{\includegraphics[scale=0.5]{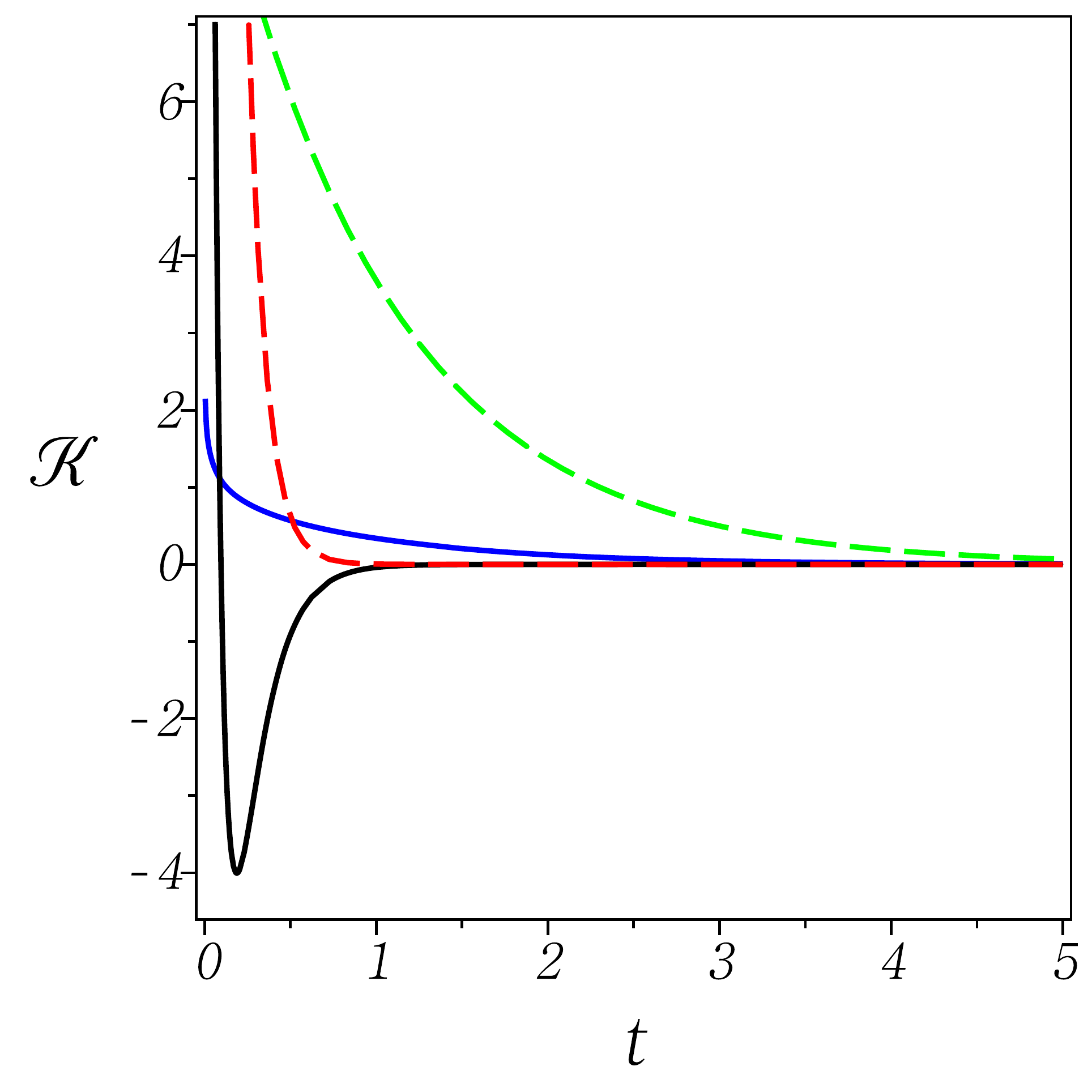}
\caption{(Color online) Bath correlation function ${\mathcal K} =
{\mathcal K}_{\mu}^{(d)}(t)$ versus time $t$, given in
(\ref{eq:identity4}). Here we set $\hbar = k_{\mbox{\tiny B}} = M =
\gamma_{\mbox{\tiny o}} = 1$; solid line plotted at $T_{\mu} = 1$
(low temperature) while dashed line at $T_{\mu} = 10$ (high
temperature). From top to bottom at $x = 0.75$, 1st: (green dash:
$\omega_d = 1$); 2nd: (blue solid: $\omega_d = 1$); 3rd: (red dash:
$\omega_d = 10$); 4th: (black solid: $\omega_d = 10$). For the 1st,
2nd and 3rd lines, we have $\omega_d < \omega_{\mu} = 2
\pi\,T_{\mu}$. For the 4th, on the other hand, we have $\omega_d >
\omega_{\mu}$, which gives rise to the low-temperature behavior of
${\mathcal K}_{\mu}^{(d)}(t)$ characterized by appearance of its
negative-valued region; all four lines diverge at $t = 0$ due to
their behavior being proportional to $\delta(t)$ with $t \to 0$,
directly obtained from
(\ref{eq:correlation-function1}).\label{fig:fig1}}}
\end{figure}

\begin{figure}[htb]
\centering\hspace*{-.0cm}{\includegraphics[scale=0.5]{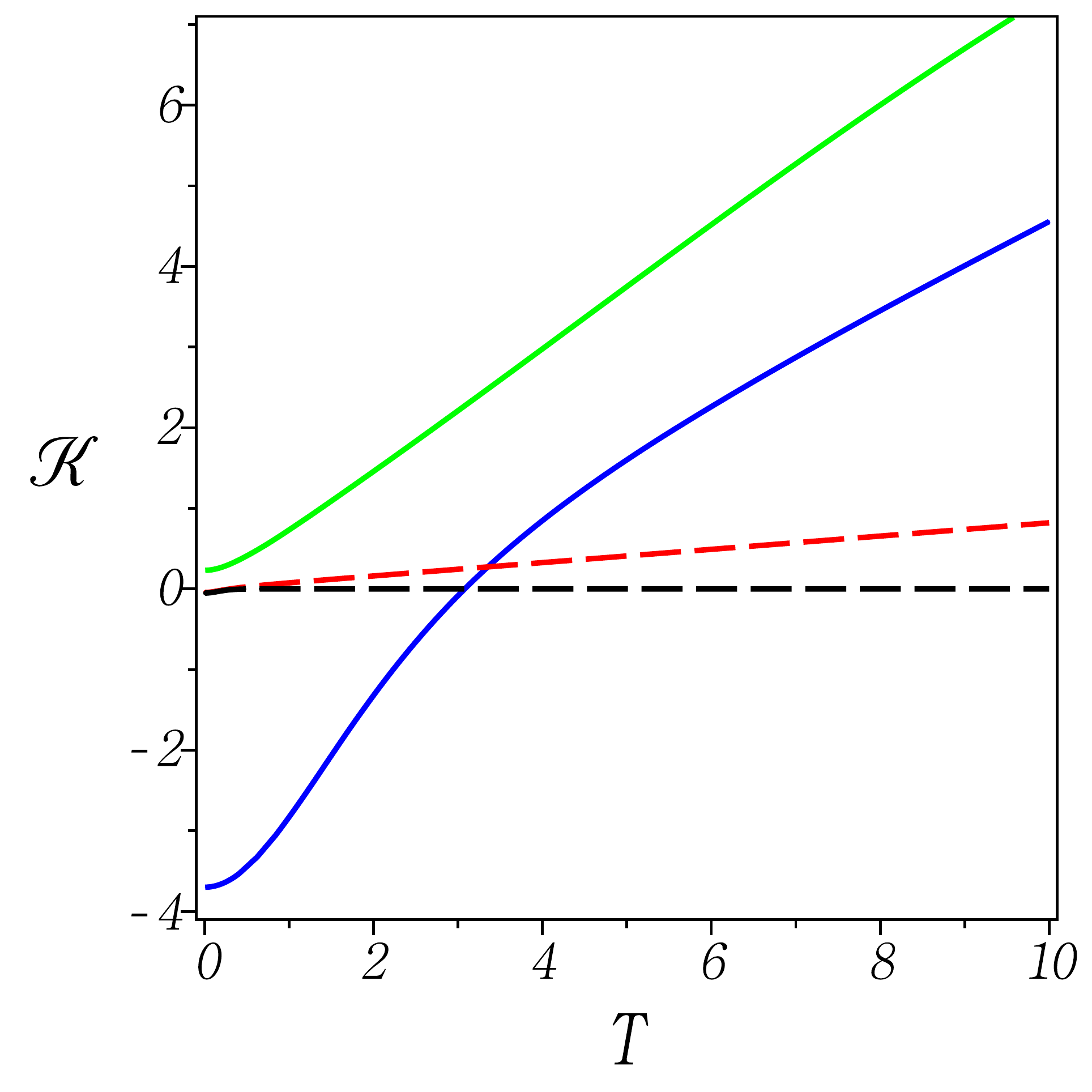}
\caption{(Color online) Bath correlation function ${\mathcal K} =
{\mathcal K}_{\mu}^{(d)}(T)$ versus temperature $T$. Here we set
$\hbar = k_{\mbox{\tiny B}} = M = \gamma_{\mbox{\tiny o}} = 1$;
solid line representing the typical early-time behavior, plotted at
time $t = 0.3$ while dashed line representing the late-time
behavior, plotted at $t = 2.5$. From top to bottom at $x = 5$, 1st:
(green solid: $\omega_d = 1$); 2nd: (blue solid: $\omega_d = 10$);
3rd: (red dash: $\omega_d = 1$); 4th: (black dash: $\omega_d = 10$).
As demonstrated, the correlation function has no singularities at
$\omega_d/\omega_{\mu} = 1, 2, \cdots$, where $\omega_{\mu} = 2 \pi
x$.\label{fig:fig2}}}
\end{figure}

For a physically realistic type of the damping kernel $\gamma(t)$,
we employ the form $\gamma_d(t) = \gamma_{\mbox{\tiny
o}}\,\omega_d\,e^{-\omega_d\,t}$ of the well-known Drude-Ullersma
model, where a cut-off frequency $\omega_d$ and a damping parameter
$\gamma_{\mbox{\tiny o}}$ \cite{ULL66}. In this model, the bath
correlation function in symmetrized form, defined as $K_{\mu}(t-t')
:=
\al\{\hat{\xi}_{b_{\mu}}(t)\,,\,\hat{\xi}_{b_{\mu}}(t')\}_+\ar_{{\scriptscriptstyle
\beta_{\mu}}}/2$ in terms of the anti-commutator $\{\,,\,\}_+$,
reduces to \cite{ING02}
\begin{eqnarray}\label{eq:correlation-function1}
    K_{\mu}^{(d)}(t-t') &=& \frac{M \gamma_{\mbox{\tiny o}}\,\omega_d^2}{\pi}
    \int_0^{\infty} d\omega\, \frac{\hbar \omega}{\omega^2 +
    \omega_d^2}\,\times\n\\
    && \coth\left(\frac{\beta_{\mu} \hbar \omega}{2}\right)\, \cos\{\omega (t-t')\}\,.
\end{eqnarray}
After some algebraic manipulations, every single step of which is
provided in detail in Appendix \ref{sec:appendix1}, we can derive an
exact expression of the correlation function $K_{\mu}^{(d)}(t-t')$,
given by
\begin{align}
    \hspace*{-.0cm}&\frac{\hbar \omega_d^2\,M
    \gamma_{\mbox{\tiny o}}}{2\pi}\,\left\{\pi\,\cot\left(\frac{\omega_d}{\omega_{\mu}}\,\pi\right)\,
    e^{-\omega_d\,|t-t'|}\, +\right.\n\\
    \hspace*{-.0cm}&\left.\Phi\left(e^{-\omega_{\mu}\,|t-t'|}, 1, \frac{\omega_d}{\omega_{\mu}}\right)\,
    +\, \Phi\left(e^{-\omega_{\mu}\,|t-t'|}, 1,
    -\frac{\omega_d}{\omega_{\mu}}\right)\right\}\,,\tag{\ref{eq:correlation-function1}a}\label{eq:identity4}
\end{align}
where we introduce the effective frequency\\$\omega_{\mu} =
2\pi\,k_{\mbox{\tiny B}} T_{\mu}/\hbar$ as well as the Lerch
function $\Phi(z,1,v) = v^{-1}\,{}_{2}\hspace*{-.04cm}F_1(1, v; 1+v;
z)$ [cf. (\ref{eq:lerch-fkt1})]. Here the singularities of $\cot(\pi
y)$ at $y = 1, 2, \cdots$ disappear due to their cancellation with
those of the two Lerch functions at the same points; behaviors of
$K_{\mu}^{(d)}(t)$ versus time and temperature are plotted in Figs.
\ref{fig:fig1} and \ref{fig:fig2}, respectively. From this closed
expression, we can straightforwardly recover, in the limit of
$\omega_{\mu} \to \infty$, its classical counterpart given by
$K_{\mbox{\tiny cl}}^{(d)}(t-t') = k_{\mbox{\tiny B}} T_{\mu}
M\,\gamma_d(t-t')$, being well-known. In addition, by letting the
cut-off frequency $\omega_d \to \infty$ corresponding to the Ohmic
(or Markovian) damping $\gamma_{\mbox{\tiny o}}(t) = 2
\gamma_{\mbox{\tiny o}}\,\delta(t)$, the classical white-noise
correlation function $K_{\mbox{\tiny cl}}^{({\mbox{\tiny
o}})}(t-t')$ immediately follows, too. In fact, Eq.
(\ref{eq:identity4}) will play critical roles in deriving analytical
expressions of the steady-state heat current in Sects.
\ref{sec:heat-current3}-\ref{sec:heat-current5}.

\section{Introduction of the steady-state heat current}\label{sec:heat-current1}
%
We consider the average heat current flowing from a hot bath at
temperature $T_1$ through the coupled harmonic chain to a cold bath
at $T_{\scriptscriptstyle N}$ in the steady state
$\hat{\rho}^{({\mbox{\tiny ss}})}(t)$, which is given by
$\lim_{t\to\infty} \hat{\rho}(t)$ in the Schr\"{o}dinger picture.
Accordingly, the steady-state energy expectation value
$\al\hat{H}_s\ar_{\hat{\rho}^{({\mbox{\tiny ss}})}}$ of the harmonic
chain is required to remain unchanged with the time,
\begin{equation}\label{eq:steady-state1}
    \frac{d}{dt} \left\al\hat{H}_s\right\ar_{\hat{\rho}^{({\mbox{\tiny ss}})}}\, =\,
    \mbox{Tr}\left\{\hat{H}_s\,\frac{d}{dt} \hat{\rho}^{({\mbox{\tiny ss}})}\right\}\, =\, 0\,.
\end{equation}
By substituting this into the Liouville equation
$d\hat{\rho}^{({\mbox{\tiny ss}})}/dt = [\hat{H},
\hat{\rho}^{({\mbox{\tiny ss}})}]/i\hbar$ and then applying the
cyclic invariance of the trace, we can easily arrive at the
expression
\begin{equation}\label{eq:steady-heat-current1}
    \mbox{Tr}\left\{\hat{J}_{{\mbox{\tiny in}}}^{({\scriptscriptstyle N})}\,\hat{\rho}^{({\mbox{\tiny ss}})}\right\}\,
    =\, -\mbox{Tr}\left\{\hat{J}_{{\mbox{\tiny out}}}^{({\scriptscriptstyle N})}\,\hat{\rho}^{({\mbox{\tiny ss}})}\right\}\,,
\end{equation}
where we have two energy current operators denoted by
$\hat{J}_{{\mbox{\tiny in}}}^{({\scriptscriptstyle N})} =
[\hat{H}_s, \hat{H}_{b_1-sb_1}]/i\hbar$ and $\hat{J}_{{\mbox{\tiny
out}}}^{({\scriptscriptstyle N})} = [\hat{H}_s,
\hat{H}_{b_{\scriptscriptstyle N}-sb_{\scriptscriptstyle
N}}]/i\hbar$. We identify the left-hand side of
(\ref{eq:steady-heat-current1}) as the steady-state input heat
current ${\mathcal J}_{{\mbox{\tiny in}}}^{({\scriptscriptstyle
N})}$ and the right-hand side as the output heat current $-{\mathcal
J}_{{\mbox{\tiny out}}}^{({\scriptscriptstyle N})}$; due to the fact
that $T_1 \geq T_{{\scriptscriptstyle N}}$, we assume that
${\mathcal J}_{{\mbox{\tiny in}}}^{({\scriptscriptstyle N})} \geq
0$. Then both heat currents ${\mathcal J}_{{\mbox{\tiny
in}}}^{({\scriptscriptstyle N})}$ and $-{\mathcal J}_{{\mbox{\tiny
out}}}^{({\scriptscriptstyle N})}$, by construction, correspond to
the input power from bath $1$ into the harmonic chain and the output
power from the chain to bath $N$, respectively.

Now let us find an explicit expression of the steady-state heat
current ${\mathcal J}_{{\mbox{\tiny in}}}^{({\scriptscriptstyle
N})}$. We first substitute
(\ref{eq:chain1})-(\ref{eq:total_hamiltonian2}) and
(\ref{eq:damping_kernel1}) into (\ref{eq:steady-heat-current1}) and
then rewrite the operator $\hat{P}_1 \otimes \hat{x}_{1,\nu}$ as
$\{\hat{P}_1\,,\,\hat{x}_{1,\nu}\}_+/2$, which will give rise to
\begin{equation}\label{eq:commutator1}
    \hat{J}_{{\mbox{\tiny in}}}^{({\scriptscriptstyle N})}\, =\,
    \frac{1}{2M} \left\{\hat{P}_1\,,\,-M \gamma(0)\,\hat{Q}_1 + \sum_{\nu}
    c_{\nu}\,\hat{x}_{1,\nu}\right\}_+\,.
\end{equation}
From the Langevin equation given in (\ref{eq:langevin-equation1})
with its damping term rewritten by integration by parts, we can also
find that for a single Brownian oscillator (i.e., with chain length
$N=1$) coupled directly to both separate baths,
\begin{eqnarray}\label{eq:bath-x-term2}
    \sum_{\nu} c_{\nu}\,\hat{x}_{\mu,\nu}(t) &=&
    \frac{M}{2}\,\ddot{\hat{Q}}_1(t) + \frac{M}{2}\,\left\{\Omega_1^2 + 2\,\gamma(0)\right\}\,\hat{Q}_1(t) +\n\\
    && \hat{\xi}_{b_{\mu}}(t) -
    \frac{1}{2}\,\left\{\hat{\xi}_{b_1}(t) + \hat{\xi}_{b_{1'}}(t)\right\}\,,
\end{eqnarray}
where $\mu = 1, 1'$ (i.e., $1' \leftarrow N$), while for $N \geq 2$,
\begin{align}
    &\sum c_{\nu}\,\hat{x}_{1,\nu}(t)\, =\, M\,\ddot{\hat{Q}}_1(t)\, +\n\\
    &\left\{M \Omega_1^2 + \kappa_1 + M \gamma(0)\right\}\,\hat{Q}_1(t) - \kappa_1\,\hat{Q}_2(t)\,.\tag{\ref{eq:bath-x-term2}a}\label{eq:hilfsequation_5}
\end{align}
Next we substitute (\ref{eq:bath-x-term2}) into
(\ref{eq:commutator1}) and then into
(\ref{eq:steady-heat-current1}), as well as apply the cyclic
invariance of the trace and $\hat{\rho}^{({\mbox{\tiny ss}})} =
\lim_{t\to\infty} \hat{U}(t)\,\hat{\rho}(0)\,\hat{U}^{\dagger}(t)$
where $\hat{U}(t) = \exp(-i\hat{H} t/\hbar)$. This allows us to
switch from the Schr\"{o}dinger picture to the Heisenberg picture.
Then the steady-state heat current turns out to be
\begin{eqnarray}\label{eq:heat-current-case-1-1}
    {\mathcal J}_{{\mbox{\tiny in}}}^{(1)} &=&
    \frac{1}{4M}\,
    \lim_{t\to\infty}\,\mbox{Tr}\left[\left\{\hat{P}_1(t)\,,\,M\,\ddot{\hat{Q}}_1(t)\,+\right.\right.\n\\
    && \left.\left. M\,\Omega_1^2\,\hat{Q}_1(t)\,+\,\hat{\xi}_{b_1}(t)\,-\,\hat{\xi}_{b_{1'}}(t)\right\}_+\,\hat{\rho}(0)\right]
\end{eqnarray}
for $N=1$. In the same way, we can also find the corresponding
expression of ${\mathcal J}_{{\mbox{\tiny out}}}^{(1)}$
independently, shown to be identical to
(\ref{eq:heat-current-case-1-1}) but with exchange of
$\hat{\xi}_{b_1}(t)$ and $\hat{\xi}_{b_{1'}}(t)$. Similarly, Eqs.
(\ref{eq:steady-heat-current1}), (\ref{eq:commutator1}) and
(\ref{eq:hilfsequation_5}) allow us to finally obtain the heat
current
\begin{align}
    \hspace*{-.0cm}{\mathcal J}_{{\mbox{\tiny in}}}^{({\scriptscriptstyle N})}\, =&\, \frac{1}{2M}\,
    \lim_{t\to\infty}\,\mbox{Tr}\left[\left\{\hat{P}_1(t)\,,\,M\,\ddot{\hat{Q}}_1(t)\, +\right.\right.\n\\
    \hspace*{-.0cm}&\, \left.\left.(M\,\Omega_1^2 +
    \kappa_1)\,\hat{Q}_1(t)\,-\,\kappa_1\,\hat{Q}_2(t)\right\}_+\,\hat{\rho}(0)\right]\tag{\ref{eq:heat-current-case-1-1}a}\label{eq:current_for_J-1}
\end{align}
for $N \geq 2$, as well as its counterpart ${\mathcal
J}_{{\mbox{\tiny out}}}^{({\scriptscriptstyle N})}$, being identical
to (\ref{eq:current_for_J-1}) but with substitution of $\kappa_1 \to
\kappa_{{\scriptscriptstyle N}-1}$ and, for all remaining
subscripts, $(1, 2) \to (N, N-1)$. As shown, the key elements to the
steady-state heat current are explicit expressions of
$\{\hat{Q}_1(t), \hat{Q}_2(t), \hat{P}_1(t)\}$ in the limit of $t
\to \infty$. We will below restrict our discussion of these
expressions, for the sake of simplicity, mainly to the case of
$\Omega_1 = \cdots = \Omega_{\scriptscriptstyle N} =: \Omega$ and
$\kappa_1 = \cdots = \kappa_{{\scriptscriptstyle N}-1} =: \kappa$.

To derive an explicit form of each individual oscillator
$\hat{Q}_j(t)$, we directly apply the Laplace transform to the
Langevin equation (\ref{eq:langevin-equation1}). Let its Laplace
transform $\underline{\hat{Q}_j}(s) := {\mathcal
L}\{\hat{Q}_j(t)\}(s)$, then giving rise to ${\mathcal
L}\{\dot{\hat{Q}}_j(t)\}(s) = s\,\underline{\hat{Q}_j}(s) -
\hat{Q}_j(0)$ and ${\mathcal L}\{\ddot{\hat{Q}}_j(t)\}(s) =
s^2\,\underline{\hat{Q}_j}(s) - s\,\hat{Q}_j(0) -
\dot{\hat{Q}}_j(0)$ \cite{ROB66}. Then we can easily obtain
\begin{eqnarray}\label{eq:laplace1}
    {\mathcal B}_{jk}(s)\,\underline{\hat{Q}_k}(s) &=& s\,\hat{Q}_j(0) + \dot{\hat{Q}}_j(0)
    +\n\\
    && \frac{1}{M}\,\left\{\underline{\hat{\xi}_{b_1}}(s)\,\delta_{j1} +
    \underline{\hat{\xi}_{b_2}}(s)\,\delta_{j{\scriptscriptstyle N}}\right\}\,,
\end{eqnarray}
where the Laplace-transformed fluctuating force
\begin{align}\tag{\ref{eq:laplace1}a}\label{eq:fourier-laplace3}
    \underline{\hat{\xi}_{b_{\mu}}}(s)\, =\, \sum_{\nu=1}^{N_b} \frac{c_{\nu}}{s^2 + \omega_{\nu}^2}\,\left\{s\,\hat{x}_{\mu,\nu}(0)
    + \frac{\hat{p}_{\mu,\nu}(0)}{m_{\nu}}\right\}\,,
\end{align}
and the symmetric tridiagonal matrix ${\mathcal B}_{jk}(s) =
s^2\,\delta_{jk} + s\,\bar{\gamma}(s)\,\Delta_{jk} + C_{jk}$
expressed in terms of\\the Laplace-transformed damping kernel
\begin{align}\tag{\ref{eq:laplace1}b}\label{eq:gamma_tilde1}
    \bar{\gamma}(s)\, =\, \frac{s}{M} \sum_{\nu=1}^{N_b}
    \frac{c_{\nu}^2}{m_{\nu}\,\omega_{\nu}^2}\,
    \frac{1}{s^2 + \omega_{\nu}^2}\,.
\end{align}
In the Drude-Ullersma model, the damping kernel $ \bar{\gamma}(s)
\to \bar{\gamma}_d(s) = \gamma_{\mbox{\tiny o}}\,\omega_d/(s +
\omega_d)$ \cite{WEI08}. Therefore, the central task to be
undertaken is the determination of an explicit form of the inverse
matrix $\hat{{\mathcal B}}^{-1}(s) =: \hat{{\mathcal A}}(s)$, which
will be performed below for individual chain lengths $N$.

\section{Steady-state heat current for the case of $N=1$}\label{sec:heat-current3}
%
We begin with the simplest case of $N=1$, in which a single
oscillator is coupled directly to two separate baths at different
temperatures. As well-known, the matrix $\hat{{\mathcal A}}(s)$ then
reduces to $M \bar{\chi}(s)$, where
\begin{equation}
    \bar{\chi}(s)\, =\, \frac{1}{M}\,\frac{1}{s^2\,+\,\Omega^2\,+\,2\,s\,\bar{\gamma}(s)}\,,\label{eq:chi_laplace_1}
\end{equation}
corresponding to the dynamic susceptibility in the frequency domain,
given by $\tilde{\chi}(\omega) \leftarrow \bar{\chi}(s)$ with $s \to
-i \omega + 0^+$ \cite{WEI08}. In the Drude-Ullersma model, Eq.
(\ref{eq:chi_laplace_1}) reduces to
\begin{align}
    \bar{\chi}_d(s)\, =\, \frac{(s + \omega_d)/M}{h_1(s)}\,,\tag{\ref{eq:chi_laplace_1}a}\label{eq:chi_laplace_2}
\end{align}
where $h_1(s) = s^3 + \omega_d\,s^2 + (\Omega^2 +
\gamma_{\mbox{\tiny o}}'\,\omega_d)\,s + \Omega^2\,\omega_d$, with
$\gamma_{\mbox{\tiny o}}' = 2 \gamma_{\mbox{\tiny o}}$ and all its
coefficients being positive-valued. Accordingly, this cubic
polynomial can be factorized as $(s + z_0) (s + z_1) (s + z_2)$,
where $\mbox{Re}(z_0), \mbox{Re}(z_1), \mbox{Re}(z_2) > 0$, through
the symmetric relations
\begin{eqnarray}\label{eq:susceptibility-2}
    \hspace*{-.0cm}z_0 + z_1 + z_2 &=& \omega_d\; ,\; \Omega^2 +
    \gamma_{\mbox{\tiny o}}'\,\omega_d\,=\,z_0\,(z_1 + z_2) +
    z_1\,z_2\; ,\n\\
    \hspace*{-.0cm}\Omega^2\,\omega_d &=& z_0\,z_1\,z_2\,;
\end{eqnarray}
$(z_0, z_1, z_2)$ can equivalently be rewritten as $({\mathbf w}_0,
z_0, \gamma)$, where \cite{FOR06}
\begin{align}
    \Omega^2\, =&\, \left({\mathbf w}_0\right)^2\; \frac{z_0}{z_0\, +\, \gamma}\;\;\; ,\;\;\; \omega_d\, =\, z_0\, +\, \gamma\;\;\;
    ,\n\\
    &\, \gamma_{\mbox{\tiny o}}'\, =\, \gamma\, \frac{z_0\, (z_0\, +\, \gamma)\,
    +\, \left({\mathbf w}_0\right)^2}{(z_0\, +\, \gamma)^2}\,;\tag{\ref{eq:susceptibility-2}a}\label{eq:parameter_change0}
\end{align}
then these lead to $z_1 = \gamma/2 + i {\mathbf w}_1$ and $z_2 =
\gamma/2 - i {\mathbf w}_1$ with ${\mathbf w}_1 = \sqrt{({\mathbf
w}_0)^2 - (\gamma/2)^2}$. The parameters $(z_0, z_1, z_2)$ will be
useful for a compact expression of the steady-state heat current
${\mathcal J}_{{\mbox{\tiny in}}}^{(1)}$ [cf.
(\ref{eq:heat-current-case-1-10-1})]. In fact, these can be
explicitly expressed in terms of $(\Omega, \omega_d,
\gamma_{\mbox{\tiny o}})$ by the cubic formula, as well-known
\cite{JAP00}.

Now let us find an explicit expression of the single oscillator
$\hat{Q}_1(t)$ in the limit of $t \to \infty$ by considering the
equation of its Laplace transform $\underline{\hat{Q}_1}(s)$ given
in (\ref{eq:laplace1}). This can be efficiently carried out with the
aid of the final value theorem of the Laplace transform
\cite{COH07}; it reads as $\lim_{t \to \infty} f(t) = \lim_{s \to 0}
s F(s)$, where $F(s) = {\mathcal L}\{f(t)\}(s)$, upon condition that
all poles of $F(s)$, except $s=0$, have negative real parts, i.e.,
if $s F(s)$ is analytic on the imaginary axis and in the right
half-plane. By noting from (\ref{eq:chi_laplace_2}) the fact that
this condition is met by $\bar{\chi}_d(s)$ and so $\lim_{s \to 0}
s\,\bar{\chi}_d(s) = \lim_{s \to 0} s^2\,\bar{\chi}_d(s) = 0$, we
can easily obtain from (\ref{eq:laplace1}) the expression
\begin{equation}\label{eq:fourier_laplace2}
    \lim_{t\to\infty} \hat{Q}_1(t)\, =\,
    \lim_{t\to\infty} \int_0^t dt'\,\chi_d(t-t')\,\left\{\hat{\xi}_{b_1}(t') +
    \hat{\xi}_{b_{1'}}(t')\right\}\,,
\end{equation}
where the response function \cite{ILK10}
\begin{align}
    \hspace*{-.0cm}&\chi_d(t)\, =\, {\mathcal L}^{-1}\{\bar{\chi}_d(s)\}\, =\, -\frac{1}{M}\,\times\tag{\ref{eq:fourier_laplace2}a}\label{eq:response_fkt_drude1}\\
    \hspace*{-.0cm}&\frac{\left(z_1^2 - z_2^2\right)\,e^{-z_0\,t}\,+\,\left(z_2^2 -
    z_0^2\right)\,e^{-z_1\,t}\,+\,\left(z_0^2 - z_1^2\right)\,e^{-z_2\,t}}{\left(z_0 -
    z_1\right)\,\left(z_1 - z_2\right)\,\left(z_2 - z_0\right)}\,.\n
\end{align}
For the sake of comparison with (\ref{eq:fourier_laplace2}), it is
also worthwhile to mention that both fluctuating forces
$\underline{\hat{\xi}_{b_1}}(s)$ and
$\underline{\hat{\xi}_{b_{1'}}}(s)$, as shown in
(\ref{eq:fourier-laplace3}), do not meet the prerequisite for
applying the final value theorem, though, and so it turns out that
\begin{align}\tag{\ref{eq:fourier_laplace2}b}\label{eq:fourier_laplace2-1}
    \lim_{t\to\infty}\, \hat{Q}_1(t)\, \ne\,
    \lim_{s\to 0}\, s\,\bar{\chi}_d(s)\,\left\{\underline{\hat{\xi}_{b_1}}(s) +
    \underline{\hat{\xi}_{b_{1'}}}(s)\right\}\,.
\end{align}
Similarly, it appears from $\hat{P}_1(t) = M \dot{\hat{Q}}_1(t)$ and
$\chi_d(0) = 0$ that
\begin{equation}\label{eq:fourier-laplace5}
    \lim_{t\to\infty} \hat{P}_1(t)\, =\,
    M\,\lim_{t\to\infty} \int_0^t dt'\,\left\{\frac{\partial}{\partial t}\chi_d(t-t')\right\}\,\left\{\hat{\xi}_{b_1}(t') +
    \hat{\xi}_{b_{1'}}(t')\right\}\,.
\end{equation}

Now we are ready to derive an explicit expression of the
steady-state current. Substituting (\ref{eq:fourier_laplace2}) and
(\ref{eq:fourier-laplace5}) into (\ref{eq:heat-current-case-1-1}),
we first obtain the steady-state expectation value
\begin{eqnarray}\label{eq:fourier-laplace6}
    \hspace*{-.3cm}&&\left\al\left\{\hat{P}_1(t)\,,\,\hat{Q}_1(t)\right\}_+\right\ar^{({\mbox{\tiny ss}})}\, =\, 2
    M\,\lim_{t\to\infty} \int_0^t d\tau \int_0^t d\tau' \chi_d(t-\tau)\n\\
    \hspace*{-.3cm}&&\times \left\{\frac{\partial}{\partial t}
    \chi_d(t-\tau')\right\}\,
    \left\{K_1^{(d)}(\tau-\tau')\,+\,K_{1'}^{(d)}(\tau-\tau')\right\}\,.
\end{eqnarray}
Plugging subsequently into this expression both
(\ref{eq:response_fkt_drude1}) and (\ref{eq:identity4}) with
(\ref{eq:lerch-fkt1}), and then using \cite{GRA07}
\begin{eqnarray}\label{eq:fourier-laplace7}
    && \lim_{t \to \infty} \int_0^t d\tau\,e^{-\alpha''\,(t-\tau)} \int_0^t d\tau'\,e^{-\alpha'\,(t-\tau')}\,e^{-\alpha\,|\tau-\tau'|}\n\\
    &=& \frac{\alpha'' + \alpha' + 2\alpha}{(\alpha'' + \alpha')\,(\alpha' + \alpha)\,(\alpha + \alpha'')}\,,
\end{eqnarray}
where $\alpha = \omega_d$ or $n \omega_{\mu}$ with $n = 0,1,2,
\cdots$, we can explicitly evaluate the double integral in
(\ref{eq:fourier-laplace6}), which turns out to vanish due to the
symmetric structure of $\chi_d(t)$ in $(z_0, z_1, z_2)$. Next,
taking into account the fact that $\ddot{\hat{Q}}_1(t) =
\dot{\hat{P}}_1(t)/M$ and $d{\chi}_d(0)/dt = 1/M$, we can also find
that
$\al\{\hat{P}_1(t)\,,\,\ddot{\hat{Q}}_1(t)\}_+\ar^{({\mbox{\tiny
ss}})} = 0$. Then the formal expression of the heat current given in
(\ref{eq:heat-current-case-1-1}) is simplified as
\begin{equation}\label{eq:heat-current-case-1-3}
    {\mathcal J}_{{\mbox{\tiny in}}}^{(1)}\, =\, \frac{1}{4M}\,
    \left\al\left\{\hat{P}_1(t)\,,\,\hat{\xi}_{b_1}(t) - \hat{\xi}_{b_{1'}}(t)\right\}_+\right\ar^{({\mbox{\tiny ss}})}\,.
\end{equation}
Along the same line, we can also obtain the expression for
${\mathcal J}_{{\mbox{\tiny out}}}^{(1)}$, identical to
(\ref{eq:heat-current-case-1-3}) but with exchange of
$\hat{\xi}_{b_1}(t)$ and $\hat{\xi}_{b_{1'}}(t)$, which immediately
verifies that ${\mathcal J}_{{\mbox{\tiny out}}}^{(1)} = -{\mathcal
J}_{{\mbox{\tiny in}}}^{(1)}$ indeed. In the equilibrium state given
by $\beta_1 = \beta_{1'}$, the heat current vanishes, as expected.
We can also expect, from (\ref{eq:fourier-laplace5}) and
(\ref{eq:heat-current-case-1-3}), the appearance of quantum
behaviors of the heat current ${\mathcal J}_{{\mbox{\tiny
in}}}^{(1)}$ due to the quantum nature of the bath correlation
function $K_{\mu}^{(d)}(t-t')$ given in
(\ref{eq:correlation-function1})-(\ref{eq:identity4}).

After some algebraic manipulations, every single step of which is
provided in detail in Appendix \ref{sec:appendix2}, Eq.
(\ref{eq:heat-current-case-1-3}) finally reduces to the exact
expression
\begin{eqnarray}\label{eq:heat-current-case-1-10}
    &&{\mathcal J}_{{\mbox{\tiny in}}}^{(1)}\; =\; \frac{\hbar\,\gamma\,\{z_0\,\omega_d + ({\mathbf
    w}_0)^2\}}{8\pi}\,\times\\
    &&\sum_{\underline{j}=0}^2 \frac{z_{\underline{j}}\cdot(\omega_d - z_{\underline{j}})}{(z_{\underline{j}} -
    z_{\underline{j+1}})\cdot(z_{\underline{j}} - z_{\underline{j+2}})}\, \left\{\Upsilon_{{\scriptscriptstyle \beta_1}}(z_{\underline{j}}) -
    \Upsilon_{{\scriptscriptstyle
    \beta_{1'}}}(z_{\underline{j}})\right\}\n
\end{eqnarray}
in terms of the parameters $({\mathbf w}_0, z_0, \gamma)$ given in
(\ref{eq:parameter_change0}), where $\underline{j} =
j\,(\mbox{mod}\; 3)$. Here,
\begin{eqnarray}\label{eq:heat-current-case-1-11}
    \Upsilon_{{\scriptscriptstyle \beta}}(z_{\underline{j}}) &:=&
    \frac{-\pi\,\cot\left(\frac{\beta \hbar \omega_d}{2}\right) + \psi\left(-\frac{\beta \hbar \omega_d}{2\pi}\right) -
    \psi\left(\frac{\beta \hbar z_{\underline{j}}}{2\pi}\right)}{\omega_d\,+\,z_{\underline{j}}}\n\\
    && -\, \frac{\psi\left(\frac{\beta \hbar \omega_d}{2\pi}\right) - \psi\left(\frac{\beta \hbar
    z_{\underline{j}}}{2\pi}\right)}{\omega_d\,-\,z_{\underline{j}}}\,,
\end{eqnarray}
where the digamma function $\psi(y) = d\,\ln\Gamma(y)/dy$
\cite{ABS74}. With the help of
(\ref{eq:susceptibility-2})-(\ref{eq:parameter_change0}) and
(\ref{eq:partial_fraction1})-(\ref{eq:summation1}) as well as the
relation given by $-\pi \cot(y) + \psi(-y/\pi) = \psi(y/\pi) +
\pi/y$, Eq. (\ref{eq:heat-current-case-1-10}) can be rewritten as a
compact expression
\begin{eqnarray}\label{eq:heat-current-case-1-10-1}
    &&{\mathcal J}_{{\mbox{\tiny in}}}^{(1)}\, =\, \frac{\hbar \omega_d^2\,\gamma_{\mbox{\tiny o}}}{2\pi}
    \left[\frac{\omega_d^2\cdot(g_{{\scriptscriptstyle \beta_1}} - g_{{\scriptscriptstyle \beta_{1'}}})}{h_1(\omega_d)}\, +\right.\n\\
    &&\left.\sum_{\underline{j}=0}^2
    \frac{z_{\underline{j}}^2\cdot\{\psi_{{\scriptscriptstyle \beta_1}}(z_{\underline{j}}) - \psi_{{\scriptscriptstyle \beta_{1'}}}(z_{\underline{j}})\}}{(z_{\underline{j}} - z_{\underline{j+1}})
    (z_{\underline{j}} - z_{\underline{j+2}}) (z_{\underline{j}} + \omega_d)}\right]\,,
\end{eqnarray}
where $\psi_{{\scriptscriptstyle \beta}}(y) := \psi(\beta \hbar
y/2\pi)$, and\\$g_{{\scriptscriptstyle \beta}} :=
-\{\psi_{{\scriptscriptstyle \beta}}(\omega_d) + 2\pi/\beta \hbar
\omega_d\}$.

Now we consider the semiclassical behavior of this heat current by
expanding the digamma functions such that in the limit of $\beta
\hbar \to 0$,
\begin{eqnarray}\label{eq:heat-current-case-1-15}
    &&{\mathcal J}_{{\mbox{\tiny in}}}^{(1)}\, =\,
    {\mathcal J}_{{\mbox{\tiny cl}}}^{(1)}\cdot\left[1\,-\,\frac{\hbar^2\,\beta_1\,\beta_{1'}}{12}\,(\Omega^2 + \gamma_{\mbox{\tiny o}}\,\omega_d)\, +\right.\n\\
    &&\frac{\hbar^4\,\beta_1\,\beta_{1'}\,(\beta_1^2 +
    \beta_1\,\beta_{1'} + \beta_{1'}^2)}{2^4\cdot 3^2\cdot 5}\, \times\\
    &&\left.\left\{\Omega^4 + \gamma_{\mbox{\tiny o}}\,\omega_d\,(3\,\Omega^2 + 2\,\gamma_{\mbox{\tiny o}}\,\omega_d +
    \omega_d^2)\right\}\,+\,{\mathcal O}\left\{(\beta\hbar)^6\right\}\right]\,,\n
\end{eqnarray}
expressed in terms of the original input parameters\\$(\Omega,
\omega_d, \gamma_{\mbox{\tiny o}})$ only, the derivation of which is
provided in Appendix \ref{sec:appendix2}). Here the leading term
\begin{align}\tag{\ref{eq:heat-current-case-1-15}a}\label{eq:classical_counterpart1}
    {\mathcal J}_{{\mbox{\tiny cl}}}^{(1)}\, =\,
    \frac{\gamma_{\mbox{\tiny o}}\,\omega_d^2}{2\,(\Omega^2 + \gamma_{\mbox{\tiny o}}\,\omega_d + \omega_d^2)}\,\left(\frac{1}{\beta_1} - \frac{1}{\beta_{1'}}\right)
\end{align}
corresponds to the classical counterpart to ${\mathcal
J}_{{\mbox{\tiny in}}}^{(1)}$, being valid in the high-temperature
limit. In the Ohmic limit $\omega_d \to \infty$, this classical
value reduces to the well-known expression given by ${\mathcal
J}_{{\mbox{\tiny cl}}}^{({\mbox{\tiny o}})} = (\gamma_{\mbox{\tiny
o}}/2)\cdot(1/\beta_1 - 1/\beta_{1'})$. On the other hand, the
steady-state heat current in (\ref{eq:heat-current-case-1-10-1})
reveals its different behavior in the low-temperature limit of
$\beta \hbar \to \infty$, explicitly given by (cf. Appendix
\ref{sec:appendix2})
\begin{eqnarray}\label{eq:stefan-B-3}
    {\mathcal J}_{{\mbox{\tiny in}}}^{(1)} &=&
    \frac{\hbar \omega_d^4\,\gamma_{\mbox{\tiny o}}}{4\pi}
    \sum_{n=0}^{\infty} \frac{B_{2n+4}}{n+2}\cdot
    \left(\frac{2\pi}{\hbar \omega_d}\right)^{2n+4}\, \times\\
    && \left(\frac{1}{\beta_1^{2n+4}}\, -\,
    \frac{1}{\beta_{1'}^{2n+4}}\right) \sum_{p=0}^{2n+1}
    \left.\frac{\omega_d^p}{p!}\cdot
    \left\{\frac{1}{h_1(s)}\right\}^{(p)}\right|_{s=0}\n
\end{eqnarray}
expressed in terms of the Bernoulli numbers $B_{2n}$, where
$\{\cdots\}^{(p)}$ denotes the $p$-th derivative. We see that in
this genuine quantum regime, the heat current is not directly
proportional to the bath-temperature difference any longer. With
$B_4 = -1/30$, the leading term $n=0$ of (\ref{eq:stefan-B-3})
easily reduces to the input power
\begin{align}\tag{\ref{eq:stefan-B-3}a}\label{eq:stefan-B-4}
    {\mathcal J}_{{\mbox{\tiny qm}}}^{(1,0)}\, =\,
    \frac{2\,\pi^3\,\gamma_{\mbox{\tiny o}}^2\,k_{\mbox{\tiny B}}^4}{15\,\hbar^3\,\Omega^4}\, \left(T_1^{4}\, -\, T_{1'}^{4}\right)\,,
\end{align}
being $\omega_d$-independent. This is the same in form of the
temperature dependency as the well-known Stefan-Boltzmann law for
the power radiated from a black-body \cite{LAN89}.

\begin{figure}[htb]
\centering\hspace*{-.7cm}{\includegraphics[scale=0.5]{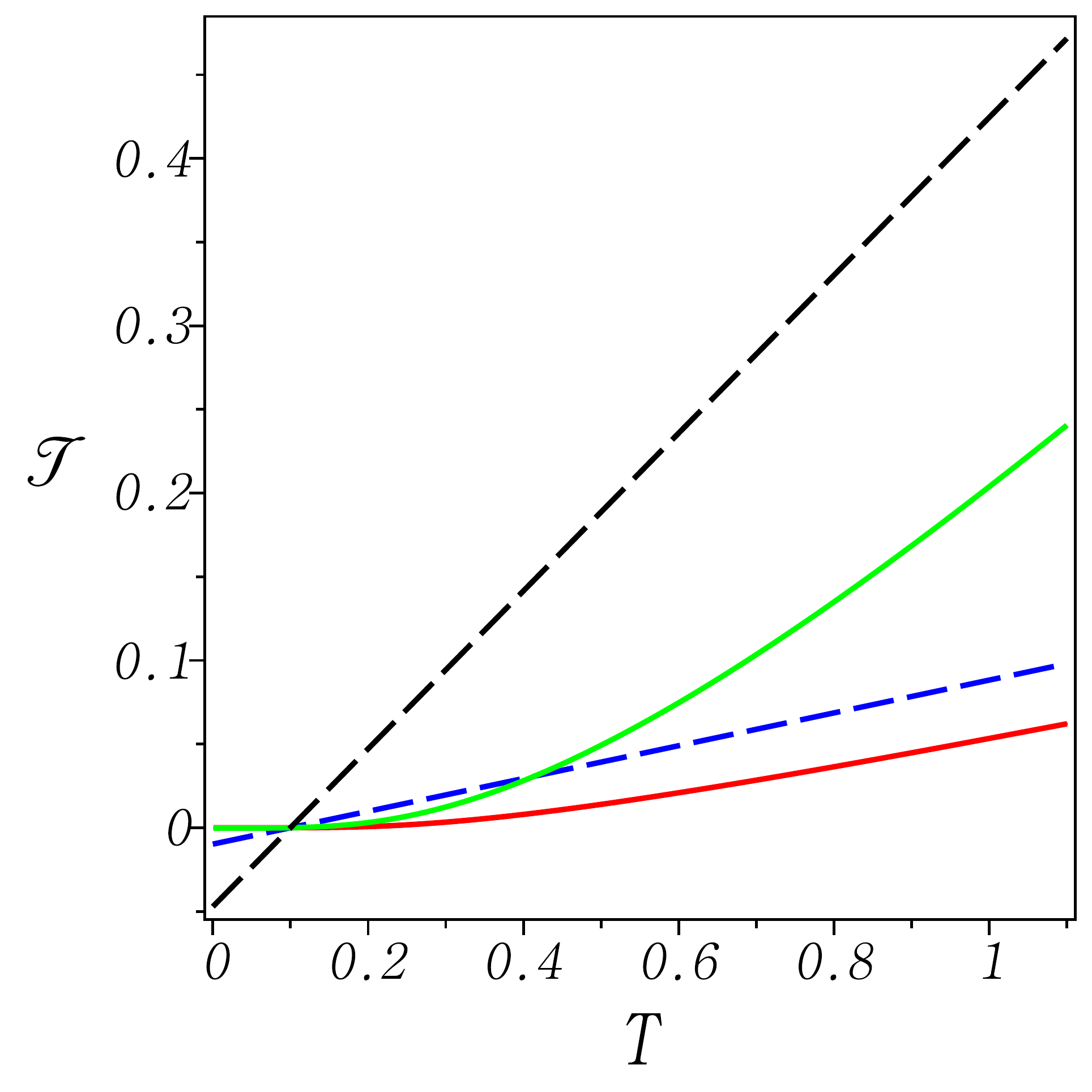}
\caption{(Color online) Heat current ${\mathcal J} = {\mathcal
J}_{{\mbox{\tiny in}}}^{(1)}(T,T_{1'})$ versus hot-bath temperature
$T = T_1$, in the low-temperature regime where the cold-bath
temperature is imposed by $T_{1'} = 0.1$. As such, we see that
${\mathcal J} = 0$ at the thermal equilibrium point, $T = 0.1$. Here
we set $\hbar = k_{\mbox{\tiny B}} = M = \Omega = 1$, and $\omega_d
= 10$; solid line plotted for the quantum-mechanical heat current
given in (\ref{eq:heat-current-case-1-10-1}) while dashed line for
its classical counterpart in (\ref{eq:classical_counterpart1}). From
top to bottom at $T = 1$, 1st: (black dash: $\gamma_{\mbox{\tiny o}}
= 1$); 2nd: (green solid: $\gamma_{\mbox{\tiny o}} = 1$); 3rd: (blue
dash: $\gamma_{\mbox{\tiny o}} = 0.2$); 4th: (red solid:
$\gamma_{\mbox{\tiny o}} = 0.2$); $\gamma_{\mbox{\tiny o}} = 0.2$
represents the weak coupling $\gamma_{\mbox{\tiny o}} \ll \Omega$
between the single oscillator and two baths.\label{fig:fig3}}}
\end{figure}

\begin{figure}[htb]
\centering\hspace*{-.0cm}{\includegraphics[scale=0.5]{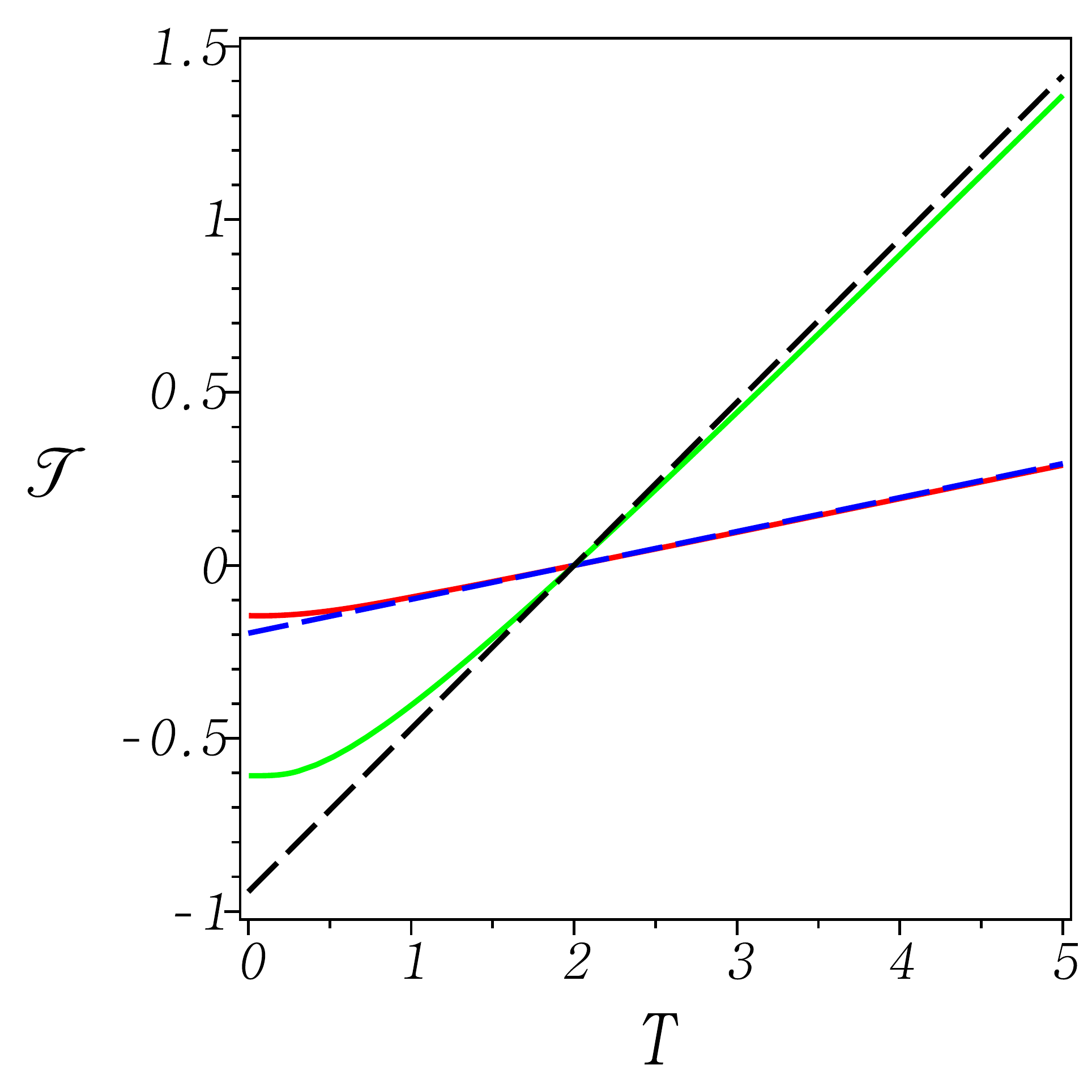}
\caption{(Color online) The same plot as in Fig.~\ref{fig:fig3}, in
the high-temperature regime where the cold-bath temperature is
imposed by $T_{1'} = 2$. As such, we see that ${\mathcal J} =
{\mathcal J}_{{\mbox{\tiny in}}}^{(1)}(T,2) = 0$ at the thermal
equilibrium point, $T = 2$. In the region of $T \geq 2$, the
quantum-mechanical and its classical values almost overlap each
other; on the other hand, ${\mathcal J} < 0$ for $T < 2$. From top
to bottom at $T = 0$, 1st: (red solid: $\gamma_{\mbox{\tiny o}} =
0.2$); 2nd: (blue dash: $\gamma_{\mbox{\tiny o}} = 0.2$); 3rd:
(green solid: $\gamma_{\mbox{\tiny o}} = 1$); 4th: (black dash:
$\gamma_{\mbox{\tiny o}} = 1$). It can further be verified that in
case that the cold-bath temperature is even higher, i.e., given by
$T_{1'} \geq 2$ (``classical regime''), the quantum-mechanical heat
current will more strongly overlap its classical
counterpart.\label{fig:fig4}}}
\end{figure}

As a result, it turns out that Fourier's law of heat conduction is
not valid even for the case of $N=1$, especially in the
low-temperature limit. The behaviors of ${\mathcal J}_{{\mbox{\tiny
in}}}^{(1)}$ versus the ``hot-bath'' temperature $T_1$ are plotted
in Figs. \ref{fig:fig3} and \ref{fig:fig4}, where two different
``cold-bath'' temperatures $T_{1'}$ are imposed in the
low-temperature and the high-temperature regime, respectively; in
the weak-coupling limit imposed by $\gamma_{\mbox{\tiny o}} \ll
\Omega$, the low-magnitude heat current is observed indeed. In the
next section, the heat current ${\mathcal J}_{{\mbox{\tiny
in}}}^{(2)}$ for $N=2$ will explicitly come out by applying its
formal expression in (\ref{eq:current_for_J-1}), valid for $N \geq
2$, rather than the one in (\ref{eq:heat-current-case-1-1}) used for
$N=1$.

\section{Steady-state heat current for the case of $N=2$}\label{sec:heat-current4}
%
We now consider the case of $N=2$. To efficiently proceed with the
determination of an explicit form of the inverse matrix
$\hat{{\mathcal B}}^{-1}(s)$, we first diagonalize the tridiagonal
matrix $\hat{{\mathcal B}}(s)$. To do so, we introduce the normal
coordinates $\{\underline{\hat{{\mathcal Q}}_j}(s)\}$ of the
isolated chain $\hat{H}_s$ given in (\ref{eq:chain1}), which satisfy
$\underline{\hat{Q}_j}(s) = O_{jk} \underline{\hat{{\mathcal
Q}}_k}(s)$ and $(\hat{O}^t\,\hat{C}\,\hat{O})_{jk} =
\bar{\Omega}_k^2\, \delta_{jk}$ with $\hat{O}^t\,\hat{O} =
\hat{O}\,\hat{O}^t = \id_{\scriptscriptstyle N}$ \cite{GAU07}. Eq.
(\ref{eq:laplace1}) is then rewritten as
\begin{eqnarray}\label{eq:fourier1-laplace1}
    {\mathfrak B}_{jk}(s)\,\underline{\hat{{\mathcal Q}}_k}(s) &=&
    s \hat{{\mathcal Q}}_j(0)\,+\,\dot{\hat{{\mathcal Q}}}_j(0)\,+\n\\
    && \frac{1}{M} \left\{\underline{\hat{\xi}_{b_1}}(s)\,O_{1j} +
    \underline{\hat{\xi}_{b_2}}(s)\,O_{{\scriptscriptstyle N}j}\right\}\,,
\end{eqnarray}
where the matrix
\begin{align}
    {\mathfrak B}_{jk}(s)\, :=&\, \left(\hat{O}^t\,\hat{\mathcal B}\,\hat{O}\right)_{jk}(s)\, =\,
    \left(s^2\,+\,\bar{\Omega}_k^2\right)\,\delta_{jk}\, +\n\\
    &\, s\,\bar{\gamma}(s)\,\left(\delta_{jk}\,-\,\sum_{n=2}^{N-1}
    O_{nj}\,O_{nk}\right)\,,\tag{\ref{eq:fourier1-laplace1}a}\label{eq:matrix_diagonal_N_2}
\end{align}
being, in fact, of diagonal form for $N=2$. For the case of
$\Omega_1 = \Omega_2 =: \Omega$ to be considered here, it easily
turns out that $\bar{\Omega}_1 = \Omega$ and $\bar{\Omega}_2 =
(\Omega^2 + 2 \kappa/M)^{1/2}$ as well as
\begin{align}\tag{\ref{eq:fourier1-laplace1}b}\label{eq:matrix_O2}
    \hat{O}\, =\, \frac{1}{\sqrt{2}} \left(\begin{array}{rr}
    1&1\\
    1&-1
    \end{array}\right)\,,
\end{align}
which is a simple constant matrix; for an isolated chain with $N
\geq 3$, in comparison, it is straightforward to verify that the
matrix elements $O_{jk}$ are not mere constants but functions of $s$
\cite{KIM13}. Then the inverse matrix $\hat{\mathfrak B}^{-1}(s)$
appears as a diagonal form with $({\mathfrak B}^{-1})_{11}(s) = M
\bar{\chi}_1(s)$ and $({\mathfrak B}^{-1})_{22}(s) = M
\bar{\chi}_2(s)$, where each of $\bar{\chi}_{\mu}(s)$ with $\mu =
1,2$ corresponds to $\bar{\chi}(s)$ given in
(\ref{eq:chi_laplace_1}) used for $N=1$, with substitution of
$\Omega \to \bar{\Omega}_{\mu}$.

We are now ready to apply to each of these two diagonal elements the
same technique as for $N=1$. Then it turns out, with the help of
(\ref{eq:matrix_O2}), that
\begin{subequations}
\begin{eqnarray}
    \lim_{t\to\infty} \hat{Q}_1(t) &=& \frac{1}{2}\,\lim_{t \to \infty} \int_0^t d\tau
    \sum_{\mu=1}^2\,\chi_{\mu}(t-\tau)\, \times\n\\
    && \left\{\hat{\xi}_{b_1}(\tau) -
    (-1)^{\mu}\,\hat{\xi}_{b_2}(\tau)\right\}\label{eq:Q_J=2-1}\\
    \lim_{t\to\infty} \hat{Q}_2(t) &=& \frac{1}{2}\,\lim_{t \to \infty} \int_0^t d\tau
    \sum_{\mu=1}^2\,\chi_{\mu}(t-\tau)\, \times\n\\
    && \left\{\hat{\xi}_{b_2}(\tau) - (-1)^{\mu}\,\hat{\xi}_{b_1}(\tau)\right\}\,.\label{eq:Q_J=2-2}
\end{eqnarray}
\end{subequations}
To explicitly evaluate the formal expression of the heat current
${\mathcal J}_{{\mbox{\tiny in}}}^{(2)}$ in
(\ref{eq:current_for_J-1}), we first focus on the steady-state
expectation value
\begin{eqnarray}\label{eq:correlation-1-J=2-1}
    \hspace*{-.5cm}&&\left\al\left\{\hat{P}_1(t)\,,\,\hat{Q}_1(t)\right\}_+\right\ar^{({\mbox{\tiny ss}})}\, =\,
    \frac{M}{2}\,\lim_{t \to \infty} \int_0^t d\tau \int_0^t
    d\tau'\\
    \hspace*{-.5cm}&&\frac{\partial}{\partial
    t}\left\{\chi_1(t-\tau)\,\chi_2(t-\tau')\right\}\,
    \left\{K_1^{(d)}(\tau-\tau')\,-\,K_2^{(d)}(\tau-\tau')\right\}\,.\n
\end{eqnarray}
By means of the same technique as for the case of $N=1$, we can
straightforwardly show that this vanishes indeed. Likewise, it also
turns out that
$\al\{\hat{P}_1(t)\,,\,\ddot{\hat{Q}}_1(t)\}_+\ar^{({\mbox{\tiny
ss}})} = 0$. Similarly, the last needed expectation value
\begin{eqnarray}\label{eq:correlation-3-J=2-1}
    &&\left\al\left\{\hat{P}_1(t)\,,\,\hat{Q}_2(t)\right\}_+\right\ar^{({\mbox{\tiny ss}})}\, =\,
    \frac{M}{2}\,\lim_{t\to\infty} \int_0^t d\tau \int_0^t
    d\tau'\n\\
    &&\left[\chi_1(t-\tau)\, \{\partial_t\,\chi_2(t-\tau')\}\, -\, \{\partial_t\,\chi_1(t-\tau)\}\, \chi_2(t-\tau')\right]\, \times\n\\
    &&\left[K_1^{(d)}(\tau-\tau')\,-\,K_2^{(d)}(\tau-\tau')\right]\, \ne\, 0
\end{eqnarray}
can be evaluated in closed form (cf. Appendix \ref{sec:appendix3}).
Substituting this form into (\ref{eq:current_for_J-1}), we can
immediately arrive at an exact expression of the steady-state heat
current
\begin{eqnarray}\label{eq:heat-current-case-1-10-1-N_20}
    &&{\mathcal J}_{{\mbox{\tiny in}}}^{(2)}\, =\,
    -\frac{\kappa}{2 M}\, \left\al\left\{\hat{P}_1(t)\,,\,\hat{Q}_2(t)\right\}_+\right\ar^{({\mbox{\tiny ss}})}\,
    =\, \frac{\hbar\,\omega_d^2\,\kappa\,\gamma_{\mbox{\tiny o}}}{4\pi M}\, \times\n\\
    &&\sum_{\underline{j}=0}^2
    \left(\left[\frac{(\omega_d^2 - z_{\underline{j}}^2)\cdot z_{\underline{j}}\cdot\{\Upsilon_{{\scriptscriptstyle \beta_1}}(z_{\underline{j}}) -
    \Upsilon_{{\scriptscriptstyle \beta_2}}(z_{\underline{j}})\}}{(z_{\underline{j}} -
    z_{\underline{j+1}})\cdot(z_{\underline{j}} - z_{\underline{j+2}})\cdot h_{12}(z_{\underline{j}})}\right]\, -\right.\n\\
    &&\left.\left[z_{\underline{j}} \to
    z_{\underline{j}}'\,;\,h_{12}(z_{\underline{j}}) \to h_{11}(z_{\underline{j}}')\right]\right)\,,
\end{eqnarray}
where the parameters $z_{\underline{j}}$'s are identical to $(z_0,
z_1, z_2)$ given in (\ref{eq:susceptibility-2}) but with
substitution of $\Omega \to \bar{\Omega}_1$, and so are
$z_{\underline{j}}'$'s with $\Omega \to \bar{\Omega}_2$. The
function $h_{11}(z_{\underline{j}}') := h_1(z_{\underline{j}}')$
with $\Omega \to \bar{\Omega}_1$, and $h_{12}(z_{\underline{j}}) :=
h_1(z_{\underline{j}})$ with $\Omega \to \bar{\Omega}_2$; by
construction, $h_{11}(-z_{\underline{j}}) =
h_{12}(-z_{\underline{j}}') = 0$. As given in
(\ref{eq:heat-current-case-1-10-1}) for $N=1$, Eq.
(\ref{eq:heat-current-case-1-10-1-N_20}) can finally be rewritten as
\begin{eqnarray}\label{eq:heat-current-case-1-10-1-N_2}
    {\mathcal J}_{{\mbox{\tiny in}}}^{(2)} &=& -\frac{\kappa^2 \gamma_{\mbox{\tiny o}}}{M^2}\,
    \left(\mbox{\={I}}_2\right)\cdot\left(\frac{1}{\beta_1} - \frac{1}{\beta_2}\right)\, +\, \frac{\hbar\,\omega_d^2\,\kappa\,\gamma_{\mbox{\tiny o}}}{2\pi M}\, \times\n\\
    && \sum_{\underline{j}=0}^2
    \left[\left\{\frac{z_{\underline{j}}^2\cdot\{\psi_{{\scriptscriptstyle \beta_1}}(z_{\underline{j}}) -
    \psi_{{\scriptscriptstyle \beta_2}}(z_{\underline{j}})\}}{(z_{\underline{j}} -
    z_{\underline{j+1}})\cdot(z_{\underline{j}} - z_{\underline{j+2}})\cdot h_{12}(z_{\underline{j}})}\right\}\, -\right.\n\\
    && \left.\left\{z_{\underline{j}} \to
    z_{\underline{j}}'\,;\,h_{12}(z_{\underline{j}}) \to h_{11}(z_{\underline{j}}')\right\}\right]\,.
\end{eqnarray}
Here $\left(\mbox{\={I}}_2\right) := \omega_d^2\cdot
f_{\kappa}/(2\,\lambda_{\kappa})$, where $f_{\kappa} = \Omega^2 +
\kappa/M + \omega_d^2$ and
\begin{align}
   \hspace*{-.0cm}&\lambda_{\kappa}\; =\; 2\,(\Omega^2 + \omega_d^2 + \gamma_{\mbox{\tiny o}}\,\omega_d) \left(\frac{\kappa}{M}\right)^3\,
   +\, \left(\Omega^4 + 2\,\Omega^2 \gamma_{\mbox{\tiny o}}\,\omega_d\, +\right.\n\\
   \hspace*{-.0cm}&\left.2\,\Omega^2\,\omega_d^2 - 2\,\gamma_{\mbox{\tiny o}}\,\omega_d^3 + \omega_d^4\right)
   \left(\frac{\kappa}{M}\right)^2\, +\, 4\,\gamma_{\mbox{\tiny o}}^2\,\omega_d^4 \left(\frac{\kappa}{M}\right)\, +\, 4\,\Omega^2\,\gamma_{\mbox{\tiny
   o}}^2\,\omega_d^4\,.\tag{\ref{eq:heat-current-case-1-10-1-N_2}a}
\end{align}

\begin{figure}[htb]
\centering\hspace*{-.7cm}{\includegraphics[scale=0.5]{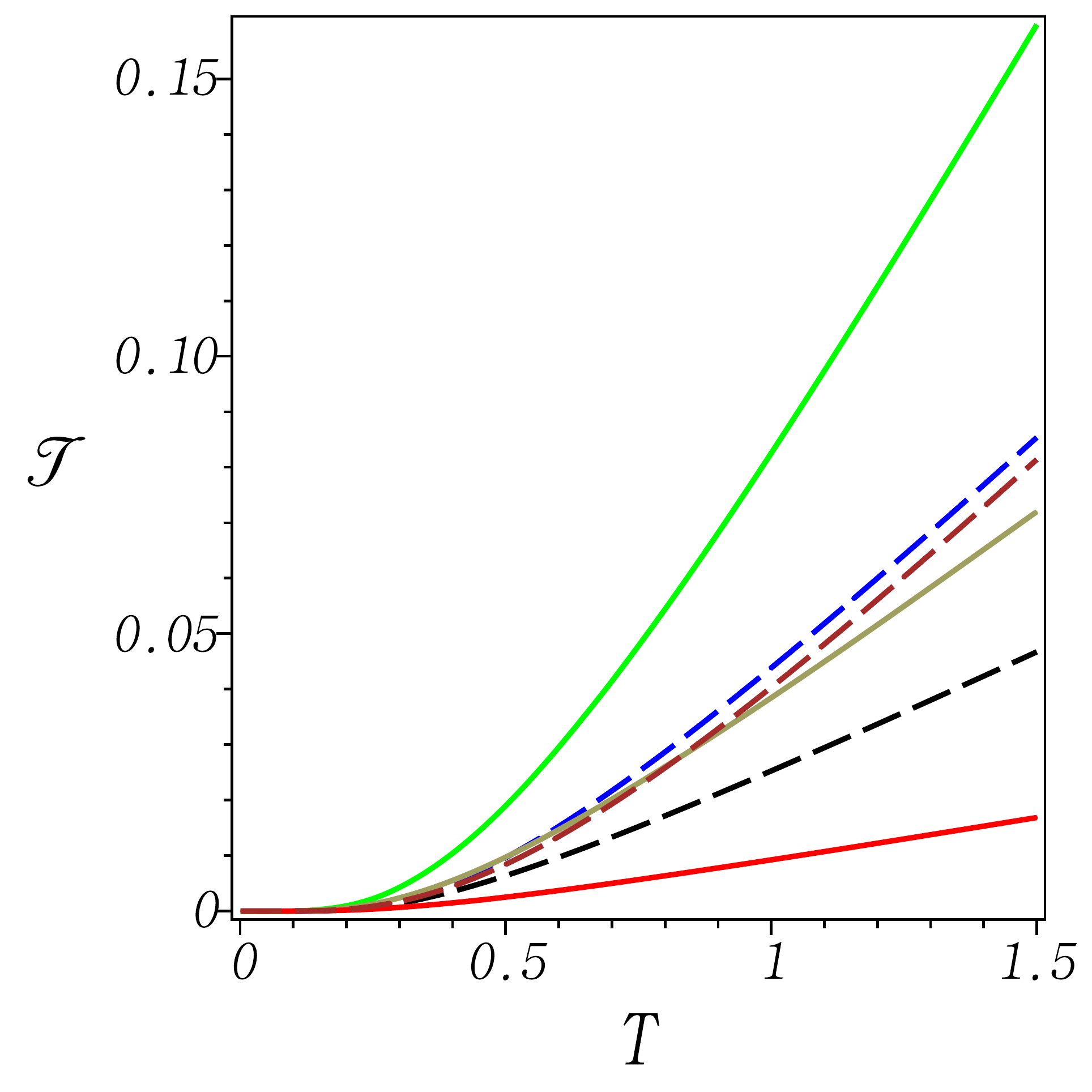}
\caption{(Color online) Heat current ${\mathcal J} = {\mathcal
J}_{{\mbox{\tiny in}}}^{(2)}(T,T_2)$ versus hot-bath temperature $T
= T_1$, given in (\ref{eq:heat-current-case-1-10-1-N_2}), in the
low-temperature regime where the cold-bath temperature is imposed by
$T_2 = 0.1$. As such, we see that ${\mathcal J} = 0$ at the thermal
equilibrium point, $T = 0.1$. Here we set $\hbar = k_{\mbox{\tiny
B}} = M = \Omega = 1$, and $\omega_d = 10$. Solid lines plotted for
$\gamma_{\mbox{\tiny o}} = 1$, from top to bottom at $T = 1.5$, 1st:
(green: $\kappa = 1$); 2nd: (khaki: $\kappa = 0.5$); 3rd: (red:
$\kappa = 0.2$). Dashed lines for $\gamma_{\mbox{\tiny o}} = 0.2$,
from top to bottom at $T = 1.5$, 1st: (blue: $\kappa = 1$); 2nd:
(brown: $\kappa = 1.5$); 3rd: (black: $\kappa = 0.2$). The 3rd
dashed line represents the weak-coupling regime, $\kappa/M,
\gamma_{\mbox{\tiny o}}^2 \ll \Omega^2$. Notably, this is of higher
value than the 3rd solid. Also, for $\gamma_{\mbox{\tiny o}} = 0.2$
the maximum (or resonant) heat current is obtained at $\kappa = 1 =
\kappa_{\mbox{\tiny R}}$ while with further increase of $\kappa >
\kappa_{\mbox{\tiny R}}$, the current decreases very
slowly.\label{fig:fig5}}}
\end{figure}

\begin{figure}[htb]
\centering\hspace*{-0cm}{\includegraphics[scale=0.5]{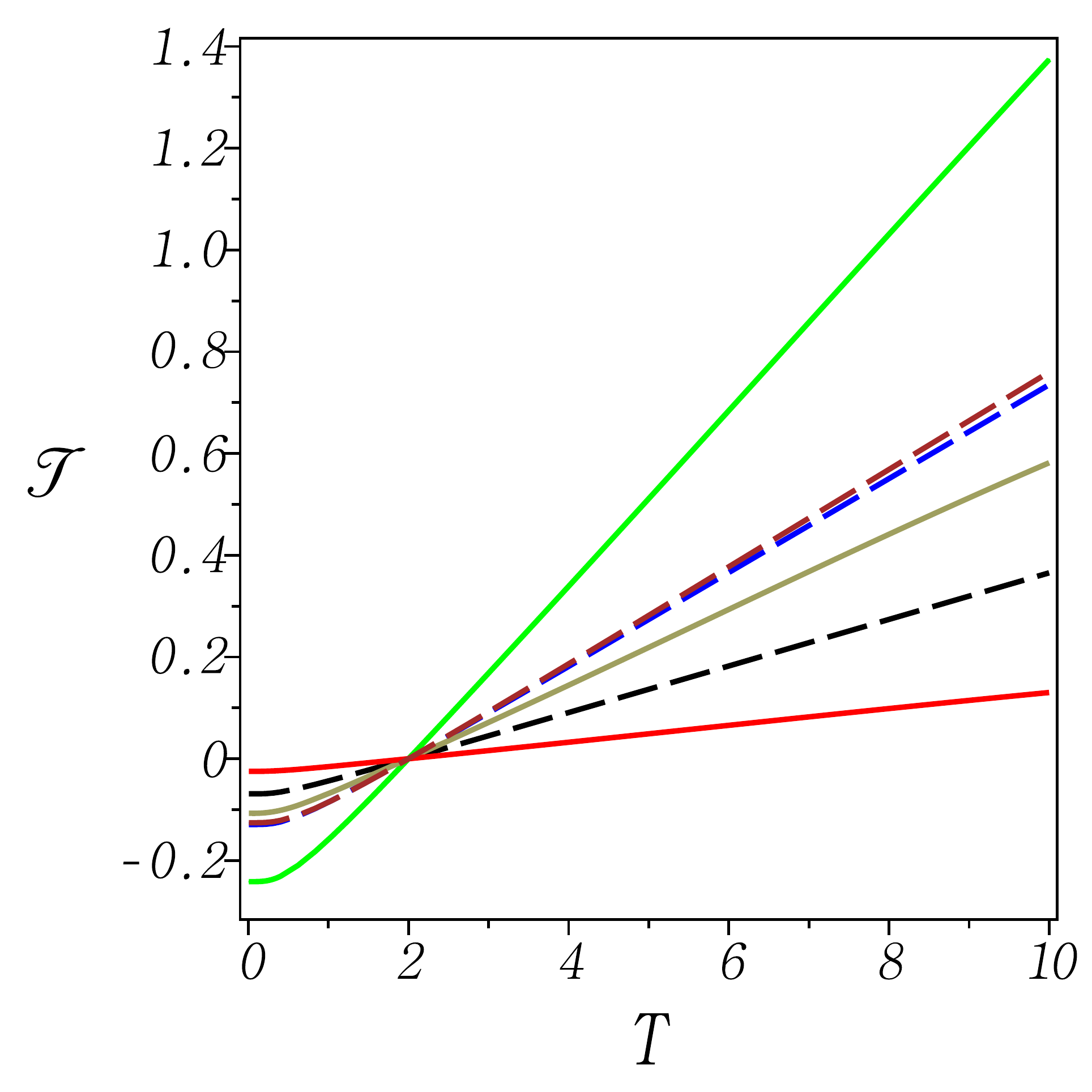}
\caption{(Color online) The same plot as in Fig.~\ref{fig:fig5}, in
the high-temperature regime where the cold-bath temperature is
imposed by $T_2 = 2$. As such, ${\mathcal J} = {\mathcal
J}_{{\mbox{\tiny in}}}^{(2)}(T,2) = 0$ at the thermal equilibrium
point, $T = 2$. In the region of $T \geq 2$ the heat current reveals
the behavior of its classical counterpart, being proportional to $T
- T_2$, while for $T < 2$ it is negative-valued. As shown, we have
the same lines for $T \geq T_2$ as in Fig.~\ref{fig:fig5}, except
that the 1st top dashed line (at $x = 10$): (brown:
$\gamma_{\mbox{\tiny o}} = 0.2$, and $\kappa = 2.2$ instead of
$1.5$), and 2nd top dash: (blue: $\gamma_{\mbox{\tiny o}} = 0.2$ and
$\kappa = 1$), i.e., for $\gamma_{\mbox{\tiny o}} = 0.2$ the maximum
heat current appears at $\kappa = 1 = \kappa_{\mbox{\tiny R}}$ while
with further increase of $\kappa > \kappa_{\mbox{\tiny R}}$, the
current decreases very slowly.\label{fig:fig6}}}
\end{figure}

Now we apply to the exact expression given in
(\ref{eq:heat-current-case-1-10-1-N_2}) the same technique as
provided for (\ref{eq:heat-current-case-1-15}), in order to study
its semiclassical behavior. Then it turns out that
\begin{eqnarray}\label{eq:heat-current-case-1-15_N_2}
    &&{\mathcal J}_{{\mbox{\tiny in}}}^{(2)}\, =\, {\mathcal J}_{{\mbox{\tiny cl}}}^{(2)}\cdot\left[1\, -\,
    \frac{\hbar^2\,\beta_1\,\beta_2}{12}\,
    \frac{v_{\kappa}}{f_{\kappa}}\, -\right.\\
    &&\left.\frac{\hbar^4\,\beta_1\,\beta_2\,(\beta_1^2 + \beta_1\,\beta_2 +
    \beta_2^2)}{2^3\cdot 3^2\cdot 5}\, \frac{\Omega^4\,\omega_d}{f_{\kappa}}\,+\,{\mathcal
    O}\left\{(\beta\hbar)^6\right\}\right]\,,\n
\end{eqnarray}
where $v_{\kappa} := (2\,\Omega^2 + 2\,\gamma_{\mbox{\tiny
o}}\,\omega_d + \omega_d^2)\cdot \kappa/M + \Omega^4 +
\Omega^2\,\omega_d\,(2\,\gamma_{\mbox{\tiny o}} + \omega_d)$. Here,
the leading term is given by the classical counterpart
\begin{align}\tag{\ref{eq:heat-current-case-1-15_N_2}a}\label{eq:classical_counterpart2}
    {\mathcal J}_{{\mbox{\tiny cl}}}^{(2)}\, =\,
    \frac{\kappa^2 \gamma_{\mbox{\tiny o}}}{M^2}\,
    \left(\mbox{\={I}}_2\right)\cdot\left(\frac{1}{\beta_1} - \frac{1}{\beta_2}\right)\,.
\end{align}
The behaviors of ${\mathcal J}_{{\mbox{\tiny in}}}^{(2)}$ in the
low-temperature and the high-temperature regime are plotted in Figs.
\ref{fig:fig5} and \ref{fig:fig6}, respectively; as demonstrated,
they are consistent with the behaviors of ${\mathcal
J}_{{\mbox{\tiny in}}}^{(1)}$. Here the weak-coupling limit is
imposed by $\kappa/M, \gamma_{\mbox{\tiny o}}^2 \ll \Omega^2$. Along
the same line as
(\ref{eq:correlation-1-J=2-1})-(\ref{eq:correlation-3-J=2-1}), we
can also obtain that
\begin{subequations}
\begin{eqnarray}
    \hspace*{-.7cm}\left\al\left\{\hat{P}_2(t)\,,\,\hat{Q}_2(t)\right\}_+\right\ar^{({\mbox{\tiny ss}})} &=&
    -\left\al\left\{\hat{P}_1(t)\,,\,\hat{Q}_1(t)\right\}_+\right\ar^{({\mbox{\tiny ss}})}\\
    \hspace*{-.7cm}\left\al\left\{\hat{P}_2(t)\,,\,\ddot{\hat{Q}}_2(t)\right\}_+\right\ar^{({\mbox{\tiny ss}})} &=&
    -\left\al\left\{\hat{P}_1(t)\,,\,\ddot{\hat{Q}}_1(t)\right\}_+\right\ar^{({\mbox{\tiny ss}})}\\
    \hspace*{-.7cm}\left\al\left\{\hat{P}_2(t)\,,\,\hat{Q}_1(t)\right\}_+\right\ar^{({\mbox{\tiny ss}})} &=&
    -\left\al\left\{\hat{P}_1(t)\,,\,\hat{Q}_2(t)\right\}_+\right\ar^{({\mbox{\tiny ss}})}\,.
\end{eqnarray}
\end{subequations}
From this, it follows that ${\mathcal J}_{{\mbox{\tiny out}}}^{(2)}
= -{\mathcal J}_{{\mbox{\tiny in}}}^{(2)}$ indeed. We see from
(\ref{eq:heat-current-case-1-15_N_2})-(\ref{eq:classical_counterpart2})
that ${\mathcal J}_{{\mbox{\tiny in}}}^{(2)} \to 0$ for $\kappa \to
0$, and so there can be no sufficiently high output power in the
weak-coupling regime, as expected. Finally we remark for a later
purpose that the steady-state heat current ${\mathcal
J}_{{\mbox{\tiny in}}}^{(2)}$ was rigorously treated based on the
two uncoupled normal modes, each of which was thoroughly studied for
${\mathcal J}_{{\mbox{\tiny in}}}^{(1)}$ already in Sect.
\ref{sec:heat-current3}.

\section{Steady-state heat current for the case of $N \geq 3$}\label{sec:heat-current5}
%
We first need to point out that the matrix $\hat{{\mathfrak B}}(s)$
given in (\ref{eq:matrix_diagonal_N_2}) is not of diagonal form for
$N \geq 3$ and neither is its inverse. Therefore, the
normal-coordinate technique provided for $N = 2$ cannot
straightforwardly be applied any longer. Instead, we adopt a
different approach to the determination of an explicit form of
$\hat{{\mathcal B}}^{-1}(s) = \hat{{\mathcal A}}(s)$, developed in
\cite{USM94,YUE06}; given an $N \times N$ symmetric tridiagonal
matrix
\begin{equation}\label{eq:matrix_B1-1}
    \hat{{\mathcal B}}(s)\; =\; \left(\begin{array}{ccccccc}
    a&c&0&0&\cdots&0&0\\
    c&b&c&0&\cdots&0&0\\
    0&c&b&c&\cdots&0&0\\
    0&0&c&b&\cdots&0&0\\
    \cdots&&\cdots&&\cdots&&\cdots\\
    0&0&0&\cdots&c&b&c\\
    0&0&0&\cdots&0&c&a
    \end{array}\right)\,,
\end{equation}
where
\begin{align}
    &a\, :=\, s^2 + s\,\bar{\gamma}(s) + \Omega^2 + \frac{\kappa}{M}\n\\
    &b\, :=\, s^2 + \Omega^2 +
    \frac{2\kappa}{M}\;\;\; ,\;\;\; c\, :=\, -\frac{\kappa}{M}\,.\tag{\ref{eq:matrix_B1-1}a}\label{eq:matrix_B1-2}
\end{align}
Let $\delta := s\,\bar{\gamma}(s) + c$, and so $a = b + \delta$.
Then its inverse is explicitly given by a symmetric form,
\begin{equation}\label{eq:matrix_B_inverse_1}
    {\mathcal A}_{jk}(s)\, =\,  \frac{1}{c\,\sin \phi}\,
    \frac{\mbox{Num}_{jk}(s;N)}{\mbox{Den}(s;N)}
\end{equation}
for $j \leq k$, where both numerator and denominator are
\begin{align}
    \hspace*{-.0cm}&\mbox{Num}_{jk}(s;N)\, =\, \left[c\,\{\sin j\phi\} - \delta\,\{\sin
    (j-1)\phi\}\right]\, \times\n\\
    \hspace*{-.0cm}&\left[\delta\,\{\sin (N-k)\phi\} - c\,\{\sin
    (N-k+1)\phi\}\right]\tag{\ref{eq:matrix_B_inverse_1}a}\label{eq:matrix_B_inverse_21}\\
    \hspace*{-.0cm}&\mbox{Den}(s;N)\, =\, c^2\,\{\sin (N+1)\phi\} -
    2\,c\,\delta\,\{\sin N\phi\}\, +\n\\
    \hspace*{-.0cm}&\delta^2\,\{\sin (N-1)\phi\}\,,\tag{\ref{eq:matrix_B_inverse_1}b}\label{eq:matrix_B_inverse_22}
\end{align}
respectively. Here,
\begin{align}\tag{\ref{eq:matrix_B_inverse_1}c}\label{eq:matrix_B1-3}
    \sin \phi\, =\, -\frac{i}{2\,c} \left(b^2 - 4\,c^2\right)^{1/2}\;\;\; ,\;\;\; \cos \phi\, =\, -\frac{b}{2\,c}\,.
\end{align}
For $s \in {\mathbb R}$, the functions, $\sin\phi$ and $\cos \phi$
are rewritten as $i\sinh r$ and $\cosh r$, respectively, in terms of
a real number $r = -i\phi > 0$.

Next let us express the matrix elements ${\mathcal A}_{jk}(s)$
explicitly in terms of $s$. By using\\$\sin (n+1) \phi =
2\,(\cos\phi) (\sin n\phi) - \sin (n-1) \phi$ and
\begin{equation}\label{eq:sine_identity1}
    \sin n\phi\, =\, \sum_{\nu=0}^n \binom{n}{\nu}\, (\cos \phi)^{\nu}\, (\sin
    \phi)^{n-\nu}\, \sin\left\{\frac{(n-\nu)\,\pi}{2}\right\}
\end{equation}
as well as $\sin (n\pi/2) = i\,\{(-i)^n - i^n\}/2$ \cite{ABS74}, we
can rewrite Eqs. (\ref{eq:matrix_B_inverse_21}) and
(\ref{eq:matrix_B_inverse_22}) as
\begin{subequations}\label{eq:num-den1}
\begin{eqnarray}
    &&\mbox{Num}_{jk}(s;N)\, =\, \frac{(-2 c)^{k-j+1-{\scriptscriptstyle N}}}{16}\, \times\label{eq:matrix_B_inverse_31}\\
    &&\left\{F_j(s) + 2\,\delta\,F_{j-1}(s)\right\}
    \left\{F_{{\scriptscriptstyle N}-k+1}(s) + 2\,\delta\,F_{{\scriptscriptstyle N}-k}(s)\right\}\n\\
    &&\mbox{Den}(s;N)\, =\, \frac{i\, (-2 c)^{1-{\scriptscriptstyle N}}}{4}\, \times\n\\
    &&\left\{(a + \delta)\,F_{\scriptscriptstyle N}(s) - 2\,(c^2 -
    \delta^2)\,F_{{\scriptscriptstyle N}-1}(s)\right\}\,,\label{eq:matrix_B_inverse_32}
\end{eqnarray}
\end{subequations}
respectively. Here $F_n(s) = G^n(s) - H^n(s)$, where
\begin{align}\tag{\ref{eq:num-den1}c}\label{eq:matrix_B_inverse_3}
    \hspace*{-.0cm}G(s)\, :=\, b + (b^2 - 4\,c^2)^{1/2}\; ,\; H(s)\, :=\, b - (b^2 - 4\,c^2)^{1/2}\,.
\end{align}
From this, we see that $F_1(s) = 4ic\,\sin\phi$, and $G =
-2c\,e^{-i\phi}$ and $H = -2c\,e^{i\phi}$, as well as
\begin{align}
    F_n(s)\, =&\, -2 i\,(-2 c)^n\,\sin(n\phi)\tag{\ref{eq:num-den1}d}\label{eq:matrix_B_inverse_30}\\
    F_{n+1}(s)/F_n(s)\, =&\, -2c\, \{\sin (n+1)\phi\}/\{\sin n\phi\}\,.\n
\end{align}
Substituting into (\ref{eq:matrix_B_inverse_1}) the expressions
given in (\ref{eq:matrix_B_inverse_31}) and
(\ref{eq:matrix_B_inverse_32}) as well as $\bar{\gamma}(s) \to
\bar{\gamma}_d(s) = \gamma_{\mbox{\tiny o}}\,\omega_d/(s +
\omega_d)$, we can explicitly obtain
\begin{eqnarray}\label{eq:A_11_12_1J_2J-1}
    &&{\mathcal A}_{11}(s)\, =\, \frac{(s + \omega_d)\, (s^{2{\scriptscriptstyle N}-1} + \omega_d\,s^{2{\scriptscriptstyle N}-2} +
    \cdots)}{h_{\scriptscriptstyle N}(s)}\n\\
    &&{\mathcal A}_{12}(s)\, =\, -c\cdot\frac{(s + \omega_d)\, (s^{2{\scriptscriptstyle N}-3} + \omega_d\,s^{2{\scriptscriptstyle N}-4} +
    \cdots)}{h_{\scriptscriptstyle N}(s)}\n\\
    &&{\mathcal A}_{1{\scriptscriptstyle N}}(s)\, =\, (-c)^{{\scriptscriptstyle N}-1}\,
    (s + \omega_d)^2/h_{\scriptscriptstyle N}(s)\n\\
    &&{\mathcal A}_{2{\scriptscriptstyle N}}(s)\, =\, (-c)^{{\scriptscriptstyle
    N}-2}\, (s + \omega_d)\cdot\bar{h}_1(s)/h_{\scriptscriptstyle N}(s)\,,
\end{eqnarray}
where the cubic polynomial $\bar{h}_1(s)$ appears from $h_1(s)$
given in (\ref{eq:chi_laplace_2}) with $\Omega \to (\Omega^2 +
\kappa/M)^{1/2}$. Here we have the $(2N+2)$th-degree polynomial,
$h_{\scriptscriptstyle N}(s) = \mbox{den}(s;N) \cdot (s +
\omega_d)^2 = a_{2{\scriptscriptstyle N}+2}\cdot
s^{2{\scriptscriptstyle N}+2} + 2\,\omega_d\,s^{2{\scriptscriptstyle
N}+1} + \cdots$, where the leading coefficient
$a_{2{\scriptscriptstyle N}+2} = 1$, and the factor $\mbox{den}(s;N)
= -4\,i\cdot\mbox{Den}(s;N)/\{(-c)^{1-{\scriptscriptstyle
N}}\,F_1(s)\}$; $\mbox{den}(s;N)$ is not a polynomial, due to the
fractional form of the damping kernel $\bar{\gamma}_d(s)$, and its
completely explicit expression is provided in Appendix
\ref{sec:appendix4}. With the aid of (\ref{eq:matrix_B_inverse_3})
and (\ref{eq:matrix_B_inverse_30}), it can also be shown that
$h_{\scriptscriptstyle N}(-\omega_d) \ne 0$.

Owing to these explicit expressions of ${\mathcal A}_{jk}(s)$, we
are now in position to straightforwardly proceed to obtain
$A_{jk}(t) = {\mathcal L}^{-1}\{{\mathcal A}_{jk}(s)\}(t)$ in time
domain. By applying the initial value theorem given by $\lim_{t\to
0} f(t) = \lim_{s\to\infty} s\,F(s)$ \cite{COH07}, we can easily
find that $A_{jk}(0) = 0$ and, e.g., $\dot{A}_{11}(0) =
\lim_{s\to\infty} s\,\{s\,{\mathcal A}_{11}(s) - A_{11}(0)\} = 1$ as
well as $\dot{A}_{12}(0) = 0$, etc. It can also be verified that
$h_{\scriptscriptstyle N}(s) = \prod_{j=0}^{2{\scriptscriptstyle
N}+1} (s + z_j)$ meets the condition of $\mbox{Re}(z_j)
> 0$ indeed (cf. Appendix \ref{sec:appendix4}), as the cubic polynomial $h_1(s)$ in
(\ref{eq:chi_laplace_2}) did. Due to this fact, we are also allowed
to apply the final value theorem, then giving rise to
$\lim_{t\to\infty} A_{jk}(t) = \lim_{t\to\infty} \dot{A}_{jk}(t) =
0$ to be needed below.

Now we explicitly consider the formal expression of the steady-state
heat current given in (\ref{eq:current_for_J-1}) based on the above
result. To do so, we begin with
\begin{subequations}
\begin{eqnarray}
    &&\lim_{t\to\infty} \hat{Q}_1(t) = \frac{1}{M} \lim_{t\to\infty}
    \int_0^t d\tau \left\{A_{11}(t-\tau)\cdot\hat{\xi}_{b_1}(\tau)\,
    +\right.\n\\
    &&\left.A_{1{\scriptscriptstyle N}}(t-\tau)\cdot\hat{\xi}_{b_{\scriptscriptstyle N}}(\tau)\right\}\label{eq:matrix_B_inverse_41}\\
    &&\lim_{t\to\infty} \hat{Q}_2(t) = \frac{1}{M} \lim_{t\to\infty}
    \int_0^t d\tau \left\{A_{21}(t-\tau)\cdot\hat{\xi}_{b_1}(\tau)\,
    +\right.\n\\
    &&\left.A_{2{\scriptscriptstyle N}}(t-\tau)\cdot\hat{\xi}_{b_{\scriptscriptstyle N}}(\tau)\right\}\label{eq:matrix_B_inverse_42}\\
    &&\lim_{t\to\infty} \hat{P}_1(t) = \lim_{t\to\infty}
    \int_0^t d\tau\, \partial_t \left\{A_{11}(t-\tau)\cdot\hat{\xi}_{b_1}(\tau)\,
    +\right.\n\\
    &&\left.A_{1{\scriptscriptstyle N}}(t-\tau)\cdot\hat{\xi}_{b_{\scriptscriptstyle N}}(\tau)\right\}\label{eq:matrix_B_inverse_43}\\
    &&\lim_{t\to\infty} \ddot{\hat{Q}}_1(t) = \frac{1}{M} \lim_{t\to\infty} \int_0^t
    d\tau\, \partial_t^2 \left\{A_{11}(t-\tau)\cdot\hat{\xi}_{b_1}(\tau)\,
    +\right.\n\\
    &&\left.A_{1{\scriptscriptstyle N}}(t-\tau)\cdot\hat{\xi}_{b_{\scriptscriptstyle N}}(\tau)\right\}\, +\n\\
    &&\frac{1}{M} \lim_{t\to\infty} \left\{\dot{A}_{11}(0)\cdot\hat{\xi}_{b_1}(t)\, +\,
    \dot{A}_{1{\scriptscriptstyle N}}(0)\cdot\hat{\xi}_{b_{\scriptscriptstyle N}}(t)\right\}\,,\label{eq:matrix_B_inverse_44}
\end{eqnarray}
\end{subequations}
which can be found from the equation of\\Laplace-transform in
(\ref{eq:laplace1}). Substituting (\ref{eq:matrix_B_inverse_42}) and
(\ref{eq:matrix_B_inverse_43}) into (\ref{eq:current_for_J-1}), we
acquire the first expectation value
\begin{eqnarray}\label{eq:matrix_B_inverse_5}
    \hspace*{-.5cm}&&\lim_{t\to\infty} \mbox{Tr}\left[\left\{\hat{P}_1(t)\,,\,\hat{Q}_2(t)\right\}_+\cdot\hat{\rho}(0)\right]\,
    =\, \frac{2}{M} \lim_{t\to\infty} \int_0^t d\tau \int_0^t d\tau'\n\\
    \hspace*{-.5cm}&&\left[\left\{\partial_t\,A_{11}(t-\tau)\right\}\cdot A_{21}(t-\tau')\cdot K_1^{(d)}(\tau-\tau')\, +\right.\\
    \hspace*{-.5cm}&&\left.\{\partial_t\,A_{1{\scriptscriptstyle
    N}}(t-\tau)\}\cdot A_{2{\scriptscriptstyle N}}(t-\tau')\cdot K_{\scriptscriptstyle N}^{(d)}(\tau-\tau')\right]\, \ne\,
    0\,.\n
\end{eqnarray}
This integral can be evaluated explicitly in the same way as in
(\ref{eq:correlation-3-J=2-1}) valid for a chain with $N=2$ only
[cf. (\ref{eq:heat-current-N-geq-3-1}) and
(\ref{eq:heat-current081})]; the detail of this evaluating process
is provided in Appendix \ref{sec:appendix5}, where we also verify
the equality
\begin{eqnarray}\label{eq;matrix_B_inverse_53}
    \hspace*{-.1cm}&&\lim_{t\to\infty} \int_0^t d\tau \int_0^t
    d\tau'\, \left\{\partial_t\,A_{11}(t-\tau)\right\} A_{21}(t-\tau') e^{-\alpha\,|\tau-\tau'|}\n\\
    \hspace*{-.1cm}&&= -\lim_{t\to\infty} \int_0^t d\tau \int_0^t
    d\tau'\, \left\{\partial_t\,A_{1{\scriptscriptstyle
    N}}(t-\tau)\right\}\, \times\n\\
    \hspace*{-.1cm}&&A_{2{\scriptscriptstyle N}}(t-\tau')\, e^{-\alpha\,|\tau-\tau'|}\,.
\end{eqnarray}
Here the constant $\alpha$ equals the cut-off frequency $\omega_d$
or the effective frequency given by $n \omega_1$ or $n
\omega_{\scriptscriptstyle N}$, where $n = 0, 1, 2, \cdots$. Next we
consider the second steady-state expectation value,
$\lim_{t\to\infty}
\mbox{Tr}[\{\hat{P}_1(t),\hat{Q}_1(t)\}_+\cdot\hat{\rho}(0)]$ by
applying the same technique to (\ref{eq:matrix_B_inverse_41}) and
(\ref{eq:matrix_B_inverse_43}); with the help of
(\ref{eq;matrix_B_inverse_53}), this will straightforwardly give
rise to
\begin{eqnarray}\label{eq:matrix_B_inverse_5-0}
    \hspace*{-.1cm}&&\lim_{t\to\infty} \int_0^t d\tau \int_0^t d\tau'\, \{\partial_t
    A_{11}(t-\tau)\} A_{11}(t-\tau') K_1^{(d)}(\tau-\tau')\n\\
    \hspace*{-.1cm}&&= \lim_{t\to\infty} \int_0^t d\tau \int_0^t d\tau'\,\,
    \{\partial_t A_{1{\scriptscriptstyle N}}(t-\tau)\}\times\n\\
    \hspace*{-.1cm}&&A_{1{\scriptscriptstyle N}}(t-\tau')\cdot K_{\scriptscriptstyle N}^{(d)}(\tau-\tau')\,
    =\, 0\,,
\end{eqnarray}
hence leading to the second expectation value vanishing. Along the
same line, we can find, too, that the last expectation value,
$\lim_{t\to\infty}
\mbox{Tr}[\{\hat{P}_1(t),\ddot{\hat{Q}}_1(t)\}_+\cdot\hat{\rho}(0)]
= 0$. As a result, the steady-state heat current given in
(\ref{eq:current_for_J-1}) reduces to
\begin{equation}\label{eq:heat-current-N-geq-3-1}
    {\mathcal J}_{{\mbox{\tiny in}}}^{({\scriptscriptstyle N})}\; =\;
    -\frac{\kappa}{2 M}\, \lim_{t\to\infty}
    \mbox{Tr}\left[\left\{\hat{P}_1(t)\,,\,\hat{Q}_2(t)\right\}_+\cdot\hat{\rho}(0)\right]\,.
\end{equation}

By applying the same technique, we can also arrive at the expression
\begin{equation}\label{eq:matrix_B_inverse_51}
     {\mathcal J}_{{\mbox{\tiny out}}}^{({\scriptscriptstyle N})}\; =\; -\frac{\kappa}{2 M}\, \lim_{t\to\infty}
    \mbox{Tr}\left[\left\{\hat{P}_{\scriptscriptstyle N}(t)\,,\,\hat{Q}_{{\scriptscriptstyle N}-1}(t)\right\}_+\cdot\hat{\rho}(0)\right]\,.
\end{equation}
We can then find that $\lim_{t\to\infty}
\hat{Q}_{{\scriptscriptstyle N}-1}(t)$ appears directly from
$\lim_{t\to\infty} \hat{Q}_2(t)$ given in
(\ref{eq:matrix_B_inverse_42}) with substitution of both $A_{21} \to
A_{{\scriptscriptstyle N}-1,1}$ and $A_{2{\scriptscriptstyle N}} \to
A_{{\scriptscriptstyle N}-1,{\scriptscriptstyle N}}$ as well as
$\lim_{t\to\infty} \hat{P}_{\scriptscriptstyle N}(t)$ comes from
$\lim_{t\to\infty} \hat{P}_1(t)$ given in
(\ref{eq:matrix_B_inverse_43}) with $A_{11} \to
A_{{\scriptscriptstyle N}1}$ and $A_{1{\scriptscriptstyle N}} \to
A_{{\scriptscriptstyle N}{\scriptscriptstyle N}}$. Further, Eq.
(\ref{eq:matrix_B_inverse_1}) straightforwardly gives rise to
${\mathcal A}_{{\scriptscriptstyle N}-1,1}(s) = {\mathcal
A}_{2{\scriptscriptstyle N}}(s)$ and ${\mathcal
A}_{{\scriptscriptstyle N}-1,{\scriptscriptstyle N}}(s) = {\mathcal
A}_{21}(s)$ as well as ${\mathcal A}_{{\scriptscriptstyle N}1}(s) =
{\mathcal A}_{1{\scriptscriptstyle N}}(s)$ and ${\mathcal
A}_{{\scriptscriptstyle N}{\scriptscriptstyle N}}(s) = {\mathcal
A}_{11}(s)$. Substituting all these results into
(\ref{eq:matrix_B_inverse_51}) and then with the aid of
(\ref{eq;matrix_B_inverse_53}), we can verify that ${\mathcal
J}_{{\mbox{\tiny out}}}^{({\scriptscriptstyle N})} = -{\mathcal
J}_{{\mbox{\tiny in}}}^{({\scriptscriptstyle N})}$ indeed.

\begin{figure}[htb]
\centering\hspace*{-.7cm}{\includegraphics[scale=0.5]{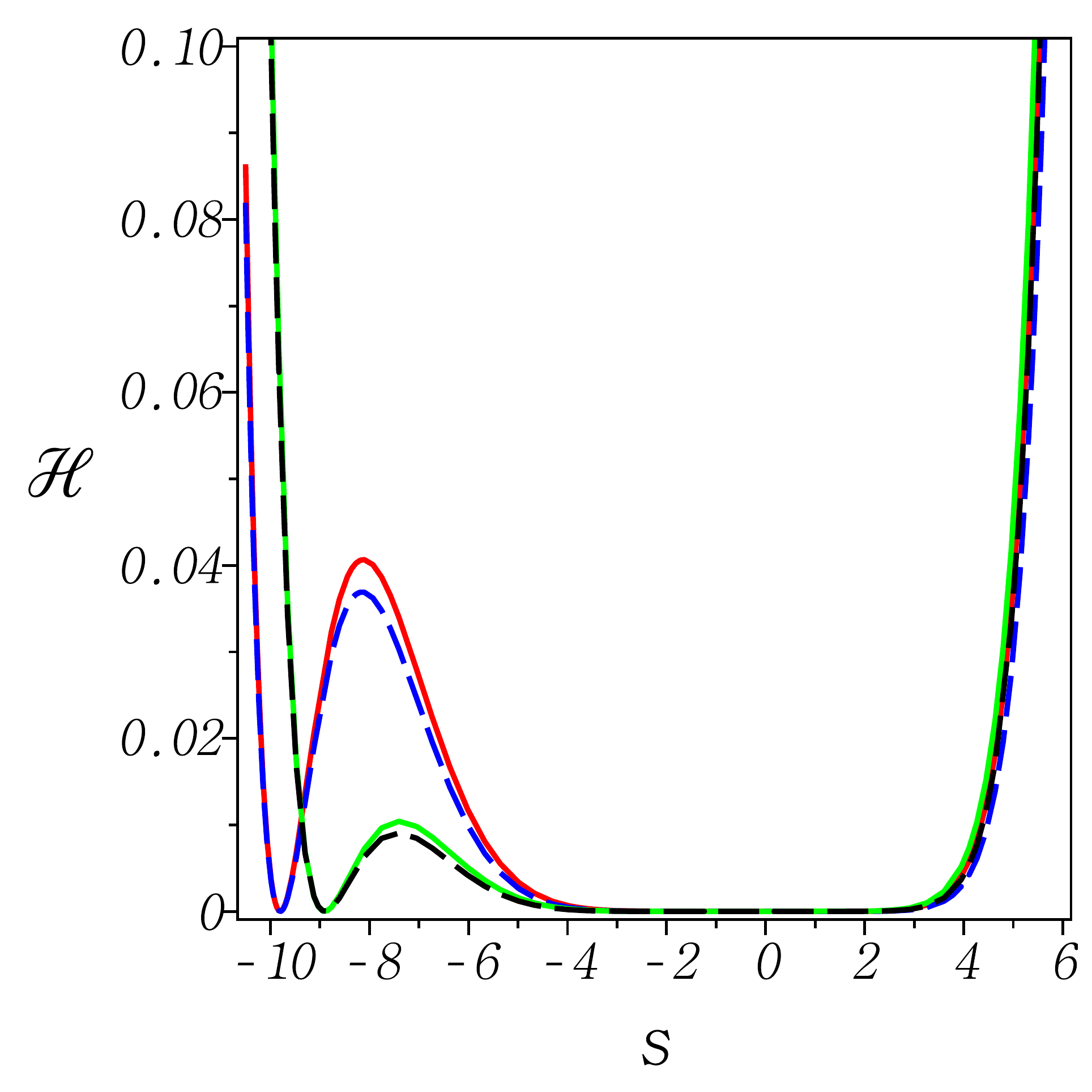}
\caption{(Color online) ${\mathcal H} = h_{\scriptscriptstyle
N}(s)\cdot 10^{-11}$ versus $s$, given in
(\ref{eq:A_11_12_1J_2J-1}). Here we set $\Omega = M = 1$ and
$\omega_d = 10$ as well as $N = 5$. From top to bottom at $s =
-7.5$, 1st) $h_{5,1}(s)\cdot 10^{-11}$: (red solid:
$\gamma_{\mbox{\tiny o}} = 0.2$ and $\kappa = 0.2$) with a single
multiple root, $s = -9.79840398$; 2nd) $h_{5,2}(s)\cdot 10^{-11}$:
(blue dash: $\gamma_{\mbox{\tiny o}} = 0.2$ and $\kappa = 1$) with a
multiple root, $s = -9.80006223$; 3rd) $h_{5,3}(s)\cdot 10^{-11}$:
(green solid: $\gamma_{\mbox{\tiny o}} = 1$ and $\kappa = 0.2$) with
a multiple root, $s = -8.89222703$; 4th) $h_{5,4}(s)\cdot 10^{-11}$:
(black dash: $\gamma_{\mbox{\tiny o}} = 1$ and $\kappa = 1$) with a
multiple root, $s = -8.90443052$. All the multiple roots have
degeneracy $m = 2$. We also have $h_{5,1}(0) = 0.00000361 \ne 0$;
$h_{5,2}(0) = 0.00005500$; $h_{5,3}(0) = 0.00000361$; $h_{5,1}(0) =
0.00005500$; in fact, $h_{\scriptscriptstyle N}(0)$ increases with
increase of $N$. The numerical analysis of the exact expression of
$h_{\scriptscriptstyle N}(s)$ reveals that there is no additional
complex-valued multiple root of the above four functions. These
properties of $h_{\scriptscriptstyle N}(s)$ are verified to be valid
for many different choices of $(\gamma_{\mbox{\tiny o}}, \kappa)$,
and $N = 3, 4, \cdots, 20$.\label{fig:fig7}}}
\end{figure}

Now we are ready to derive an explicit expression of the
steady-state heat current, based on the results obtained from the
previous paragraphs. Then it turns out that (cf. Appendix
\ref{sec:appendix5})
\begin{eqnarray}\label{eq:heat-current081}
    &&{\mathcal J}_{{\mbox{\tiny in}}}^{({\scriptscriptstyle N})}\; =\;
    \frac{\hbar\,\omega_d^4\,\gamma_{\mbox{\tiny o}}^2}{\pi}\,\left(\frac{\kappa}{M}\right)^{2{\scriptscriptstyle
    N}-2}\, \times\n\\
    &&\sideset{}{'}\sum_{j=0}^{2N+1}\,
    \left[\frac{(\omega_d^2 - z_j^2)\cdot z_j^2\cdot\{\Upsilon_{{\scriptscriptstyle \beta_1}}(z_j) -
    \Upsilon_{{\scriptscriptstyle \beta_{\scriptscriptstyle N}}}(z_j)\}}{h_{\scriptscriptstyle N}'(-z_j)\cdot h_{\scriptscriptstyle N}(z_j)}\right]\,,
\end{eqnarray}
which is, in fact, valid for arbitrary values of the input
parameters $(\Omega, \kappa, \omega_d, \gamma_{\mbox{\tiny o}})$.
Here the primed sum denoted by $\sum_j\hspace*{-.25cm}{}^{'}\;
[\cdots]$ means that if one of $z_j$'s is repeated, say,
$z_{2{\scriptscriptstyle N}} = z_{2{\scriptscriptstyle N}+1}$ and so
$h_{\scriptscriptstyle N}(s) = (s + z_{2{\scriptscriptstyle
N}})^2\cdot f_{\scriptscriptstyle N}(s)$ where
$f_{\scriptscriptstyle N}(s) = \prod_{j=0}^{2{\scriptscriptstyle
N}-1} (s + z_j)$ and $h_{\scriptscriptstyle
N}'(-z_{2{\scriptscriptstyle N}}) = 0$, then it is needed to
consider this primed sum split into two parts such as
$\sum_{j=0}^{2{\scriptscriptstyle N}-1}\,[\cdots]\,+\,\Lambda(2N,
2N+1)$, where $h_{\scriptscriptstyle N}'(-z_j) = (-z_j +
z_{2{\scriptscriptstyle N}})^2\cdot f_{\scriptscriptstyle N}'(-z_j)
\ne 0$; the extra part denoted by $\Lambda(2N, 2N+1)$ is contributed
solely by the multiple root, $s = -z_{2{\scriptscriptstyle N}}$ [cf.
(\ref{eq:heat-current0812})-(\ref{eq:heat-current08120})]. Fig.
\ref{fig:fig7} plots the typical behaviors of $h_{\scriptscriptstyle
N}(s)$. We also point out that the expression of heat current in
(\ref{eq:heat-current081}), valid for $N \geq 3$, corresponds to the
heat current ${\mathcal J}_{{\mbox{\tiny in}}}^{(2)}$ given in
(\ref{eq:heat-current-case-1-10-1-N_20}); note, however, that the
denominator of each summand is here given by $h_{\scriptscriptstyle
N}'(-z_j)\cdot h_{\scriptscriptstyle N}(z_j)$ whereas it is in form
of $h_1'(-z_j)\cdot h_1(z_j)$ for $N=2$.

To simplify the expression in (\ref{eq:heat-current081}), we
substitute (\ref{eq:heat-current-case-1-11}) into this and then
apply the same technique as for $N = 1, 2$. Then we can
straightforwardly obtain the exact closed expression
\begin{eqnarray}\label{eq:heat-current0811}
    &&{\mathcal J}_{{\mbox{\tiny in}}}^{({\scriptscriptstyle N})}\,
    =\, -\frac{\kappa^2 \gamma_{\mbox{\tiny o}}}{M^2}\,
    \left(\mbox{\={I}}_{\scriptscriptstyle N}\right)\cdot\left(\frac{1}{\beta_1} -
    \frac{1}{\beta_{\scriptscriptstyle N}}\right)\,
    +\, \frac{2\,\hbar \omega_d^4\,\gamma_{\mbox{\tiny o}}^2}{\pi}\, \times\n\\
    &&\left(\frac{\kappa}{M}\right)^{2{\scriptscriptstyle N}-2}\;
    \sideset{}{'}\sum_{j=0}^{2N+1}\, \frac{z_j^3\cdot\{\psi_{{\scriptscriptstyle \beta_1}}(z_j) -
    \psi_{{\scriptscriptstyle \beta_{\scriptscriptstyle N}}}(z_j)\}}{h_{\scriptscriptstyle N}'(-z_j)\cdot h_{\scriptscriptstyle N}(z_j)}\,,
\end{eqnarray}
where
\begin{align}\tag{\ref{eq:heat-current0811}a}\label{eq:heat-current-case-N-15-00}
    \left(\mbox{\={I}}_{\scriptscriptstyle N}\right)\, :=\,
    2\,\omega_d^4\,\gamma_{\mbox{\tiny o}}\,\left(\frac{\kappa}{M}\right)^{2{\scriptscriptstyle N}-4}\;
    \sideset{}{'}\sum_j^{2N+1}\, \frac{-z_j^2}{h_{\scriptscriptstyle N}'(-z_j)\cdot h_{\scriptscriptstyle
    N}(z_j)}\,;
\end{align}
note that this is in the unit of $1/(\mbox{frequency})^4$ as is the
case for $(\mbox{\={I}}_2)$ given in
(\ref{eq:classical_counterpart2}). Here we also employed the sum
rule given by $\sum_j\hspace*{-.25cm}{}^{'}\;
z_j^n/\{h_{\scriptscriptstyle N}'(-z_j)\cdot h_{\scriptscriptstyle
N}(z_j)\} = 0$ for $n$ odd (cf. Appendix \ref{sec:appendix5}). If
$z_{2{\scriptscriptstyle N}} = z_{2{\scriptscriptstyle N}+1}$, then
the heat current ${\mathcal J}_{{\mbox{\tiny
in}}}^{({\scriptscriptstyle N})}$ contains the terms contributed
solely by this multiple root, explicitly given by ${\mathcal
J}_{{\scriptscriptstyle \Lambda}}^{({\scriptscriptstyle N})} =
J_{{\scriptscriptstyle \Lambda}}^{({\scriptscriptstyle N})}(\beta_1)
- J_{{\scriptscriptstyle \Lambda}}^{({\scriptscriptstyle
N})}(\beta_{{\scriptscriptstyle N}})$, where
\begin{align}\tag{\ref{eq:heat-current0811}b}\label{eq:heat-current0812}
    \hspace*{-.0cm}J_{{\scriptscriptstyle \Lambda}}^{({\scriptscriptstyle N})}(\beta_{\mu})\, =\,
    \frac{\hbar\,\omega_d^4\, \gamma_{\mbox{\tiny o}}^2\, (\kappa/M)^{2{\scriptscriptstyle N}-2}}{2
    \pi\cdot f_{\scriptscriptstyle N}(-z_{2{\scriptscriptstyle N}})\cdot f_{\scriptscriptstyle N}(z_{2{\scriptscriptstyle N}})\cdot
    z_{2{\scriptscriptstyle N}}}\, \times\,
    \Xi_{{\scriptscriptstyle \Lambda}}^{({\scriptscriptstyle N})}(\beta_{\mu})\,.
\end{align}
Here $\mu = 1,N$, and $f_{\scriptscriptstyle
N}(-z_{2{\scriptscriptstyle N}}) = h_{\scriptscriptstyle
N}''(-z_{2{\scriptscriptstyle N}})/2$, as well as
\begin{align}
    &\Xi_{{\scriptscriptstyle \Lambda}}^{({\scriptscriptstyle N})}(\beta_{\mu})\; =\; \frac{\omega_{\mu}}{2}\, -\, \frac{z_{2{\scriptscriptstyle
    N}}^2}{\omega_{\mu}}\cdot\psi_{{\scriptscriptstyle \beta_{\mu}}}^{(1)}(z_{2{\scriptscriptstyle
    N}})\, +\n\\
    &\frac{1}{\omega_d}\,\left(1\, +\, \sum_{j=0}^{2N-1} \frac{z_{2{\scriptscriptstyle N}}^2}{z_j^2\, -\, z_{2{\scriptscriptstyle N}}^2}\right)\, \times\tag{\ref{eq:heat-current0811}c}\label{eq:heat-current08120}\\
    &\left[\omega_{\mu}\, (z_{2{\scriptscriptstyle N}}\, -\, \omega_d)\, -\, 2\, z_{2{\scriptscriptstyle
    N}}\, \omega_d\, \{\psi_{{\scriptscriptstyle \beta_{\mu}}}(z_{2{\scriptscriptstyle N}})\, -\, \psi_{{\scriptscriptstyle
    \beta_{\mu}}}(\omega_d)\}\right]\,,\n
\end{align}
where $\omega_{\mu} = 2\pi/\beta_{\mu}\hbar$, and the trigamma
function\\$\psi_{{\scriptscriptstyle
\beta_{\mu}}}^{(1)}(z_{2{\scriptscriptstyle N}}) =
\psi^{(1)}(z_{2{\scriptscriptstyle N}}/\omega_{\mu})$. If one of
$z_j$'s is repeated over than twice, we can straightforwardly
generalize the result given in (\ref{eq:heat-current08120}) with the
help of (\ref{eq:partial_fraction1}) and
(\ref{eq:degeneracy-case1})-(\ref{eq:degeneracy-case2})
\cite{ILK13}. It is also worthwhile to mention that the compact
expression given in (\ref{eq:heat-current0811}) can be rewritten in
terms of the input parameters $(\Omega, \kappa, \omega_d,
\gamma_{\mbox{\tiny o}})$ with the aid of all coefficients of
$h_{\scriptscriptstyle N}(s)$ explicitly given in
(\ref{eq:matrix_B_inverse_10}) and their symmetric properties like
given in (\ref{eq:susceptibility-2}) used for $N =1, 2$; the
resultant expression will, however, be highly complicated even for
$N=3$. Eq. (\ref{eq:heat-current0811}) is, in fact, the central
result of this paper.

Next we consider the semiclassical behavior of the heat current
${\mathcal J}_{{\mbox{\tiny in}}}^{({\scriptscriptstyle N})}$. To do
so, we expand the digamma function and its derivative given in
(\ref{eq:heat-current0811})-(\ref{eq:heat-current08120}) [cf.
Appendix \ref{sec:appendix2}], which will, in the semiclassical
limit of $\beta_{\mu}\hbar \to 0$, give rise to
\begin{equation}\label{eq:heat-current-case-N-15-0}
    {\mathcal J}_{{\mbox{\tiny in}}}^{({\scriptscriptstyle N})}\, =\,
    {\mathcal J}_{{\mbox{\tiny cl}}}^{({\scriptscriptstyle N})}(\hbar^0)\, +\,
    {\mathcal J}_{{\mbox{\tiny q2}}}^{({\scriptscriptstyle N})}(\hbar^2)\, +\,
    {\mathcal J}_{{\mbox{\tiny q3}}}^{({\scriptscriptstyle N})}(\hbar^3)\, +\,
    {\mathcal O}(\hbar^4)\,.
\end{equation}
Here the leading term is given by the classical heat current
\begin{equation}\label{eq:semiclassical_N_current}
    {\mathcal J}_{{\mbox{\tiny cl}}}^{({\scriptscriptstyle N})}(\hbar^0)\,
    =\, \frac{\kappa^2 \gamma_{\mbox{\tiny o}}}{M^2}\,\left(\mbox{\={I}}_{\scriptscriptstyle N}\right)\cdot\left(\frac{1}{\beta_1}\,
    -\, \frac{1}{\beta_{\scriptscriptstyle N}}\right)\,.
\end{equation}
If $z_{2{\scriptscriptstyle N}} = z_{2{\scriptscriptstyle N}+1}$,
then this classical value contains ${\mathcal
J}_{{\scriptscriptstyle \Lambda},{\mbox{\tiny
cl}}}^{({\scriptscriptstyle N})} = J_{{\scriptscriptstyle
\Lambda},{\mbox{\tiny cl}}}^{({\scriptscriptstyle N})}(\beta_1) -
J_{{\scriptscriptstyle \Lambda},{\mbox{\tiny
cl}}}^{({\scriptscriptstyle N})}(\beta_{\scriptscriptstyle N})$,
where
\begin{align}
    &J_{{\scriptscriptstyle \Lambda},{\mbox{\tiny cl}}}^{({\scriptscriptstyle N})}(\beta_{\mu})\,
    =\, \frac{-\omega_d^3\, \gamma_{\mbox{\tiny o}}^2\,
    (\kappa/M)^{2{\scriptscriptstyle N}-2}}{2\,\beta_{\mu}\cdot
    f_{\scriptscriptstyle N}(-z_{2{\scriptscriptstyle N}})\cdot
    f_{\scriptscriptstyle N}(z_{2{\scriptscriptstyle N}})\cdot
    z_{2{\scriptscriptstyle N}}}\, \times\n\\
    &\left\{\omega_d\, -\, 6\,(z_{2{\scriptscriptstyle N}}\, -\, \omega_d)\,\left(1\, +\,
    \sum_{j=0}^{2N-1} \frac{z_{2{\scriptscriptstyle N}}^2}{z_j^2\, -\,
    z_{2{\scriptscriptstyle N}}^2}\right)\right\}\,.\tag{\ref{eq:semiclassical_N_current}a}\label{eq:heat-current-classical-N-01}
\end{align}
The first quantum correction is given by
\begin{eqnarray}\label{eq:q-correction_N_current1}
    &&{\mathcal J}_{{\mbox{\tiny q2}}}^{({\scriptscriptstyle
    N})}(\hbar^2)\, =\, \frac{1}{6}\,\hbar^2\,\omega_d^4\, \gamma_{\mbox{\tiny o}}^2\, \left(\frac{\kappa}{M}\right)^{2{\scriptscriptstyle N}-2}\,
    \left(\beta_1\, -\, \beta_{\scriptscriptstyle N}\right)\,
    \times\n\\
    &&\sideset{}{'}\sum_j^{2N+1}\, \frac{z_j^4}{h_{\scriptscriptstyle N}'(-z_j)\cdot h_{\scriptscriptstyle N}(z_j)}\,,
\end{eqnarray}
if necessary, with
\begin{align}
    &J_{{\scriptscriptstyle \Lambda},{\mbox{\tiny q2}}}^{({\scriptscriptstyle N})}(\beta_{\mu})\,
    =\, \frac{-\hbar^2\,\omega_d^4\, \gamma_{\mbox{\tiny o}}^2\,
    (\kappa/M)^{2{\scriptscriptstyle N}-2}\; \beta_{\mu}}{24\, f_{\scriptscriptstyle N}(-z_{2{\scriptscriptstyle N}})\cdot
    f_{\scriptscriptstyle N}(z_{2{\scriptscriptstyle N}})}\,
    \times\n\\
    &\left\{z_{2{\scriptscriptstyle N}}\, +\, 2\,(z_{2{\scriptscriptstyle N}}\, -\, \omega_d)\,\left(1\, +\,
    \sum_{j=0}^{2N-1}\,\frac{z_{2{\scriptscriptstyle N}}^2}{z_j^2\, -\, z_{2{\scriptscriptstyle N}}^2}\right)\right\}\,.\tag{\ref{eq:q-correction_N_current1}a}\label{eq:q-correction_Lambda_N_current1}
\end{align}
The next quantum correction in ${\mathcal O}(\hbar^3)$ is
non-vanishing only if $z_{2{\scriptscriptstyle N}} =
z_{2{\scriptscriptstyle N}+1}$ in such a way that
\begin{eqnarray}\label{eq:q-correction_3_N_current1}
    \hspace*{-.0cm}&&J_{{\scriptscriptstyle \Lambda},{\mbox{\tiny q3}}}^{({\scriptscriptstyle N})}(\beta_{\mu})\, =\, \frac{2\,\zeta(3)\,\hbar^3\,\omega_d^4\, \gamma_{\mbox{\tiny o}}^2\,
    (\kappa/M)^{2{\scriptscriptstyle N}-2}\; \beta_{\mu}^2}{(2\pi)^3\,
    f_{\scriptscriptstyle N}(-z_{2{\scriptscriptstyle N}})\cdot
    f_{\scriptscriptstyle N}(z_{2{\scriptscriptstyle N}})}\,
    \times\n\\
    \hspace*{-.0cm}&&\left\{z_{2{\scriptscriptstyle N}}^2\, +\, (z_{2{\scriptscriptstyle N}}^2\, -\, \omega_d^2)\,\left(1\, +\,
    \sum_{j=0}^{2N-1}\,\frac{z_{2{\scriptscriptstyle N}}^2}{z_j^2\, -\,
    z_{2{\scriptscriptstyle N}}^2}\right)\right\}\,,
\end{eqnarray}
where the symbol $\zeta(n)$ denotes the Riemann zeta function. In
fact, all higher-order quantum corrections in closed form will
straightforwardly come out.

It is now interesting to directly compare the classical result given
in (\ref{eq:semiclassical_N_current}) with Fourier's law of heat
conduction. This allows us to identify the classical heat
conductivity $\kappa_{{\mbox{\tiny F}}}$ as $(N\,\kappa^2
\gamma_{\mbox{\tiny o}}/M^2)\cdot(\mbox{\={I}}_{\scriptscriptstyle
N})$, which depends on chain length $N$ hence violating Fourier's
law already. From the same comparison of the quantum result given in
(\ref{eq:heat-current-case-N-15-0})-(\ref{eq:q-correction_3_N_current1}),
we can easily find the ``effective'' heat conductivity, which
depends even on temperature due to the quantum-correction
contributions. Therefore, we may argue that non-universal behaviors
of the (effective) heat conductivity, especially in low-dimensional
lattices lying in the low-temperature regime (not only the harmonic
chain under our investigation, as briefly stated in Sect.
\ref{sec:introduction}), are ascribed by the non-classical
contributions, as explicitly given in
(\ref{eq:q-correction_N_current1})-(\ref{eq:q-correction_3_N_current1})
for the harmonic chain, which are, in fact, not proportional to $T_1
- T_{{\scriptscriptstyle N}}$ any longer.

\begin{figure}[htb]
\centering\hspace*{-0cm}{\includegraphics[scale=0.5]{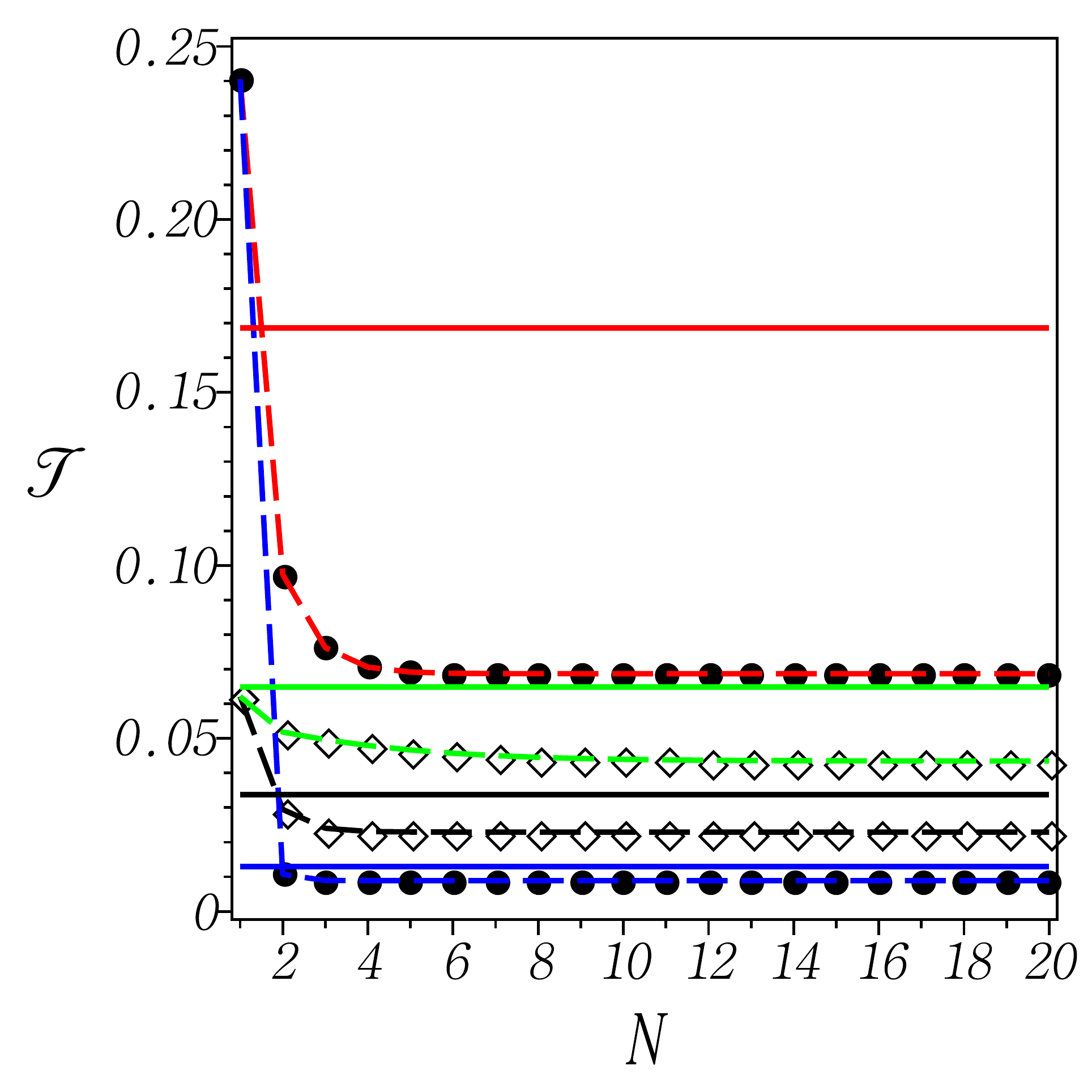}
\caption{(Color online) Heat current ${\mathcal J} = {\mathcal
J}_{{\mbox{\tiny in}}}^{({\scriptscriptstyle
N})}(T_1,T_{\scriptscriptstyle N})$ versus chain length $N$, given
in (\ref{eq:heat-current0811})-(\ref{eq:heat-current08120}), in the
low-temperature regime imposed by $T_1 = 1.1$ and
$T_{\scriptscriptstyle N} = 0.1$. Here we set $\hbar =
k_{\mbox{\tiny B}} = M = \Omega = 1$, and $\omega_d = 10$. Dashed
lines, from top to bottom at $N = 20$, 1st: (red with solid circles:
$\gamma_{\mbox{\tiny o}} = \kappa = 1$); 2nd: (green  with diamonds:
$\gamma_{\mbox{\tiny o}} = 0.2$ and $\kappa = 1$); 3rd: (black with
diamonds: $\gamma_{\mbox{\tiny o}} = \kappa = 0.2$); 4th: (blue with
solid circles: $\gamma_{\mbox{\tiny o}} = 1$ and $\kappa = 0.2$). In
comparison, solid lines are inserted for ${\mathcal J}_{{\mbox{\tiny
B-M}}}^{({\scriptscriptstyle N})}$ given in (\ref{eq:briegel1}),
from top to bottom, in the same order as for the dashed lines. As
demonstrated, this represents a good approximation in the
weak-coupling regime.\label{fig:fig8}}}
\end{figure}

\begin{figure}[htb]
\centering\hspace*{-.7cm}{
\includegraphics[scale=0.5]{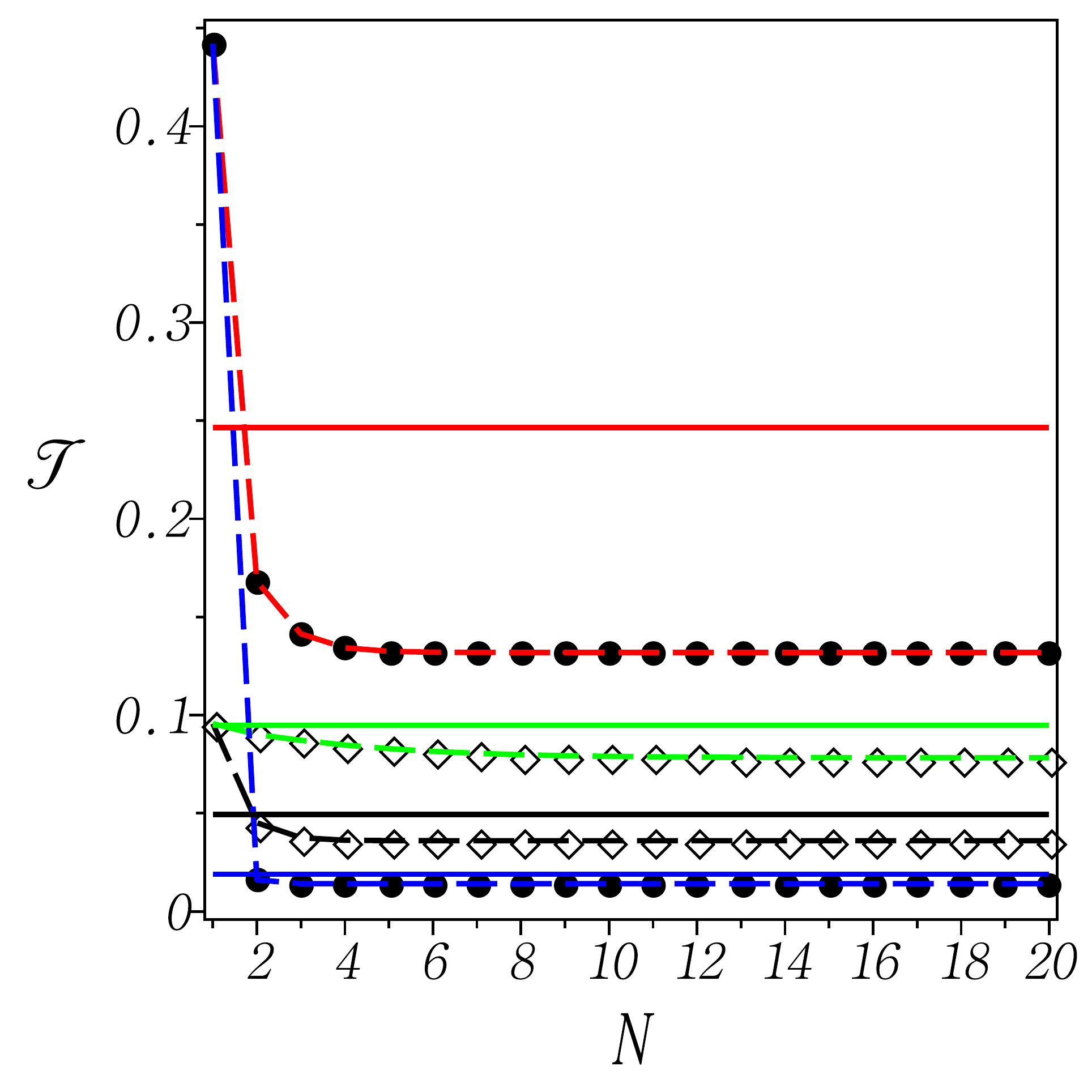}
\caption{(Color online) The same plot as in Fig.~\ref{fig:fig8}, in
the high-temperature regime imposed by $T_1 = 3$ and
$T_{\scriptscriptstyle N} = 2$.\label{fig:fig9}}}
\end{figure}

\begin{figure}[htb]
\centering\hspace*{-0cm}{
\includegraphics[scale=0.5]{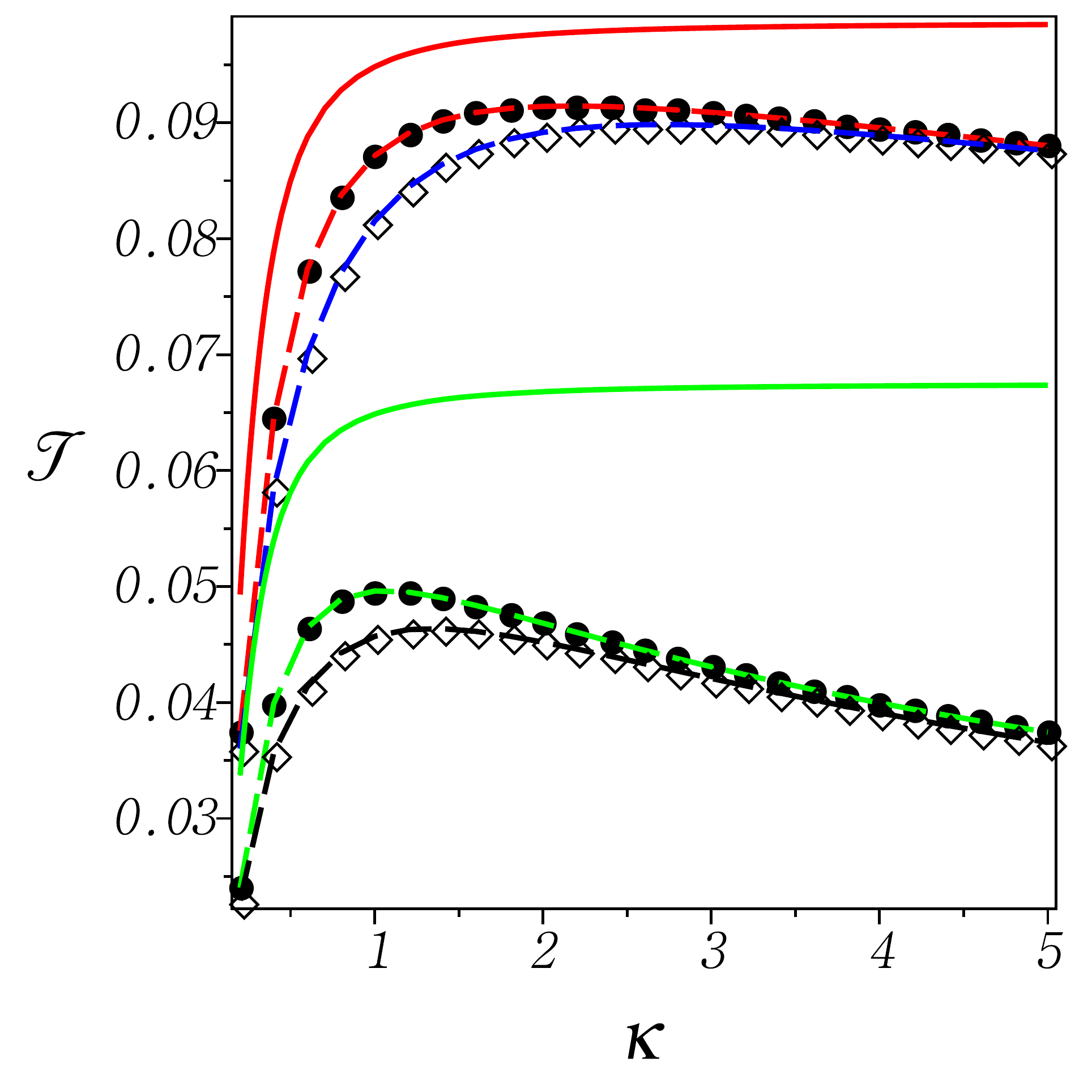}
\caption{(Color online) Heat current ${\mathcal J} = {\mathcal
J}_{{\mbox{\tiny in}}}^{({\scriptscriptstyle
N})}(T_1,T_{\scriptscriptstyle N};\kappa)$ versus intra-coupling
strength $\kappa$, given in
(\ref{eq:heat-current0811})-(\ref{eq:heat-current08120}). Here we
set $\hbar = k_{\mbox{\tiny B}} = M = \Omega = 1$, and $\omega_d =
10$, as well as $\gamma_{\mbox{\tiny o}} = 0.2$. Dashed lines, from
top to bottom at $\kappa = 1$, 1st: (red with solid circles: $N=3$
as well as $T_1 = 3$ and $T_{\scriptscriptstyle N} = 2$, with its
maximum at $\kappa = \kappa_{\mbox{\tiny R}} = 2.2$); 2nd: (blue
with diamonds: $N=6$ as well as $T_1 = 3$ and $T_{\scriptscriptstyle
N} = 2$, with $\kappa_{\mbox{\tiny R}} = 2.8$); 3rd: (green with
solid circles: $N=3$ as well as $T_1 = 1.1$ and
$T_{\scriptscriptstyle N} = 0.1$, with $\kappa_{\mbox{\tiny R}} =
1$); 4th: (black with diamonds: $N=6$ as well as $T_1 = 1.1$ and
$T_{\scriptscriptstyle N} = 0.1$, with $\kappa_{\mbox{\tiny R}} =
1.4$). This shape of the heat current with respect to the
intra-coupling strength is verified to be true for different choices
of the temperature range and all other input parameters. In
comparison, two solid lines are inserted for ${\mathcal
J}_{{\mbox{\tiny B-M}}}^{({\scriptscriptstyle N})}$ given in
(\ref{eq:briegel1}); the red upper for $T_1 = 3$ and
$T_{\scriptscriptstyle N} = 2$, asymptotically approaching
$0.09862324$ with $\kappa \to \infty$, and the green lower for $T_1
= 1.1$ and $T_{\scriptscriptstyle N} = 0.1$, approaching
$0.06746888$ with $\kappa \to \infty$.\label{fig:fig10}}}
\end{figure}

The behaviors of heat current ${\mathcal J}_{{\mbox{\tiny
in}}}^{({\scriptscriptstyle N})}$ versus chain length $N$ are
plotted in Figs. \ref{fig:fig8}-\ref{fig:fig10} for various input
parameters. First, it turns out that in the weak-coupling regime
imposed by $\kappa/M, \gamma_{\mbox{\tiny o}}^2 \ll \Omega^2$, the
low-magnitude heat currents are typically acquired, as expected from
the results for $N = 1,2$. They also reveal the almost
$N$-independent behaviors (for $N \geq 2$). This can be understood
from the forms of the matrix elements ${\mathcal A}_{11}(s)$ and
${\mathcal A}_{12}(s)$ in the weak-coupling limit; with the aid of
$\sin N\phi = \sin\phi\cdot\cos(N-1)\phi +
\cos\phi\cdot\sin(N-1)\phi$, we can exactly rewrite ${\mathcal
A}_{11}(s)$ in (\ref{eq:matrix_B_inverse_1}) as
\begin{equation}\label{eq:note_01}
    \frac{\delta\, -\, c\cdot\cos\phi\,
    -\, c\cdot\sin\phi\cdot\cot(N-1)\phi}{{\mathcal A}_{{\mbox{\tiny den}}}(s)}\,,
\end{equation}
where
\begin{align}
    &{\mathcal A}_{{\mbox{\tiny den}}}(s)\, :=\, \delta^2\, -\, c^2\, +\, 2 c\cdot(c\,\cos\phi\,
    -\, \delta)\cdot\cos\phi\, +\n\\
    &2 c\cdot(c\,\cos\phi\, -\,
    \delta)\cdot\sin\phi\cdot\cot(N-1)\phi\,,\tag{\ref{eq:note_01}a}\label{eq:note_01_1}
\end{align}
as well as $\cot(N-1)\phi = -i \coth(N-1)r$, expressed in terms of
the real number $r = -i \phi > 0$, with
\begin{align}\tag{\ref{eq:note_01}b}\label{eq:note_02}
    \coth (N-1)r\, =\, \frac{2}{1\, -\, (e^{-r})^{2({\scriptscriptstyle N}-1)}}\, -\, 1\,.
\end{align}
In the weak-coupling limit leading to $|c| \ll b + 2 c$, Eqs.
(\ref{eq:matrix_B_inverse_3}) and (\ref{eq:matrix_B_inverse_30})
allow us to easily have
\begin{align}
    \hspace*{-.0cm}&e^{-r}\, =\, -\frac{1}{2 c}\,H(s)\, =\tag{\ref{eq:note_01}c}\label{eq:note_021}\\
    \hspace*{-.0cm}&-\frac{c}{s^2\, +\, \Omega^2}\, -\, 2 \left(\frac{c}{s^2\, +\,
    \Omega^2}\right)^2\, +\, {\mathcal O}\left\{\left(\frac{c}{s^2\, +\,
    \Omega^2}\right)^3\right\}\,,\n
\end{align}
which gives rise to $\coth (N-1)r \to 1$ for $N$ large enough (a
fairly good approximation even for $N=3$). Accordingly, the matrix
element ${\mathcal A}_{11}(s)$ reduces to be\\$N$-independent. Along
the same line, we can do the same job for ${\mathcal A}_{12}(s),
{\mathcal A}_{1{\scriptscriptstyle N}}(s)$ and ${\mathcal
A}_{2{\scriptscriptstyle N}}(s)$, respectively. Consequently, the
heat current ${\mathcal J}_{{\mbox{\tiny in}}}^{({\scriptscriptstyle
N})}$ reduces to be\\$N$-independent in this regime. In this
context, it is also worthwhile to mention that this behavior of heat
current is consistent to the result by Asadian {\em et al.} in
\cite{BRI13}, which was obtained from the consideration of a
harmonic chain restricted to the rotating wave approximation of the
isolated chain $\hat{H}_s$ given in (\ref{eq:chain1}) as well as to
the Born-Markovian regime imposed by the weak-coupling and the Ohmic
damping ($\omega_d \to \infty$). Their result for steady-state heat
current can be rewritten in terms of our notation, i.e., with $V \to
\kappa/2$ as well as $\Gamma_1, \Gamma_{\scriptscriptstyle N} \to
\gamma_{\mbox{\tiny o}}$ in their equation (25), as the
$N$-independent expression
\begin{equation}\label{eq:briegel1}
    {\mathcal J}_{{\mbox{\tiny B-M}}}^{({\scriptscriptstyle N})}\, =\,
    \frac{\kappa^2 \gamma_{\mbox{\tiny o}}}{M^2}\,\left(\mbox{\={I}}_{\mbox{\tiny B-M}}\right)\cdot\hbar\,\Omega\,
    \left(\al\hat{n}\ar_{{\scriptscriptstyle \beta_1}}\, -\, \al\hat{n}\ar_{\scriptscriptstyle \beta_{\scriptscriptstyle N}}\right)\,,
\end{equation}
where $(\mbox{\={I}}_{\mbox{\tiny B-M}}) := (1/2)\,\{(\kappa/M)^2\,
+\, (\Omega\,\gamma_{\mbox{\tiny o}})^2\}^{-1}$, and the average
excitation number $\al\hat{n}\ar_{{\scriptscriptstyle \beta_{\mu}}}
= 1/(e^{{\scriptscriptstyle \beta_{\mu}} \hbar \Omega}\, -\, 1)$. In
Figs. \ref{fig:fig8}-\ref{fig:fig10}, this approximation is compared
with our exact result denoted by ${\mathcal J}_{{\mbox{\tiny
in}}}^{({\scriptscriptstyle N})}$. It is then shown that this may be
a good approximation in the weak-coupling regime, as expected,
whereas it is not case beyond the weak-coupling regime.

Next we pay attention to the above behavior of heat current beyond
the weak-coupling regime, which has so far not been systematically
explored. As demonstrated in the figures, the heat current increases
with increase of the intra-coupling strength $\kappa$ for a given
chain-bath coupling strength characterized by the imposed damping
parameter $\gamma_{\mbox{\tiny o}}$, and reaches its maximum value
at some specific coupling strength $\kappa_{\mbox{\tiny R}}$
``resonant'' to the chain-bath coupling strength. With further
increase of the intra-coupling strength, the heat current decreases
very slowly, whereas this behavior cannot be found from the
Born-Markovian result given in (\ref{eq:briegel1}). Also, the heat
current typically behaves in such a way that its magnitude is at the
maximum with $N = 1$, and then gradually decreases with increase of
chain length $N$, being in fact almost $N$-independent in the range
of $N$ large enough. This may already be qualitatively underwood
from the behavior of $\coth (N-1)r$ given in (\ref{eq:note_01}) with
respect to $N$. As a result, Fourier's law proves violated also in
this regime.

\section{Conclusion}\label{sec:conclusion}
In summary, we derived an exact closed expression of the
steady-state heat current through a chain of quantum Brownian
oscillators coupled to two separate baths. It was obtained, without
any approximation indeed, for arbitrary coupling strengths both in
the intra-couplings between two nearest-neighboring chain elements
and in the chain-baths couplings, as well as in the Drude-Ullersma
damping model in order to look at the behavior of this heat current
beyond the Born-Markovian regime. Then we systematically observed
that in the weak-coupling regime, the heat current with its
low-magnitude simply reduces to be almost independent of the chain
length while in the regime beyond the weak-coupling, the magnitude
of heat current can be raised up by appropriate manipulation of the
chain-baths coupling strengths and the intra-chain coupling
strengths as control parameters; in fact, the largest currents
result if both couplings are ``resonant'' in a sense of comparable
strengths (cf. Figs. \ref{fig:fig8}-\ref{fig:fig9}).

As a result, this rigorous study carried out from the fundamental
side contains the previous results of heat transport in the same
type of harmonic chains, as cited in Sect. \ref{sec:introduction},
as the corresponding limiting cases. By doing so, we could also
explore the relevance between the coupling strengths as input
parameters and the magnitude of the output heat-current, which may
be considered a fundamental issue for building a quantum
thermodynamic engine with high power. We believe that our finding
will provide a useful starting point for the analytical approach to
the steady-state heat current through a more general type of quantum
Brownian chains beyond the weak-coupling regime.

\begin{acknowledgements}
The author thanks G. Mahler\\(Stuttgart), G.J. Iafrate (NC State),
and J. Kim (KIAS) for helpful remarks.
\end{acknowledgements}

\appendix\section{Derivation of correlation function in Eq. (\ref{eq:identity4})}
\label{sec:appendix1}
%
To derive a closed form of the correlation function in
(\ref{eq:correlation-function1}), we first employ the identity
\begin{equation}\label{eq:hyperbolic-cotangent1}
    \coth\left(\frac{\beta \hbar \omega}{2}\right)\, =\, \frac{2}{\beta \hbar
    \omega} \left(1 + 2 \sum_{n=1}^{\infty} \frac{\omega^2}{\omega^2 + \nu_n^2}\right)\,,
\end{equation}
where the so-called Matsubara frequencies $\nu_n = 2\pi n/(\beta
\hbar)$ \cite{WEI08}. Substituting (\ref{eq:hyperbolic-cotangent1})
into (\ref{eq:correlation-function1}) and then applying the integral
identities \cite{GRA07}
\begin{eqnarray}\label{eq:identity1}
    \int_0^{\infty} dy\,\frac{\cos(ay)}{y^2 + b^2} &=& \frac{\pi}{2}\,
    \frac{e^{-ab}}{b}\;,\n\\
    \int_0^{\infty} \frac{dy\,\cos(ay)}{(y^2 + b^2)\,(y^2 +
    c^2)} &=& \frac{\pi}{2}\, \frac{b\,e^{-ac} - c\,e^{-ab}}{b c\,(b^2 - c^2)}\,,
\end{eqnarray}
we can straightforwardly obtain
\begin{eqnarray}\label{eq:correlation-function2}
    && \int_0^{\infty} d\omega\,\frac{\omega}{\omega^2 + \omega_d^2}\,
    \coth\left(\frac{\beta \hbar \omega}{2}\right)\, \cos\{\omega(t-t')\}\\
    &=& \frac{\pi}{\beta \hbar \omega_d}\,
    e^{-\omega_d\,|t-t'|}\,+\,\frac{2\pi}{\beta \hbar} \sum_{n=1}^{\infty}
    \frac{\nu_n\,e^{-\nu_n\,|t-t'|} - \omega_d\,e^{-\omega_d\,|t-t'|}}{\nu_n^2 -
    \omega_d^2}\,.\n
\end{eqnarray}
We next apply to this expression both sum rules
\begin{equation}\label{eq:identity2}
    \sum_{n=1}^{\infty} \frac{1}{n^2 - y^2}\, =\, \frac{1 - \pi y\,\cot(\pi y)}{2 y^2}
\end{equation}
and
\begin{equation}\label{eq:identity3}
    \sum_{n=1}^{\infty} \frac{2 n\,e^{-a n}}{n^2 - y^2}\, =\, \Phi(e^{-a}, 1, y) + \Phi(e^{-a}, 1, -y)
\end{equation}
expressed in terms of the Lerch function \cite{GRA07}
\begin{equation}\label{eq:lerch-fkt1}
    \Phi(z,s,v)\, =\, \sum_{n=0}^{\infty} \frac{z^n}{(n + v)^s}\,,
\end{equation}
which finally gives rise to the closed expression in
(\ref{eq:identity4}).

\section{Derivation of Eqs. (\ref{eq:heat-current-case-1-10})-(\ref{eq:classical_counterpart1})}\label{sec:appendix2}
%
We first substitute (\ref{eq:fourier-laplace5}) into
(\ref{eq:heat-current-case-1-3}), which immediately yields
\begin{eqnarray}\label{eq:heat-current-case-1-4}
    &&\left\al\left\{\hat{P}_1(t)\,,\,\xi_{b_1}(t)\right\}_+\right\ar^{({\mbox{\tiny ss}})}\,
    =\n\\
    &&2 M \lim_{t \to \infty} \int_0^t d\tau\,\left\{\frac{\partial}{\partial
    t}\chi_d(t-\tau)\right\}\,K_1^{(d)}(t-\tau)\,.
\end{eqnarray}
Here the bath correlation function is explicitly given by
\begin{eqnarray}\label{eq:correlation-fkt-appendix1}
    K_{\mu}^{(d)}(t-\tau) &=& \frac{\hbar \omega_d^2\,M
    \gamma_{\mbox{\tiny o}}}{2\pi}\,\left\{\pi\,\cot\left(\frac{\beta_{\mu} \hbar \omega_d}{2}\right)\,
    e^{-\omega_d\,|t-\tau|}\, +\right.\n\\
    && \left.\sum_{n=0}^{\infty}\,\frac{2\,n\cdot e^{-n\,\omega_{\mu}\,|t-\tau|}}{(n + \omega_d/\omega_{\mu})\cdot(n - \omega_d/\omega_{\mu})}\right\}
\end{eqnarray}
[cf. (\ref{eq:lerch-fkt1})]. We can easily evaluate the integral in
(\ref{eq:heat-current-case-1-4}) explicitly, which leads to
\begin{equation}\label{eq:heat-current-case-1-5}
    \hspace*{-.2cm}\left\al\left\{\hat{P}_1(t)\,,\,\xi_{b_1}(t)\right\}_+\right\ar^{({\mbox{\tiny ss}})}\, =\,
    \frac{\hbar \omega_d^2\,\gamma_{\mbox{\tiny o}}\,M\cdot {\mathcal Y}_{{\scriptscriptstyle \beta_1}}}{\pi\, (z_0 - z_1) (z_1 - z_2) (z_2 - z_0)}\,.
\end{equation}
Here
\begin{eqnarray}\label{eq:heat-current-case-1-6}
    {\mathcal Y}_{{\scriptscriptstyle \beta_1}} &=& \pi\,\cot\left(\frac{\beta_1 \hbar \omega_d}{2}\right)\cdot Y(\omega_d)\,
    +\n\\
    && \sum_{n=0}^{\infty} \left(\frac{1}{n + \omega_d/\omega_1}\, +\, \frac{1}{n - \omega_d/\omega_1}\right)\cdot
    Y\left(n\,\omega_1\right)\,,
\end{eqnarray}
where
\begin{equation}\label{eq:heat-current-case-1-7}
    Y(\omega)\, =\, \sum_{\underline{j}=0}^2
    \frac{(z_{\underline{j+1}}^2 - z_{\underline{j+2}}^2)\,z_{\underline{j}}}{\omega + z_{\underline{j}}}\,.
\end{equation}
By means of the identity of the digamma function \cite{ABS74}
\begin{equation}\label{eq:identity_digamma1}
    \sum_{n=0}^{\infty} \frac{1}{(n + a)\,(n + b)}\, =\, \frac{\psi(a) - \psi(b)}{a - b}\,,
\end{equation}
we can easily rewrite the summation in
(\ref{eq:heat-current-case-1-6}) as
\begin{equation}\label{eq:heat-current-case-1-8}
    -\left[Y(\omega_d)\cdot\sum_{\underline{j}=0}^2 \left\{\psi\left(-\frac{\beta_1 \hbar \omega_d}{2\pi}\right) -
    \psi\left(\frac{\beta_1 \hbar z_{\underline{j}}}{2\pi}\right)\right\}\right]\, -\, \left[\omega_d \to
    -\omega_d\right]\,.
\end{equation}
Substituting now the expression in (\ref{eq:heat-current-case-1-6})
into (\ref{eq:heat-current-case-1-5}) and subsequently into
(\ref{eq:heat-current-case-1-3}), we can finally arrive at the
result given in (\ref{eq:heat-current-case-1-10}).

Next let us derive the expression of heat current given in
(\ref{eq:heat-current-case-1-15}) in terms of the input parameters
$(\Omega, \omega_d, \gamma_{\mbox{\tiny o}})$ only, expanded in the
semiclassical limit. To do so, we first plug into
(\ref{eq:heat-current-case-1-10}) the expansions given by\\$\cot(y)
= \sum_{n=0}^{\infty}\,(-1)^n\,\{2^{2n}/(2n)!\}\,B_{2n}\,y^{2n-1}$
for $0 < |y| < \pi$ and $\psi(y) = -1/y - \gamma_e +
\sum_{n=1}^{\infty}\,(-1)^{n+1}\,\zeta(n+1)\,y^n$; here the
Bernoulli numbers $B_{2n}$, the Euler constant $\gamma_e =
0.5772\cdots$, and the Riemann zeta function $\zeta(n+1)$, as well
as $B_{2n} = 2\,(-1)^{n-1} \{(2n)!/(2\pi)^{2n}\}\,\zeta(2n)$
\cite{GRA07}. After some steps of algebraic manipulation, this gives
rise to
\begin{eqnarray}\label{eq:heat-current-case-1-12}
    {\mathcal J}_{{\mbox{\tiny in}}}^{(1)} &=& {\mathcal J}_{{\mbox{\tiny cl}}}^{(1)}(\hbar^0)\, +\, {\mathcal J}_{{\mbox{\tiny q1}}}^{(1)}(\hbar^1)\, +\,
    {\mathcal J}_{{\mbox{\tiny q2}}}^{(1)}(\hbar^2)\, +\, {\mathcal J}_{{\mbox{\tiny q3}}}^{(1)}(\hbar^3)\,
    +\n\\
    && {\mathcal J}_{{\mbox{\tiny q4}}}^{(1)}(\hbar^4)\, +\, {\mathcal J}_{{\mbox{\tiny q5}}}^{(1)}(\hbar^5)\, +\, {\mathcal
    O}(\hbar^6)\,.
\end{eqnarray}
Here we have the leading term
\begin{eqnarray}\label{eq:heat-current-case-1-15_appendix}
    {\mathcal J}_{{\mbox{\tiny cl}}}^{(1)}(\hbar^0) &=& \frac{\gamma_{\mbox{\tiny o}}\,\omega_d}{2} \left(\frac{1}{\beta_1} -
    \frac{1}{\beta_{1'}}\right)\, \times\n\\
    && \sum_{\underline{j}=0}^2 \frac{z_{\underline{j}}\,(z_{\underline{j}} - \omega_d)}{(z_{\underline{j}} -
    z_{\underline{j+1}}) (z_{\underline{j}} -
    z_{\underline{j+2}}) (z_{\underline{j}} + \omega_d)}\,.
\end{eqnarray}
To evaluate this summation explicitly, we take into account the
technique of partial fraction for two polynomials $P(s)$ and $Q(s)$
with $\mbox{deg}\,Q(s) < \mbox{deg}\,P(s) = n$, where $P(s) = (s -
b_1) (s - b_2) \cdots (s - b_n)$ with $b_j \ne b_k$ for $j \ne k$.
This is explicitly given by \cite{COH07}
\begin{equation}\label{eq:partial_fraction1}
    \bar{f}(s)\; :=\; \frac{Q(s)}{P(s)}\; =\; \sum_{\nu=1}^n \frac{Q(b_{\nu})}{P'(b_{\nu})\cdot(s -
    b_{\nu})}\,.
\end{equation}
Applying this relation, the summation in
(\ref{eq:heat-current-case-1-15_appendix}) easily reduces to
\begin{equation}\label{eq:summation1}
    \left.\frac{s\,\left(s + \omega_d\right)}{\left(s +
    z_0\right) \left(s + z_1\right) \left(s + z_2\right)}\right|_{s \to
    \omega_d}\,,
\end{equation}
which allows us to have the classical result in
(\ref{eq:classical_counterpart1}). Here we also used the relations
in (\ref{eq:susceptibility-2}). Next, we consider the quantum
corrections
\begin{eqnarray}\label{eq:summation2}
    {\mathcal J}_{{\mbox{\tiny q2}}}^{(1)}(\hbar^2) &=& \frac{\hbar^2 \omega_d^2\,\gamma_{\mbox{\tiny o}}}{24}\,\left(\beta_1 -
    \beta_{1'}\right)\, \times\\
    && \sum_{\underline{j}=0}^2 \frac{z_{\underline{j}}^2\,(z_{\underline{j}} - \omega_d)}{(z_{\underline{j}} -
    z_{\underline{j+1}}) (z_{\underline{j}} -
    z_{\underline{j+2}}) (z_{\underline{j}} +
    \omega_d)}\n\\
    {\mathcal J}_{{\mbox{\tiny q4}}}^{(1)}(\hbar^4) &=& \frac{\hbar^4 \omega_d^2\,\gamma_{\mbox{\tiny o}}}{2^5\cdot 3^2\cdot 5}\,\left(\beta_1^3 -
    \beta_{1'}^3\right)\, \times\\
    && \sum_{\underline{j}=0}^2 \frac{z_{\underline{j}}^2\,(z_{\underline{j}}^3 - \omega_d^3)}{(z_{\underline{j}} -
    z_{\underline{j+1}}) (z_{\underline{j}} -
    z_{\underline{j+2}}) (z_{\underline{j}} + \omega_d)}\,.\n
\end{eqnarray}
Here we used $\zeta(2) = \pi^2/6$ and $\zeta(4) = \pi^4/90$.
Applying again (\ref{eq:partial_fraction1}) to these two summations
and then evaluating them at $s = \omega_d$, respectively, we can
finally arrive at the result in (\ref{eq:heat-current-case-1-15}).
In fact, every quantum correction with the odd-degree $\hbar$-power
in (\ref{eq:heat-current-case-1-12}) is shown to vanish indeed by
applying the same technique. Along the same line, we can also derive
the expression of heat current given in (\ref{eq:stefan-B-3}), valid
in the low-temperature limit, by plugging into
(\ref{eq:heat-current-case-1-10-1}) the asymptotic expansion given
by $\psi(y) = \ln y - 1/2y -
\sum_{n=1}^{\infty}\,(B_{2n}/2n)/y^{2n}$ \cite{ABS74}.

Finally we point out that if one of the roots $b_{\nu}$ is repeated
$m$ times in (\ref{eq:partial_fraction1}), then the expansion for
$\bar{f}(s)$ contains the terms of form
\begin{equation}\label{eq:degeneracy-case1}
    \frac{\lambda_1}{s - b_{\nu}}\, +\, \frac{\lambda_2}{(s - b_{\nu})^2}\,
    +\, \cdots\, +\, \frac{\lambda_m}{(s - b_{\nu})^m}\,,
\end{equation}
where
\begin{equation}\label{eq:degeneracy-case2}
    \lambda_{m-r}\, =\, \lim_{s \to b_{\nu}}\,
    \left[\frac{1}{r!}\, \left(\frac{d}{ds}\right)^r\left\{(s -
    b_{\nu})^m\cdot\bar{f}(s)\right\}\right]\,.
\end{equation}
This will be used in Sect. \ref{sec:heat-current5}.

\section{Evaluation of Eqs. (\ref{eq:correlation-1-J=2-1})-(\ref{eq:heat-current-case-1-10-1-N_20})}\label{sec:appendix3}
%
We consider the double integral given in
(\ref{eq:correlation-1-J=2-1})
\begin{equation}\label{eq:correlation-N_2-appendix}
    (\mbox{I}_2)\, :=\, \lim_{t \to \infty}\,\int_0^t d\tau \int_0^t d\tau'\, \chi_1(t-\tau)\cdot\partial_t\{\chi_2(t-\tau')\}\cdot K_{\mu}^{(d)}(\tau-\tau')\,.
\end{equation}
We substitute (\ref{eq:response_fkt_drude1}) and
(\ref{eq:correlation-fkt-appendix1}) into this. In doing so, let
$\chi_d(t) \to \chi_1(t)$ expressed in terms of $\bar{z}_{1,j}$'s,
and $\chi_d(t) \to \chi_2(t)$ expressed in terms of
$\bar{z}_{2,j}$'s. Then it turns out that
\begin{eqnarray}\label{eq:correlation-N_2-appendix0}
    &&(\mbox{I}_2) = -\frac{\hbar \omega_d^2\,\gamma_{\mbox{\tiny o}}}{2\pi
    M}\, \times\n\\
    &&\frac{\sum_{\underline{j},\underline{k}=0}^2\, (\bar{z}_{1,\underline{j+1}}^2 -
    \bar{z}_{1,\underline{{j+2}}}^2)\,
    (\bar{z}_{2,\underline{k+1}}^2 - \bar{z}_{2,\underline{{k+2}}}^2)\, \bar{z}_{2,\underline{{k}}}}{\{(\bar{z}_{1,0} - \bar{z}_{1,1})\,
    (\bar{z}_{1,1} - \bar{z}_{1,2})\,
    (\bar{z}_{1,2} - \bar{z}_{1,0})\}\cdot\{(\bar{z}_{1,l} \to \bar{z}_{2,l})\}}\, \times\n\\
    &&\left\{\pi\,\cot\left(\frac{\beta_{\mu} \hbar \omega_d}{2}\right)\cdot(\mbox{I}_{2,1})_{\omega_d}\,
    +\right.\n\\
    &&\left.\sum_{n=0}^{\infty}\,\frac{2\,n}{(n + \omega_d/\omega_{\mu})\cdot(n - \omega_d/\omega_{\mu})}\cdot
    (\mbox{I}_{2,1})_{n \omega_{\mu}}\right\}\,,
\end{eqnarray}
where $l = 0, 1, 2$, and
\begin{eqnarray}\label{eq:identity_appendix3}
    (\mbox{I}_{2,1})_{\alpha} &:=& \lim_{t \to \infty}\,\int_0^t d\tau\,e^{-\bar{z}_{1,j}\,(t-\tau)} \int_0^t
    d\tau'\,e^{-\bar{z}_{2,k}\,(t-\tau')}\,e^{-\alpha\,|\tau-\tau'|}\n\\
    &=& \frac{1}{\bar{z}_{1,j} +
    \bar{z}_{2,k}} \left(\frac{1}{\bar{z}_{1,j} + \alpha}\, +\, \frac{1}{\bar{z}_{2,k} + \alpha}\right)\,.
\end{eqnarray}
Here we also used the integral identity given in
(\ref{eq:fourier-laplace7}).

Next we consider
\begin{equation}\label{eq:correlation-N_2-appendix3}
    (\mbox{II}_2)\, :=\, \lim_{t \to \infty}\,\int_0^t d\tau \int_0^t d\tau'\,
    \partial_t\{\chi_1(t-\tau)\}\cdot\chi_2(t-\tau')\cdot K_{\mu}^{(d)}(\tau-\tau')\,.
\end{equation}
Along the same line, this reduces to the expression given in
(\ref{eq:correlation-N_2-appendix0}) but with exchange of
$\bar{z}_{1,\underline{{j}}}$ and $\bar{z}_{2,\underline{{k}}}$.
Noting $\chi_1(0) = \chi_2(0) = 0$ from
(\ref{eq:response_fkt_drude1}), we can first find that $(\mbox{I}_2)
+ (\mbox{II}_2)$ vanishes indeed, and so does Eq.
(\ref{eq:correlation-1-J=2-1}). Next, $(\mbox{I}_2) - (\mbox{II}_2)$
gives rise to an explicit evaluation of the integral in
(\ref{eq:correlation-3-J=2-1}) and subsequently the exact result in
(\ref{eq:heat-current-case-1-10-1-N_20}) expressed in terms of
$\{z_{\underline{j}}|\,z_{\underline{j}} =
\bar{z}_{1,\underline{j}}\}$ and
$\{z_{\underline{j}}'|\,z_{\underline{j}}' =
\bar{z}_{2,\underline{j}}\}$.

\section{Mathematical supplements for Eq. (\ref{eq:A_11_12_1J_2J-1})}\label{sec:appendix4}
%
First, let us acquire an explicit expression of $\mbox{den}(s;N)$
which leads to the polynomial $h_{\scriptscriptstyle N}(s) =
\mbox{den}(s;N) \cdot (s + \omega_d)^2$. To do so, we rewrite Eq.
(\ref{eq:matrix_B_inverse_3}) as $F_{\scriptscriptstyle N}(s) =
F_1(s)\cdot T_{\scriptscriptstyle N}(s)$, where
\begin{eqnarray}\label{eq:matrix_B_inverse_7}
    T_{\scriptscriptstyle N}(s) &=& \sum_{\nu=0}^{N-1}\, \{G(s)\}^{{\scriptscriptstyle N}-1-\nu}\cdot\{H(s)\}^{\nu}\n\\
    &=& \sum_{j=0}^{\infty}\,\binom{N}{2j+1}\cdot b^{{\scriptscriptstyle N}-2j-1}\cdot
    \left(b^2 - 4\,c^2\right)^j\,.
\end{eqnarray}
With the aid of
(\ref{eq:matrix_B_inverse_31})-(\ref{eq:matrix_B_inverse_32}), we
can then find that
\begin{subequations}
\begin{eqnarray}
    \hspace*{-.5cm}&&{\mathcal A}_{11}(s) = \frac{2^{1-{\scriptscriptstyle N}}}{\mbox{den}(s;N)}\, \left[T_{\scriptscriptstyle N}(s)\, +\,
    2\,\{s\,\bar{\gamma}(s) + c\}\cdot T_{{\scriptscriptstyle N}-1}(s)\right]\label{eq:matrix_B_inverse_6}\\
    \hspace*{-.5cm}&&{\mathcal A}_{12}(s) = \frac{2^{2-{\scriptscriptstyle N}}\,(-c)}{\mbox{den}(s;N)}\,
    \left[T_{{\scriptscriptstyle N}-1}(s)\, +\, 2\,\{s\,\bar{\gamma}(s) + c\}\cdot
    T_{{\scriptscriptstyle N}-2}(s)\right]\\
    \hspace*{-.5cm}&&{\mathcal A}_{1{\scriptscriptstyle N}}(s) =
    \frac{(-c)^{{\scriptscriptstyle N}-1}}{\mbox{den}(s;N)}\; ,\;
    {\mathcal A}_{2{\scriptscriptstyle N}}(s) =
    \frac{a\, (-c)^{{\scriptscriptstyle N}-2}}{\mbox{den}(s;N)}\,.\label{eq:matrix_B_inverse_61}
\end{eqnarray}
\end{subequations}
Substituting (\ref{eq:matrix_B_inverse_7}) into
(\ref{eq:matrix_B_inverse_32}) and then applying Pascal's rule
$\binom{N}{k} = \binom{N-1}{k} + \binom{N-1}{k-1}$, we can finally
arrive, after some algebraic manipulations, at the expression
\begin{eqnarray}\label{eq:matrix_B_inverse_10}
    &&\mbox{den}(s;N)\, =\, 2^{1-{\scriptscriptstyle N}}
    \sum_{j,k,n=0}^{\infty} \binom{N-2j-2}{k}\,
    \left\{\binom{N-1}{2j+1}\, \times\right.\n\\
    &&\left[d^2\, +\, 2\,\{x + (-c)\}\,d\, +\, 2 x^2\right]\, +\,
    \binom{N-1}{2j}\, \times\n\\&&\left.\left[d^2\, +\, 2\,\{x + (-c)\}\,d\, +\, 4\,(-c)\,x\right]\right\}\,
    \binom{j}{n} \times\n\\
    &&2^{k+2n}\cdot d^{{\scriptscriptstyle N}-k-n-2}\cdot (-c)^{k+n}\,,
\end{eqnarray}
where $x(s) := s\,\bar{\gamma}_d(s)$ and $d(s) := s^2 + \Omega^2$.
From this, all coefficients of $\mbox{den}(s;N)$ can exactly be
determined, which are non-negative, as shown; e.g., the highest
$s$-power term is given by $s^{2{\scriptscriptstyle N}+2}/(s +
\omega_d)^2$ and the second highest $s$-power term is
$2\,\omega_d\,s^{2\scriptscriptstyle N+1}/(s + \omega_d)^2$.
Substituting (\ref{eq:matrix_B_inverse_10}) into
(\ref{eq:matrix_B_inverse_6})-(\ref{eq:matrix_B_inverse_61}), we can
get the explicit expressions in (\ref{eq:A_11_12_1J_2J-1}).

Next, we let us prove that $\mbox{Re}(z_j) > 0$ for all $z_j$'s
satisfying $h_{\scriptscriptstyle N}(-z_j) = 0$, which is needed for
applying the final value theorem of the Laplace transform. Due to
the non-negativeness of all coefficients given in
(\ref{eq:matrix_B_inverse_10}), it suffices to prove that
$\mbox{den}(s;N) \ne 0$ for any purely imaginary number $s = ir$,
where $r \in {\mathbb R}$. We assume that $\mbox{den}(ir;N) = 0$,
though. Then, its conjugate number $s = -i r$ should also satisfy
the equality, $\mbox{den}(-ir;N) \stackrel{!}{=} 0$. By applying
these two equality conditions to (\ref{eq:matrix_B_inverse_22})
simultaneously, we can obtain both
\begin{subequations}
\begin{eqnarray}
    &&c\cdot\sin N\phi = \left(c + \frac{r^2\,\gamma_{\mbox{\tiny o}}\,\omega_d}{r^2 +
    \omega_d^2}\right)\cdot\sin(N-1)\phi\label{eq:matrix_B_inverse_221}\\
    &&c\, \left(c + \frac{r^2\,\gamma_{\mbox{\tiny o}}\,\omega_d}{r^2 +
    \omega_d^2}\right)\cdot\sin(N+1)\phi\n\\
    &=& \delta\, \left\{2\,\left(c +
    \frac{r^2\,\gamma_{\mbox{\tiny o}}\,\omega_d}{r^2 +
    \omega_d^2}\right) - \delta\right\}\cdot\sin N\phi\,.\label{eq:matrix_B_inverse_222}
\end{eqnarray}
\end{subequations}
From this, we notice that $\sin N\phi \ne 0$ if $\sin\phi \ne 0$.
Combining (\ref{eq:matrix_B_inverse_221}) and
(\ref{eq:matrix_B_inverse_222}) to eliminate the sine functions
therein, we can easily acquire
\begin{equation}
    \{(\gamma_{\mbox{\tiny o}}\,\omega_d + c)^2 +
    c^2\}\,X^2\, +\, \omega_d^2\,\{(\gamma_{\mbox{\tiny o}}\,\omega_d + c)^2 + 3\,c^2\}\,X\, +\,
    2\,c^2\,\omega_d^4\, =\, 0\,,
\end{equation}
where $X := r^2 > 0$. Then we see that each root $X_q$ of this
quadratic equation would be required to meet the condition
$\mbox{Re}(X_q) < 0$, which, however, contradicts itself.
Consequently, we cannot have any numbers $z_j$'s being purely
imaginary. It is, however, nontrivial indeed to extract all
individual roots $(-z_j)$'s of the polynomial $h_{\scriptscriptstyle
N}(s)$ expressed explicitly in terms of the parameters $(\Omega,
\kappa, \omega_d, \gamma_{\mbox{\tiny o}})$, even for $N=3$ when we
need to deal with the $8$th-degree polynomial $h_3(s)$.

\section{Evaluation of Eqs. (\ref{eq:matrix_B_inverse_5})-(\ref{eq:heat-current081})}\label{sec:appendix5}
%
To evaluate the integral in (\ref{eq:matrix_B_inverse_5})
explicitly, we first consider the double integral
\begin{equation}\label{eq:heat-current01}
    (\mbox{I}_{\scriptscriptstyle N})_{\alpha}\, :=\, \lim_{t \to \infty}\,\int_0^t d\tau\, f(t-\tau) \int_0^t d\tau'\, g(t-\tau')\cdot e^{-\alpha\,|\tau-\tau'|}\,,
\end{equation}
where $f(t) = \dot{A}_{11}(t)$ and $g(t) = A_{21}(t)$. By applying
the technique in (\ref{eq:identity_appendix3}) used for $N=2$, we
can transform (\ref{eq:heat-current01}) into
\begin{equation}\label{eq:heat-current031}
    (\mbox{I}_{{\scriptscriptstyle N}})_{\alpha}\, =\, {\mathcal L}\left\{\int_0^t d\tau\, \left[f(t)\cdot g(\tau)\,
    +\, g(t)\cdot f(\tau)\right]\, e^{\alpha \tau}\right\}(\alpha)\,.
\end{equation}
Let $\bar{f}(s) = {\mathcal L}\{f(t)\}(s)$ and $\bar{g}(s) =
{\mathcal L}\{g(t)\}(s)$. Now we consider the product rule, which
reads as ${\mathcal L}\{f_1(t)\,f_2(t)\}(s) = (1/2\pi i)
\int_{c_1-i\infty}^{c_1+i\infty} du\,\bar{f}_1(u)\,\bar{f}_2(s-u)$
\cite{COH07}; here the integration is carried out along the vertical
line, $\mbox{Re}(u) = c_1$ that lies entirely within the region of
convergence of $\bar{f}_1(u)$. Then we can easily rewrite
(\ref{eq:heat-current031}) as
\begin{equation}\label{eq:matrix_B_inverse_11}
    (\mbox{I}_{{\scriptscriptstyle N}})_{\alpha}\, =\,
    \int_{c_1-i\infty}^{c_1+i\infty} \frac{du}{u-\alpha}\,
    \left\{\bar{f}(u)\,\bar{g}(-u)\, +\,
    \bar{g}(u)\,\bar{f}(-u)\right\}\,,
\end{equation}
where $\bar{f}(u) = u\,{\mathcal A}_{11}(u)$ and $\bar{g}(u) =
{\mathcal A}_{21}(u)$. Here we also used ${\mathcal L}\{\int_0^t
d\tau\,g(\tau)\}(s) = \bar{g}(s)/s$. With the aid of
(\ref{eq:matrix_B_inverse_30}), the integrand given by
$\{u\,[{\mathcal A}_{11}(u)\,{\mathcal A}_{21}(-u) - {\mathcal
A}_{21}(u)\,{\mathcal A}_{11}(-u)]\}$ can be transformed into
$\{-u\,[{\mathcal A}_{1{\scriptscriptstyle N}}(u)\,{\mathcal
A}_{2{\scriptscriptstyle N}}(-u) - {\mathcal
A}_{2{\scriptscriptstyle N}}(u)\,{\mathcal A}_{1{\scriptscriptstyle
N}}(-u)]\}$, which immediately allows us to obtain the relation
given in (\ref{eq;matrix_B_inverse_53}).

Therefore we can evaluate the integral in (\ref{eq:heat-current01})
by plugging $f(t) \to \dot{A}_{1{\scriptscriptstyle N}}(t)$ and
$g(t) \to A_{2{\scriptscriptstyle N}}(t)$ giving rise to
$-(\mbox{I}_{\scriptscriptstyle N})_{\alpha}$; in fact,
$A_{1{\scriptscriptstyle N}}(t)$ and $A_{2{\scriptscriptstyle
N}}(t)$ are simpler in form than $A_{11}(t)$ and $A_{21}(t)$,
respectively. First we rewrite the expressions in
(\ref{eq:A_11_12_1J_2J-1}) as
\begin{eqnarray}\label{eq:heat-current03}
    {\mathcal A}_{1{\scriptscriptstyle N}}(s) &=&
    (-c)^{{\scriptscriptstyle N}-1}\,\sideset{}{'}\sum_{j=0}^{2N+1}
    \frac{(-z_j + \omega_d)^2}{h_{\scriptscriptstyle N}'(-z_j)\cdot(s +
    z_j)}\n\\
    {\mathcal A}_{2{\scriptscriptstyle N}}(s) &=&
    (-c)^{{\scriptscriptstyle N}-2}\,\sideset{}{'}\sum_{j=0}^{2N+1} \frac{(-z_j + \omega_d)\cdot\bar{h}_1(-z_j)}{h_{\scriptscriptstyle N}'(-z_j)\cdot (s + z_j)}\,,
\end{eqnarray}
respectively. The meaning of the primed sum denoted by
$\sum_j\hspace*{-.25cm}{}^{'}$ is explicitly given below
(\ref{eq:heat-current081}), which must be treated with care if one
of $z_j$'s is repeated [cf. (\ref{eq:partial_fraction1}) and
(\ref{eq:degeneracy-case1})-(\ref{eq:degeneracy-case2})]. Then it
easily follows that
\begin{subequations}
\begin{eqnarray}\label{eq:heat-current0310}
    f(t) &=& (-c)^{{\scriptscriptstyle N}-1}\, \sideset{}{'}\sum_{j=0}^{2N+1}\, \frac{(-z_j)\cdot(z_j - \omega_d)^2}{h_{\scriptscriptstyle
    N}'(-z_j)}\; e^{-z_j\,t}\\
    g(t) &=& (-c)^{{\scriptscriptstyle N}-2}\,
    \sideset{}{'}\sum_{j=0}^{2N+1}\,
    \frac{(-z_j + \omega_d)\cdot\bar{h}_1(-z_j)}{h_{\scriptscriptstyle N}'(-z_j)}\; e^{-z_j\,t}\,,
\end{eqnarray}
\end{subequations}
each of which, in the $z_k$-degenerate case, contains the terms
resulting from
(\ref{eq:degeneracy-case1})-(\ref{eq:degeneracy-case2}) in such a
way that
\begin{equation}\label{eq:degeneracy-case3}
    {\mathcal L}^{-1}\left\{\frac{\lambda_{m-r}}{(s + z_k)^{m-r}}\right\}(t)\; =\; \lambda_{m-r}\, \frac{t^{m-r-1}}{(m - r -1)!}\,
    e^{-z_k\,t}\,,
\end{equation}
where $r = 0, 1, \cdots, m-1$. We now substitute this result into
(\ref{eq:heat-current01}) and evaluate the double integral
explicitly, which is the same in form as the integral in
(\ref{eq:identity_appendix3}) considered for chain length $N=2$
only. Therefore we can straightforwardly obtain
\begin{eqnarray}\label{eq:integral-appendix-N_1}
    (\mbox{I}_{\scriptscriptstyle N})_{\alpha} &=&
    (-c)^{2{\scriptscriptstyle N}-3}\,
    \sideset{}{'}\sum_{j,k}^{2N+1}\,
    \frac{z_j\cdot(z_j - \omega_d)^3\cdot\bar{h}_1(-z_k)}{h_{\scriptscriptstyle
    N}'(-z_j)\cdot h_{\scriptscriptstyle N}'(-z_k)\cdot (z_j +
    z_k)}\, \times\n\\
    && \left(\frac{1}{z_j + \alpha}\, +\, \frac{1}{z_k +
    \alpha}\right)\,.
\end{eqnarray}
Applying the technique of partial fraction given in
(\ref{eq:partial_fraction1}), this can be simplified as
\begin{equation}\label{eq:integral-appendix-N_2}
    (\mbox{I}_{\scriptscriptstyle N})_{\alpha}\, =\,
    2\,\omega_d^2\, \gamma_{\mbox{\tiny o}}\, (-c)^{2{\scriptscriptstyle N}-3}\,
    \sideset{}{'}\sum_j^{2N+1}\,
    \frac{z_j^2\cdot(z_j^2 - \omega_d^2)}{h_{\scriptscriptstyle
    N}'(-z_j)\cdot h_{\scriptscriptstyle N}(z_j)\cdot (\alpha +
    z_j)}\,.\n
\end{equation}
This allows us to have an explicit evaluation of the integral in
(\ref{eq:matrix_B_inverse_5}) and then that of the heat current
\begin{eqnarray}
    \hspace*{-.3cm}&&{\mathcal J}_{{\mbox{\tiny in}}}^{({\scriptscriptstyle N})} = \frac{\hbar \omega_d^2\,\kappa\,\gamma_{\mbox{\tiny o}}}{2\,M}\,
    \left[\left\{\cot\left(\frac{\hbar \omega_d}{2\,k_{\mbox{\tiny B}} T_1}\right)\, -\, \cot\left(\frac{\hbar \omega_d}{2\,k_{\mbox{\tiny B}}
    T_{\scriptscriptstyle N}}\right)\right\}
    \left(\mbox{I}_{\scriptscriptstyle N}\right)_{\omega_d}\right.\n\\
    \hspace*{-.3cm}&&\left.+ \frac{2}{\pi}\,\sum_{n=0}^{\infty}\,\left\{\frac{n\cdot\left(\mbox{I}_{\scriptscriptstyle N}\right)_{n\,\omega_1}}{\left(n + \frac{\omega_d}{\omega_1}\right)\left(n - \frac{\omega_d}{\omega_1}\right)}\,
    -\, \frac{n\cdot\left(\mbox{I}_{\scriptscriptstyle N}\right)_{n\,\omega_{\scriptscriptstyle N}}}{\left(n + \frac{\omega_d}{\omega_{\scriptscriptstyle N}}\right)\left(n -
    \frac{\omega_d}{\omega_{{\scriptscriptstyle
    N}}}\right)}\right\}\right]\,,\n
\end{eqnarray}
as provided in (\ref{eq:heat-current081}).

Next let us prove the sum rule given by
\begin{equation}\label{eq:sum_rule_N_1}
    \sideset{}{'}\sum_{j=0}^{2N+1}\, \frac{z_j^n}{h_{\scriptscriptstyle N}'(-z_j)\cdot h_{\scriptscriptstyle N}(z_j)}\, =\, 0
\end{equation}
for $n$ odd, which is used for (\ref{eq:heat-current0811}). First we
rewrite $h_{\scriptscriptstyle N}(z_j)$ as
$\prod_{k=0}^{2{\scriptscriptstyle N}+1} (z_j + z_k) =
\prod_{k=0}^{2{\scriptscriptstyle N}+1} (-z_j +
z_{2{\scriptscriptstyle N}+2+k})$, where we introduce
$z_{2{\scriptscriptstyle N}+2+k} := -z_k$. Then it turns out that
$h_{\scriptscriptstyle N}'(-z_j)\cdot h_{\scriptscriptstyle N}(z_j)
= \prod_{k=0}^{4{\scriptscriptstyle N}+3} (-z_j + z_k)$ where $k \ne
j$. Next let $H_{\scriptscriptstyle N}(s) := h_{\scriptscriptstyle
N}(s)\cdot h_{\scriptscriptstyle N}(-s) =
\prod_{k=0}^{4{\scriptscriptstyle N}+3} (s + z_k)$, and we consider
\begin{equation}
    F(s)\; :=\; \frac{s^{n+1}}{H_{\scriptscriptstyle N}(s)}\; =\;
    \sideset{}{'}\sum_{k=0}^{4N+3}\, \frac{(-z_k)^{n+1}}{H_{\scriptscriptstyle N}'(-z_k)\cdot(s + z_k)}\,.
\end{equation}
Then we can easily obtain
\begin{equation}\label{eq:sum_rule_N_2}
    F(0)\; =\; 0\; =\;
    (-1)^{n+1}\, \{1 - (-1)^n\}\, \sideset{}{'}\sum_{j=0}^{2N+1}\, \frac{z_j^n}{h_{\scriptscriptstyle N}'(-z_j)\cdot h_{\scriptscriptstyle
    N}(z_j)}\,,
\end{equation}
which immediately gives rise to the sum rule in
(\ref{eq:sum_rule_N_1}). In case that one of $z_j$'s is repeated, it
is also straightforward to verify this result.

%
%
%
\newpage
\newpage
\newpage
\newpage
\newpage
%
\newpage
\newpage
%
\newpage
\newpage
\newpage
%

\begin{thebibliography}{}
%
\bibitem{MAH04} J. Gemmer, M. Michel and G. Mahler, {\em Quantum
Thermodynamics} (Springer, Berlin, 2004), and references therein.
%
\bibitem{GRO84} S.R. de Groot and P. Mazur, {\em Nonequilibrium Thermodynamics} (Dover, New York,
1984).
%
\bibitem{RIE67} Z. Rieder, J.L. Lebowitz and E. Lieb, J. Math. Phys. {\bf 8}, 1073 (1967).
%
\bibitem{RUB71} R.J. Rubin and W.L. Greer, J. Math. Phys. {\bf 12}, 1686 (1971).
%
\bibitem{LEB71} A. Casher and J.L. Lebowitz, J. Math. Phys. {\bf 12}, 1701 (1971).
%
\bibitem{PRO00} T. Prosen and D.K. Campbell, Phys. Rev. Lett. {\bf 84}, 2857 (2000).
%
\bibitem{LEP03} S. Lepri, R. Livi, and A. Politi, Phys. Rep. {\bf 377}, 1 (2003).
%
\bibitem{POL03} S. Lepri, R. Livi, and A. Politi, Phys. Rev. E {\bf 68}, 067102 (2003).
%
\bibitem{DHA08} A. Dhar, Adv. Phys. {\bf 57}, 457 (2008).
%
\bibitem{BRI13} A. Asadian, D. Manzano, M. Tiersch, and H.J. Briegel, Phys. Rev. E {\bf 87}, 012109
(2013).
%
\bibitem{REG98} L.G.C. Rego and G. Kirczenow, Phys. Rev. Lett. {\bf
81}, 232 (1998).
%
\bibitem{BLE99} M.P. Blencowe, Phys. Rev. B {\bf 59}, 4992 (1999).
%
\bibitem{SAI03} K. Saito, Europhys. Lett. {\bf 61}, 34 (2003).
%
\bibitem{DHA03} A. Dhar and B.S. Shastry, Phys. Rev. B {\bf 67}, 195405 (2003).
%
\bibitem{HAE03} D. Segal, A. Nitzan, and P. H\"{a}nggi, J. Chem. Phys. {\bf 119}, 6840 (2003).
%
\bibitem{MAH05} M. Michel, G. Mahler, and J. Gemmer, Phys. Rev. Lett. {\bf 95}, 180602 (2005).
%
\bibitem{DHA06} A. Dhar and D. Roy, J. Stat. Phys. {\bf 125}, 805
(2006).
%
\bibitem{WAN06} J.-S. Wang, J. Wang, and N. Zeng, Phys. Rev. B {\bf 74}, 033408 (2006).
%
\bibitem{YAM06} T. Yamamoto and K. Watanabe, Phys. Rev. Lett. {\bf 96}, 255503
(2006).
%
\bibitem{GAU07} Ch. Gaul and H. B\"{u}ttner, Phys. Rev. E {\bf 76}, 011111 (2007).
%
\bibitem{DUB09} Y. Dubi and M. Di Ventra, Phys. Rev. E {\bf 79}, 042101 (2009).
%
\bibitem{MAN12} D. Manzano, M. Tiersch, A. Asadian, and H.J. Briegel, Phys. Rev. E {\bf 86}, 061118 (2012).
%
\bibitem{COM13} Some formal expressions of the steady-state heat curent in different types of
quantum harmonic chains were obtained, e.g., using the quantum
Langevin approach \cite{DHA03,HAE03,DHA06} and the Keldysh formalism
\cite{WAN06,YAM06}. However, none of them has systematically and
rigorously treated the heat transport beyond the weak-coupling
regime in the chain-baths couplings as well as the intra-chain
couplings, leading to the exact results in closed form and their
numerical evaluations, which is, in fact, the central subject of the
current paper.
%
\bibitem{ROU05} K.C. Schwab and M.L. Roukes, Phys. Today {\bf 58}, 36 (2005).
%
\bibitem{WEI08} U. Weiss, {\em Quantum Dissipative Systems}, 3rd ed. (World Scientific, Singapore, 2008).
%
\bibitem{MAR88} M. Marcus and H. Minc, {\em Introduction to Linear Algebra} (Dover, New York, 1988).
%
\bibitem{ULL66} P. Ullersma, Physica {\bf 32}, 27, 56, 74, 90 (1966).
%
\bibitem{ING02} G.-L. Ingold, in {\em
Coherent Evolution in Noisy Environments}, edited by A. Buchleitner
and K. Hornberger, Lecture Notes in Physics {\bf 611} (Springer,
Berlin, 2002).
%
\bibitem{ROB66} G.E. Roberts, H. Kaufman, {\em Table of Laplace Transforms} (W.B.
Saunders, Philadelphia, 1966).
%
\bibitem{FOR06} G.W. Ford and R.F. O'Connell, Phys. Rev. Lett. {\bf 96}, 020402 (2006).
%
\bibitem{JAP00} Math. Soc. of Japan, {\em Encyclopedic Dictionary of Mathematics}, 2nd
ed., edited by K. It\^{o} (MIT Press, Cambridge, MA, 2000).
%
\bibitem{COH07} A.M. Cohen, {\em Numerical Methods for Laplace Transform
Inversion} (Springer, New York, 2007).
%
\bibitem{ILK10} I. Kim, Phys. Lett. A {\bf 374}, 3828 (2010).
%
\bibitem{GRA07} I.S. Gradshteyn and I.M. Ryzhik, {\em Table of Integrals, Series,
and Products}, 7th ed. (Academic Press, San Diego, 2007).
%
\bibitem{ABS74} M. Abramowitz and I. Stegun, {\em Handbook of Mathematical Functions
with Formulas, Graphs, and Mathematical Tables} (Dover, New York,
1974).
%
\bibitem{LAN89} P.T. Landsberg and A.D. Vos, J. Phys. A: Math. Gen.
{\bf 22}, 1073 (1989).
%
\bibitem{KIM13} Due to this, the normal
coordinates $\{\underline{\hat{{\mathcal Q}}_j}(s)\}$ in the case of
$N \geq 3$ would place the additional $\tau$-dependency into the
integrand in (\ref{eq:Q_J=2-1})-(\ref{eq:Q_J=2-2}) for an explicit
evaluation of $\lim_{t\to\infty} \hat{Q}_k(t)$, which is, in
general, nontrivial to treat.
%
\bibitem{USM94} R.A. Usmani, Linear Algebra and its Applications {\bf 212/213}, 413 (1994).
%
\bibitem{YUE06} W-Ch. Yueh, Appl. Math. E-Notes {\bf 6}, 74 (2006).
%
\bibitem{ILK13} The numerical anyalsis of the exact expression of $h_{\scriptscriptstyle
N}(s)$ demonstrates that if there is a degeneracy of its roots
$(-z_j)$'s, then typically a single real-valued root is repeated
with the order of degeneracy $m = 2$ as given in Fig.
\ref{fig:fig7}.
%
\end{thebibliography}
\end{document}